\documentclass[
 aip,
 amsmath,amssymb,
 reprint,
]{revtex4-1}
\usepackage[utf8]{inputenc}
\usepackage[T1]{fontenc}
\usepackage{mathptmx}
\usepackage{graphicx}
\usepackage{dcolumn}% Align table columns on decimal point
\usepackage{makecell}
\usepackage{bm}
\usepackage{hyperref}
\hypersetup{colorlinks=true}

\usepackage{xcolor}
\definecolor{tab_green}{RGB}{44, 160, 44}
\definecolor{tab_purple}{RGB}{148, 103, 189}

\begin{document}

\title{Anisotropic properties of monolayer 2D materials: an overview from the C2DB database}

\author{Luca Vannucci}
\email{lucav@dtu.dk}
\thanks{These authors contributed equally}
\affiliation{CAMD, Department of Physics, Technical University of Denmark, 2800 Kongens Lyngby, Denmark}

\author{Urko Petralanda}
\email{upeho@dtu.dk}
\thanks{These authors contributed equally}
\affiliation{CAMD, Department of Physics, Technical University of Denmark, 2800 Kongens Lyngby, Denmark}

\author{Asbjørn Rasmussen}
%\email{asbra@fysik.dtu.dk}
\affiliation{CAMD, Department of Physics, Technical University of Denmark, 2800 Kongens Lyngby, Denmark}
\affiliation{Center for Nanostructured Graphene (CNG), Technical University of Denmark, 2800 Kongens Lyngby, Denmark}

\author{Thomas Olsen}
%\email{tolsen@fysik.dtu.dk}
\affiliation{CAMD, Department of Physics, Technical University of Denmark, 2800 Kongens Lyngby, Denmark}

\author{Kristian S. Thygesen}
\email{thygesen@fysik.dtu.dk}
\affiliation{CAMD, Department of Physics, Technical University of Denmark, 2800 Kongens Lyngby, Denmark}
\affiliation{Center for Nanostructured Graphene (CNG), Technical University of Denmark, 2800 Kongens Lyngby, Denmark}

\date{\today}

\begin{abstract}

We analyze the occurrence of in-plane anisotropy in the electronic, magnetic, elastic and transport properties of more than one thousand 2D materials from the C2DB database. We identify hundreds of anisotropic materials and classify them according to their point group symmetry and degree of anisotropy. A statistical analysis reveals that a lower point group symmetry and a larger amount of different elements in the structure favour all types of anisotropies, which could be relevant for future materials design approaches. Besides, we identify novel compounds, predicted to be easily exfoliable from a parent bulk compound, with anisotropies that largely outscore those of already known 2D materials. Our findings provide a comprehensive reference for future studies of anisotropic response in atomically-thin crystals and point to new previously unexplored materials for the next generation of anisotropic 2D devices.

\end{abstract}

\maketitle

\section{Introduction}

Anisotropy is the characteristic of a material whereby it displays different physical properties along different directions. It is intrinsic to the atomic structure and can therefore influence the electric, magnetic, optical or mechanical response of a material to an external perturbation. In fact, anisotropic materials have become increasingly present in modern devices, finding applications in diverse fields. One paradigmatic example of anisotropc material is an optical polarizer, which is transparent to electromagnetic radiation polarized along a well defined axis, while blocking or deviating light that is polarized along a different direction.

Layered van der Waals (vdW) materials represent an interesting class of naturally anisotropic materials. In vdW materials, the anisotropy derives from the dispersive nature of the bonds between the two-dimensional (2D) atomic layers, which is much weaker than the covalent bonds existing between atoms within the 2D layers. This intrinsic anisotropy can be exploited for various applications. For example, in certain layered materials the interplay between the structural and electronic properties is so strong that it changes the iso-frequency surfaces of light from elliptic to hyperbolic\cite{Gjerding17_hyperbolic} with fascinating perspectives for sub-wavelength imaging and radative emission control\cite{jacob2006optical,lu2014enhancing}. 

Individual 2D atomic layers are obviously anisotropic due to the missing third dimension, but they can exhibit in-plane anisotropy as well. However, the most widely studied 2D materials --- graphene \cite{Novoselov04}, hexagonal boron nitride (hBN) \cite{Ci10_BN} and the family of transition metal dichalcogenides (TMDCs) \cite{Mak10_MoS2} --- have in-plane isotropic properties due their highly symmetric crystal structure. Graphene, for instance, has a six-fold rotational symmetry and three mirror planes, while hBN and TMDCs such as MoS$_2$ have a 6-fold roto-inversion symmetry with two mirror planes.
Such large sets of crystal symmetries turn out to inhibit any form of anisotropic response.

The prototypical example of in-plane anisotropic 2D material is phosphorene \cite{Li14_BP}. Phosphorene is obtained by mechanical exfoliation of black phosphorus down to the monolayer limit, and exhibits a highly-anisotropic puckered structure, which differs along the zigzag and armchair direction (as shown in Fig.\ \ref{fig:overview}). This strong anisotropy has motivated a large number of theoretical and experimental studies of phosphorene, which have revealed the effect of the structural anisotropy on its electronic, optoelectronic, electro-mechanical, thermal, and excitonic properties.\cite{Fei14_phospho, Xia14_phospho, Wei14, Low14, Wang15_phospho, Jain15_phospho, Wang15, Tian16_BP_synapses, Yang17_birefringence, Wang20_optical_anisotropy}.

Other notable examples of in-plane anisotropic 2D materials are TMDCs in the distorted 1T'-phase (such as WTe$_2$ \cite{Ma16, Torun16, Zhang19, Zhang19_WTe2, Gjerding17_hyperbolic, Wang20_WTe2_hyperbolic}), titanium trichalcogenides (most notably TiS$_3$ \cite{Island14, Island15, Kang15, Jin15, SilvaGuillen17, Khatibi19}), ReS$_2$ and ReSe$_2$ \cite{Liu15_ReS2_FET, Zhang16_ReS2, Yang17_birefringence, Echeverry18, Liu20_ReSe2}, GaTe \cite{Wang19_anisotropic_resistance}, and pentagonal structures such as PdSe$_2$ \cite{Oyedele17, Lu20_anisotropy_PdSe2}. Such materials exhibit anisotropy in their mechanical, electrical, optical and magnetic properties, with intriguing applications for optical devices (such as birefringent wave plate or hyperbolic plasmonic surfaces), high mobility transistors, ultra-thin memory devices and controllable magnetic devices among others.

As of today, more than fifty different 2D materials have been identified and synthesized or exfoliated in monolayer form\cite{Haastrup18}, but they represent only a small fraction of all the possible stable 2D materials that have been predicted by computations\cite{Mounet2017, Haastrup18, Ashton17, Choudhary17, Cheon17, Zhou2019_2DMatPedia}. It is therefore reasonable to expect that the above mentioned examples of anisotropic 2D materials will be soon complemented by additional atomically-thin layers with highly direction-dependent properties.

Here we take a first step in this direction by presenting an extensive analysis of the occurrence of in-plane anisotropic features in the magnetic, elastic, transport and optical properties of more than 1000 predicted stable 2D materials from the Computational 2D Materials Database (C2DB)\cite{Haastrup18}.
We discuss trends and similarities in the atomic and electronic structure of anisotropic monolayer materials by classifying them according to their point symmetry group, and highlight the most interesting candidates for different applications.

After introducing the C2DB database and the criteria used to assess the stability of the materials in Section \ref{sec:overview}, we analyze the occurrence of anisotropy in the magnetic easy-axis direction, elastic response, effective masses and polarizability in four separate sub-sections of Section \ref{sec:results}.
We have tried to make these sections as self-contained as possible so they can be read independently, with a separate introduction to the formalism used and the relevant literature for each of them.
We conclude by summarizing the main results in Section \ref{sec:conclusions}, where we highlight the most interesting anisotropic and potentially exfoliable 2D materials identified in the study.

\section{Overview of the C2DB database}
\label{sec:overview}

The Computational 2D Materials Database (C2DB) is an open database containing thermodynamic, elastic, electronic, magnetic, and optical properties of 2D monolayer materials \cite{Haastrup18}.
All properties were calculated with the electronic structure code GPAW \cite{Enkovaara10} using additional software packages for atomic simulation and workflows handling such as ASE \cite{HjorthLarsen17}, ASR \cite{ASR_docs} and MyQueue \cite{Mortensen20}. Unless explicitly stated, all properties reported in this work were calculated with the PBE exchange-correlation functional \cite{Perdew96}.

The latest development version of C2DB contains 4262 fully relaxed structures at the time of writing, which are categorized in terms of their dynamic and thermodynamic stability.
The dynamic stability determines whether a material is stable with respect to distortions of the atomic positions or the unit cell, and is established from phonon frequencies (at the $\Gamma$-point and high-symmetry points of the Brillouin zone boundary) and the stiffness tensor. A material is dynamically stable only if all phonon frequencies are real and the stiffness tensor eigenvalues are positive. 
On the other hand, the thermodynamic stability of a given 2D material is assessed in terms of its heat of formation and total energy with respect to other competing phases (taken as the most stable elemental and binary compounds from the OQMD database\cite{saal2013materials}) --- also known as \textit{energy above the convex hull} \cite{Haastrup18}.
A material's thermodynamic stability is classified as \emph{high} if the heat of formation is negative and the energy above the convex hull is below 0.2 eV/atom.

Materials with high thermodynamic and dynamic stability are the most likely to be exfoliated or synthesized in the lab.
Although these criteria are not sufficient to ensure experimental realization, we note that they have been determined from a detailed analysis of more than 50 already synthesized monolayers\cite{Haastrup18}. We will therefore focus on the subset of highly stable materials (according to the criteria used in the C2DB) in the remainder of this work.
For further details on the stability assessment and a complete overview of the C2DB, we refer the reader to Ref.\ \onlinecite{Haastrup18}.

\begin{figure*}
    \includegraphics[width=\linewidth]{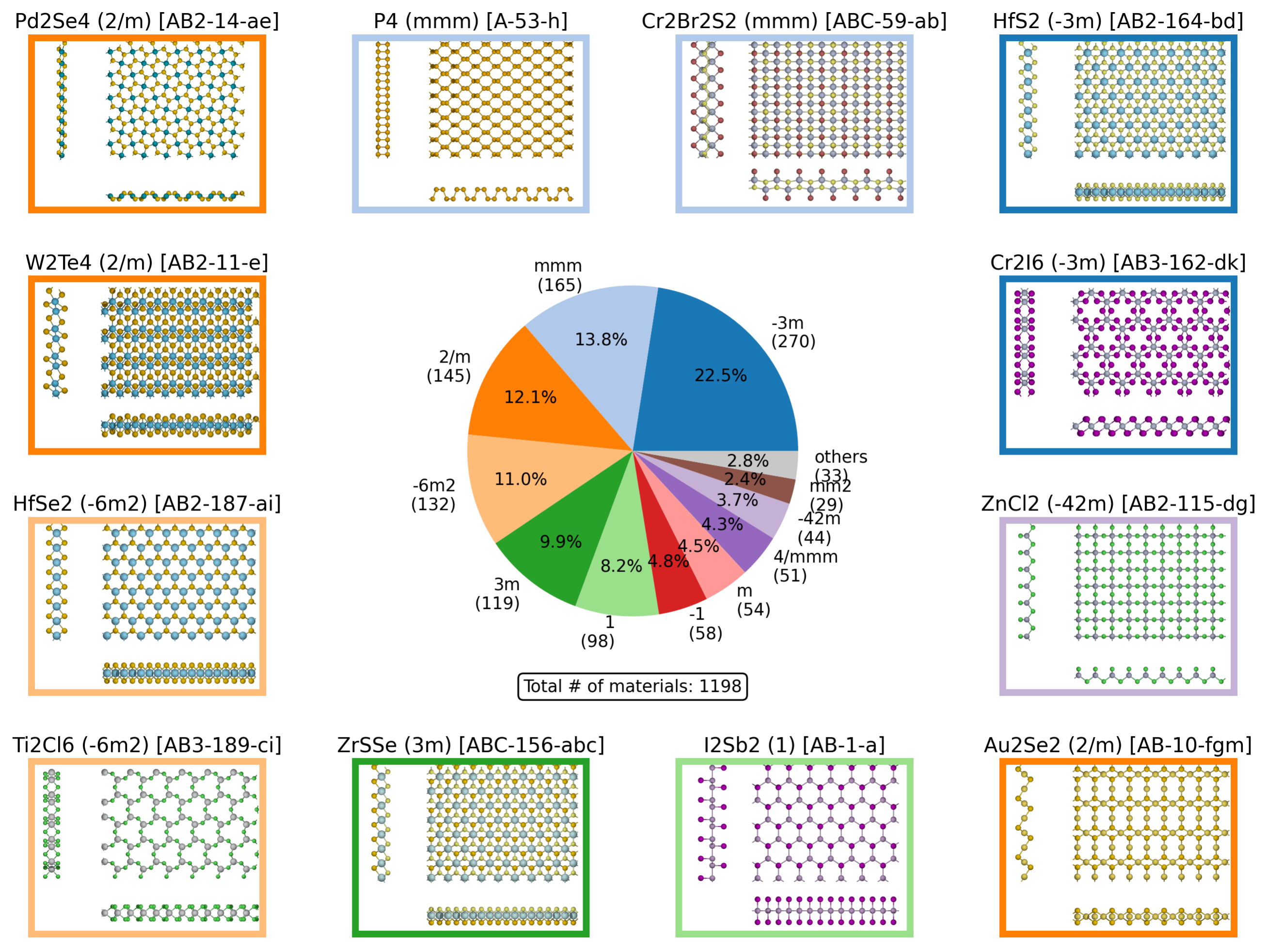} % <--- HIGH RESOLUTION
    \caption{
        \label{fig:overview}
        Overview of the point group distribution of highly stable materials in C2DB, with some examples of corresponding crystal structures.
        Each example is accompanied by the crystal structure prototype, which is a label of the form \textit{S-n-p} with \textit{S} the stoichiometry, \textit{n} the space group number and \textit{p} the set of occupied Wyckoff positions. The following point groups representing less than 2\% of stable materials are omitted in the Figure: \textit{2} (17\%, 20 materials); \textit{-3} (0.8\%, 8 materials), \textit{3} (0.2\%, 2 materials); \textit{32} (0.2\%, 2 materials); \textit{6/mmm} (0.1\%, 1 material).
        }
\end{figure*}

As shown by Neumann more than one century ago, the symmetries of any physical property of a material must include the symmetry operations of the point group to which the crystalline lattice belongs \cite{Neumann1885}. Fig.\ \ref{fig:overview} provides an overview of the 1198 thermodynamically and dynamically stable materials in C2DB grouped according to their point group symmetry, with the latter obtained from the Spglib library \cite{Togo18_Spglib}. Some specific examples of materials are shown with their point group indicated by the color of the frame. The selected materials represent some of the most interesting anisotropic 2D materials identified in this work and discussed in the following.

From Fig. \ref{fig:overview}, we make the following general observations:

\begin{itemize}
    \item
    Materials with trigonal symmetry, that is, materials belonging to the point groups \textit{-3m, 3m, -3, 3, 32} in the international notation\footnote{Note that we will substitute the common overline notation with a dash (that is, we will use $-n$ instead of $\overline n$)}, are the most frequently occurring ($33\%$ of the total). These include, among others, TMDCs in the 1T phase such as HfS$_2$ \cite{Xu15_HfS2}, group IV monolayers \cite{Zhu15_stanene}, hydrogenated graphene (i.e.\ graphane \cite{Elias09}), MXY Janus structures \cite{Lu17_Janus, Fulop18_Janus, Riis-Jensen20} such as ZrSSe, and monolayer magnetic materials such as CrI$_3$ \cite{Huang2017_FM}. 

    \item
    Monoclinic materials (groups \textit{2, m, 2/m}) account for 18\% of the total.
    They include TMDCs in the 1T' phase, such as WTe$_2$ \cite{Tang17_WTe2}, TiS$_3$ \cite{Island14, Island15}, and the pentagonal PdSe$_2$ \cite{Oyedele17}. Of the 219 monoclinic materials investigated in this work, 140 of them bear orthogonal structure, while the remaining 79 have an inclined crystal system.

    \item
    The orthorhombic structures comprise 16\% of the total (groups \textit{mmm, mm2}). Notable examples are the highly anisotropic puckered phosphorene (that is, monolayer black phosphorus \cite{Li14_BP}) and puckered arsenene \cite{Kamal15_As}. We also point out the magnetic ternary compound CrSBr, which has been recently exfoliated from the layered bulk structure\cite{Telford2020_arxiv_CrSBr, Lee2020_arxiv_CrSBr} and whose crystal prototype is largely recurrent in C2DB among orthorhombic structures.

    \item
    Triclinic materials (groups \textit{1, -1}) account for 13\% of the total. This group include materials with low symmetry, such as TMDC alloys \cite{Komsa12, Xie15_alloys}, the topological insulator SbI \cite{Song14_BiX_SbX, Vannucci20}, and other potentially exfoliable materials such as AuSe.

    \item
    11\% of structures have hexagonal point group symmetry (groups \textit{-6m2, 6/mmm}). They include TMDCs in the H phase such as HfSe$_2$, hexagonal boron nitride (hBN), graphene (which is the only stable representative of the point group \textit{6/mmm}) and other less common structures possessing 6-fold rotation symmetry such as TiCl$_3$.

    \item
    The remaining 8\% correspond to tetragonal structures (groups \textit{-42m, 4/mmm}) such as ZnCl$_2$, which is predicted to be easily exfoliable\cite{Mounet2017}.
\end{itemize}

This set of 1198 known or potentially exfoliable/synthesizable materials forms the basis for the anisotropy analysis presented in this work.

\section{Results and discussion}
\label{sec:results}

\subsection{Magnetic easy axis}

Magnetic anisotropy is defined in a material as the dependence of its properties on the direction of its magnetization. The main manifestation of magnetic anisotropy is the existence of an easy axis, along which it takes the least energy to magnetize the crystal, and a hard axis, where it takes the most. In order to quantify the degree of anisotropy, the magnetic anisotropy energy (MAE) is defined, which accounts for the energy necessary to deflect the magnetization from the easy to the hard direction. In general, the MAE may have contributions from different features of a crystal such as strain or defects. In this work we will consider perfect crystals, wherein only the so-called magnetocrystalline anisotropy, given by the coupling of the lattice to the spin magnetic moment, contributes to the MAE.

In 2D materials magnetic anisotropy gains a special importance due to the Mermin–Wagner theorem \cite{Mermin1966}, which prohibits a broken symmetry phase at finite temperatures. This means that for a magnetic order to emerge, the spin rotational symmetry has to be broken explicitly by magnetic anisotropy. This has attracted a wide interest on the topic in the recent years, both in the light of new fundamental questions \cite{Huang2017_FM,Fei2018,Zhang2015,Rossier2017,Torelli2018, Torelli2019} and applications \cite{Gong2019,Cardoso2018,Zhong2017,Burch2018,WangMorpurgo2018}.

In this work we will focus on the in-plane MAE and define the $x$ and $y$ axes to span the atomic plane of the material.
We define the in-plane MAE, $\Delta_{xy}$, as:
\begin{equation}
    \Delta_{xy}=|E(\vec{M}\parallel y)-E(\vec{M}\parallel x)|,
\end{equation}
where $E(\vec{M}\parallel x)$ and $E(\vec{M}\parallel y)$ are the electronic energies including spin-orbit coupling (SOC) with magnetization parallel to the $x$ and $y$ axes, respectively.
In the electronic structure code GPAW, SOC is added non self-consistently on top of a converged PBE calculation as described in Ref.\ \onlinecite{Olsen16_SOC}. This method yields very accurate results, as evidenced, for instance, by the excellent agreement of MoS$_2$ SOC-induced band splitting with other first-principles approaches and with experiments \cite{Olsen16_SOC}, or by the correct description of topological physics governed by SOC \cite{Olsen19_topological, Vannucci20}. The MAE obtained with this method has been used in the calculation of critical temperature of magnetic 2D materials, providing good agreement with experiments \cite{Torelli2018, Torelli2019}.

Let us note that the definition of $x$ and $y$ axes is of course arbitrary. In C2DB, the $x$ axis is systematically chosen to be parallel to one of the lattice vectors, with the $y$ axis orthogonal to it. All properties are then calculated with respect to this reference frame.
With this choice, the anisotropy is assessed with respect to the high symmetry axes for most of the materials in C2DB, except only for low symmetric obliques structures such as the ones in point group \textit{1} and \textit{-1} (which represent only a 13\% of the total). For such cases, the degree of anisotropy for some properties might be slightly underestimated.

\begin{figure*}
    \includegraphics[width=\linewidth]{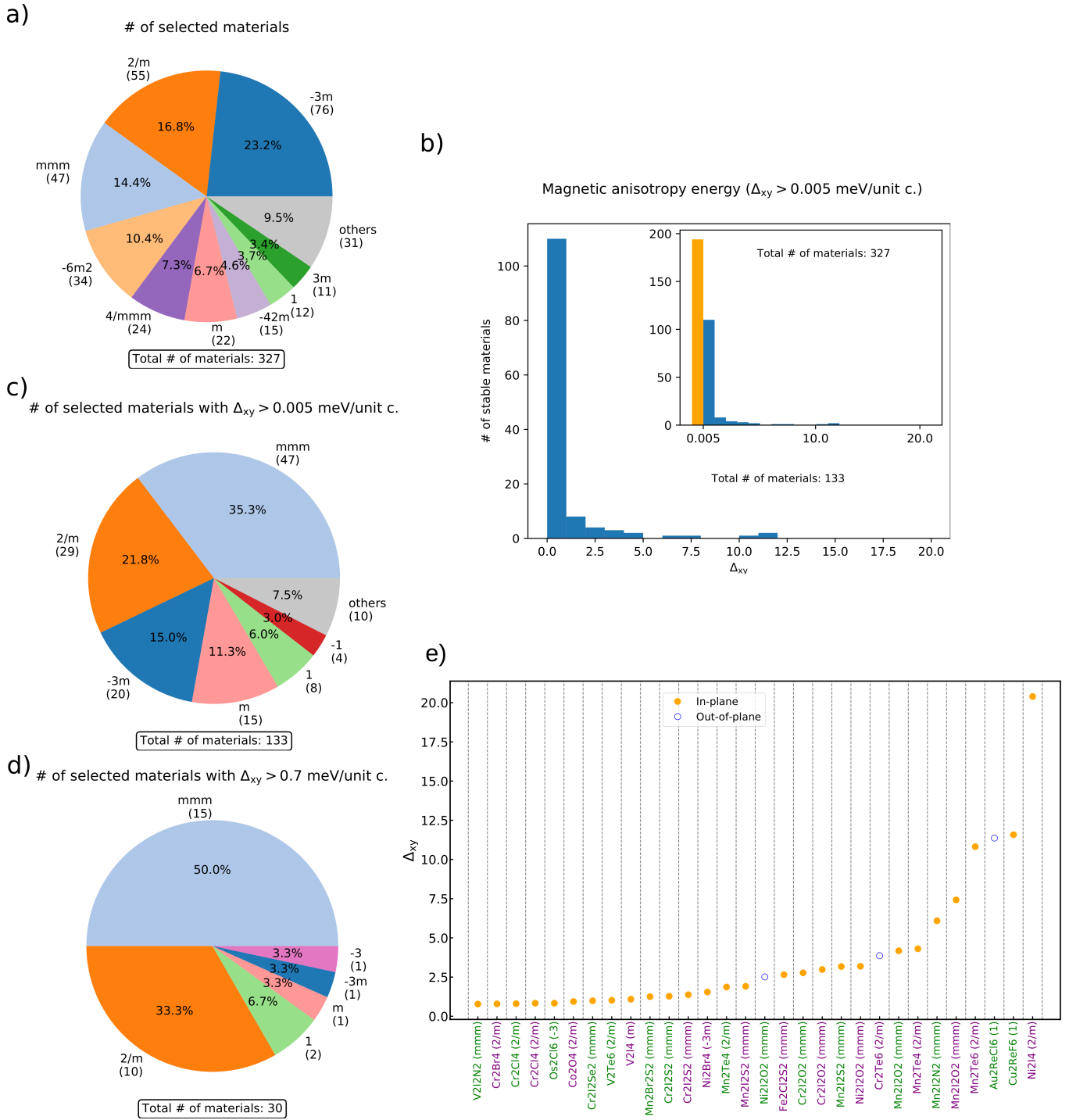}
    \caption{
        \label{fig:mag.pdf}
        a) Distribution of predicted stable magnetic materials (FM or AFM) from the C2DB database grouped according to their point group symmetry. The percentage of materials represented by each point group and their number of occurrences in the database is explicitly shown, except for those point groups representing less than 3$\%$ of materials. Panels (b) and (c) show the distribution of magnetic materials in the database with a low (0.005 meV/unit cell) in-plane magnetic anisotropy (MA) threshold. The main chart in (b) shows the distribution of the materials above the threshold, while the orange bin in the inset represents all magnetic materials below the threshold for comparison. Panels (d) and (e) show the distribution of the materials with a high MA (>0.7 meV/unit cell). In (e) the materials are sorted by the size of their MA and their magnetic state is indicated by the font color (green for FM, purple for AFM). The marker color corresponds to the signs of $\Delta_{zx}$ and $\Delta_{zy}$: empty marker with blue edge for $\Delta_{zx}<0$ and $\Delta_{zy}<0$ (out-of-plane easy axis) and full orange if any of the two is above 0 (in-plane easy axis).
        }
\end{figure*}

In Figure \ref{fig:mag.pdf} we show the distribution of the magnetic materials in the C2DB database, sorted by their point group symmetry and the value of their in-plane magnetic anisotropy $\Delta_{xy}$. 
In comparison with Fig.\ \ref{fig:overview}, we find a similar landscape once we filter for ferromagnetic (FM) or anti-ferromagnetic (AFM) materials, as shown in Fig \ref{fig:mag.pdf}a. This indicates that being magnetic or not is not strongly correlated to the point group, but to the presence of magnetic atoms in the structure. However, once anisotropy comes into play we do observe, in Fig \ref{fig:mag.pdf}b, important structural features that condition it. In fact, one can expect a magnetic easy axis to appear in the direction where magnetic atoms are packed more loosely, creating an anisotropy in the magnetic properties.
For example, as we set a very low (0.005 meV/unit cell) threshold for $\Delta_{xy}$, all hexagonal and tetragonal point groups vanish and only point groups not restricting the in-plane perpendicular directions by symmetry hold an in-plane magnetic anisotropy. Another feature we observe from Figure \ref{fig:mag.pdf}c, is the prevalence of the orthorhombic (\textit{mmm}) and, to a lower degree monoclinic (\textit{2/m}), systems with crystals of \textit{mmm} point group symmetry representing over 35$\%$ of the materials with a low $\Delta_{xy}$ threshold.
When we increase the $\Delta_{xy}$ threshold to 0.7 meV/unit cell we see that the trend is enhanced: \textit{mmm} dominates with half of the materials and \textit{2/m} comprises a third of the materials (Figure \ref{fig:mag.pdf}d). The materials above this threshold are classified and sorted according to their anisotropy in Figure \ref{fig:mag.pdf}e. We see that both FM and AFM magnetic orders are equally represented, indicating little influence of the type of magnetic order on the anisotropy.
In Figure \ref{fig:mag.pdf}e we also show the direction of the magnetic easy axis, indicated by a full orange marker if it lies within the plane and an empty blue one if it is oriented out-of-plane. It is clear that most of the selected anisotropic materials indeed present an in-plane easy axis.

Among the 113 materials with $\Delta_{xy}$>0.005 meV/unit cell we find that the ternary compound structure prototype of orthorhombic symmetry \textit{ABC-59-ab} \cite{Haastrup18} is the most frequently occurring with 47 entries (see CrSBr in Fig.\ \ref{fig:overview} for an example of this structure). The main reason for this might be the mentioned lack of symmetry between the $x$ and $y$ directions in the plane, along with the fact that it is more likely to contain a magnetic atom due to its ternary nature (most other crystals in C2DB are binary). To the best of our knowledge, the only 2D material from this group that has been successfully synthesized and exfoliated is CrSBr, whose FM order down to the monolayer limit has been very recently confirmed in experiments \cite{Telford2020_arxiv_CrSBr, Lee2020_arxiv_CrSBr}.
We note that several materials bearing the very same structure prototype are listed as easily exfoliable in Ref \onlinecite{Mounet2017}, e.g. CrOBr, CrOCl, FeOCl, VOBr and VOCl.
Finally, the monoclinic T' phase of transition metal dichalcogenides occurs 15 times, followed by the trigonal MoS$_2$ \cite{Kappera2014} type with 10 occurrences.

We also cross-checked the rest of our selected anisotropic materials against the list of exfoliable 2D materials in Ref \onlinecite{Mounet2017}. We found, out of the 113 materials with $\Delta_{xy}$>0.005 meV/unit cell, over 20 materials whose stoichiometry match entries in the list of exfoliable materials. Among these, perhaps the most promising material with regard to a potential experimental realization is the AFM T' di-halide V$_2$I$_4$, which lies at the  convex hull according to the C2DB database \cite{Haastrup18}. V$_2$I$_4$ shows an in-plane magnetic easy axis and $\Delta_{xy}=1.09$ meV/unit cell, that competes with the highest out-of-plane anisotropies known to date \cite{Torelli2018}. In addition, we find several materials that are only a few meV above the convex hull and show remarkably high anisotropies. Among these the AFM Ni$_2$I$_4$ compound stands out with an exceptional in-plane anisotropy of over 20 meV/unit cell and an in-plane easy axis. Other materials in the same stability category, such as Ni$_2$Br$_4$, Co$_2$O$_4$ and CrBr$_2$, also show large $\Delta_{xy}$ values and are listed in Table \ref{tab:magani}.

\begin{table}[htb]
   \centering
    \begin{tabular}{c|c|c|c|c|c}
         & Sym.&Mag.&$E_{\mathrm{hull}}$(meV) & \makecell{Lowest $E_{\mathrm{hull}}$ \\ monolayer?} &$\Delta_{xy}$(meV/unit c.)\\
         \hline
         Cr$_2$Br$_4$& P2$_1$/m&FM&54.4&No&0.79\\
         %\hline
         Co$_2$O$_4$&C2/m&FM&7.1&No&0.94\\
         %\hline
         V$_2$I$_4$&Pm&AFM&0.0&Yes&1.09\\
         %\hline
         Ni$_2$Br$_4$&P$\bar{3}$m1&AFM&8.8&No&1.55\\
         %\hline
         Ni$_2$I$_4$&C2/m&AFM&10.3&No&20.40
    \end{tabular}
    \caption{Monolayers predicted stable and with the highest in-plane magnetic anisotropy in the C2DB database and in-plane magnetic easy axis, whose stoichiometry matches that of entries in the list of easily exfoliable 2D materials in Ref \onlinecite{Mounet2017}. The table shows the chemical formula, the space group symmetry, energy above the convex hull, magnetic state and in-plane magnetic anisotropy.}
    \label{tab:magani}
\end{table}

\subsection{Elastic response and auxetic effect}

The elastic response of 2D materials to strains and deformations is usually expressed in terms of the Young modulus $E$ and Poisson ratio $\nu$ \cite{Akinwande17, Androulidakis18}. The former measures the response along a direction that is parallel to the applied strain, while the latter describes how the material reacts along orthogonal directions.
For anisotropic materials, both the Young modulus and the Poisson ratio depend on the directions of stresses and strains. Assuming that the 2D material lies in the $xy$ plane, and neglecting the elastic response along the out-of-plane axis $z$, we will denote the axis-dependent Young modulus with $E_i$, $i = \{x, y\}$. Similarly, the coefficient relating the stress along the $i$ axis to an applied strain in the perpendicular $j$ direction will be quantified by the Poisson ratio $\nu_{ij}$, with $i \ne j$.

More generally, the elastic response of a continuous 2D medium is quantified in terms of the 2D stiffness tensor $C$, which is a linear map between the strain tensor $\varepsilon$ and the stress tensor $\sigma$\cite{Landau_elasticity}:
\begin{equation}
    \sigma_{ij} = \sum_{kl} C_{ijkl} \varepsilon_{kl} .
\end{equation}
Here we have $i = \{x, y\}$, since we restrict to in-plane stresses and strains.
A generic matrix element $\sigma_{ij}$ represents the $i$ component of the stress acting on a plane perpendicular to the $j$ direction, while the strain components $\varepsilon_{ij}$ are given by $\varepsilon_{ij} = (\partial_i u_j + \partial_j u_i)/2$ in terms of the infinitesimal deformations $u_i$.

Being a linear map between two 2nd rank tensors, the stiffness tensor is naturally a 4th rank tensor. However, one can make use of the symmetric properties of both $\sigma$ and $\varepsilon$ at equilibrium to write both of them as one-dimensional vectors, namely
\begin{align}
    \tilde \sigma & = \left(\sigma_{xx}, \sigma_{yy}, \sigma_{xy} \right)^T := \left(\sigma_{1}, \sigma_{2}, \sigma_{3} \right)^T , \\
    \tilde \varepsilon & = \left(\varepsilon_{xx}, \varepsilon_{yy}, 2\varepsilon_{xy} \right)^T := \left(\varepsilon_{1}, \varepsilon_{2}, \varepsilon_{3} \right)^T .
\end{align}
Such a notation is often called \textit{Voigt} notation. Then, the stiffness tensor becomes a 2nd rank symmetric tensor with only 6 independent components,
\begin{equation}
    \tilde \sigma =
    \begin{pmatrix}
        C_{11} & C_{12} & C_{13} \\
        C_{12} & C_{22} & C_{23} \\
        C_{13} & C_{23} & C_{33} 
    \end{pmatrix} \tilde \varepsilon .
\end{equation}
We will restrict the following analysis to the class of \textit{orthotropic materials}, that is, materials having three mutually-orthogonal planes of reflection symmetry.
In such a case, the stiffness tensor takes the form 
\begin{equation}
    C =
    \begin{pmatrix}
        C_{11} & C_{12} & 0 \\
        C_{12} & C_{22} & 0 \\
        0 & 0 & C_{33} 
    \end{pmatrix} .
\end{equation}
In practice, this means that we restrict attention to materials where the shear deformations $\varepsilon_{xy}$ are decoupled from $xx$ and $yy$ stresses. This allows us to straightforwardly relate the components $C_{ij}$ of the stiffness tensor to the in-plane Young modulus $E_i$ and in-plane Poisson ratio $\nu_{ij}$ via the following relations: 
\begin{subequations}
\begin{align}    
    E_x & = \frac{C_{11} C_{22} - C_{12}^2}{C_{22}} , \\
    E_y & = \frac{C_{11} C_{22} - C_{12}^2}{C_{11}} , \\
    \nu_{xy} & = \frac{C_{12}}{C_{11}} , \\
    \nu_{yx} & = \frac{C_{12}}{C_{22}} .
\end{align}
\label{eq:def_Young_Poisson}
\end{subequations}

In C2DB, each component of the 2D stiffness tensor is calculated by straining a material along a given direction ($x$ or $y$) and calculating the forces acting on the unit cell after relaxing the position of the atoms within the fixed unit cell \cite{Haastrup18}.
To restrict to orthotropic materials only, we have discarded all materials whose stiffness tensor components $C_{13}$ or $C_{23}$ exceed a certain tolerance $C_\mathrm{max}$, which we set to $C_\mathrm{max} = 0.01$ N/m. With this method, we have obtained a subset of 555 materials (roughly 50\% of all the stable materials) that we analyze in the following. 

\begin{figure*}
    \includegraphics[width=\linewidth]{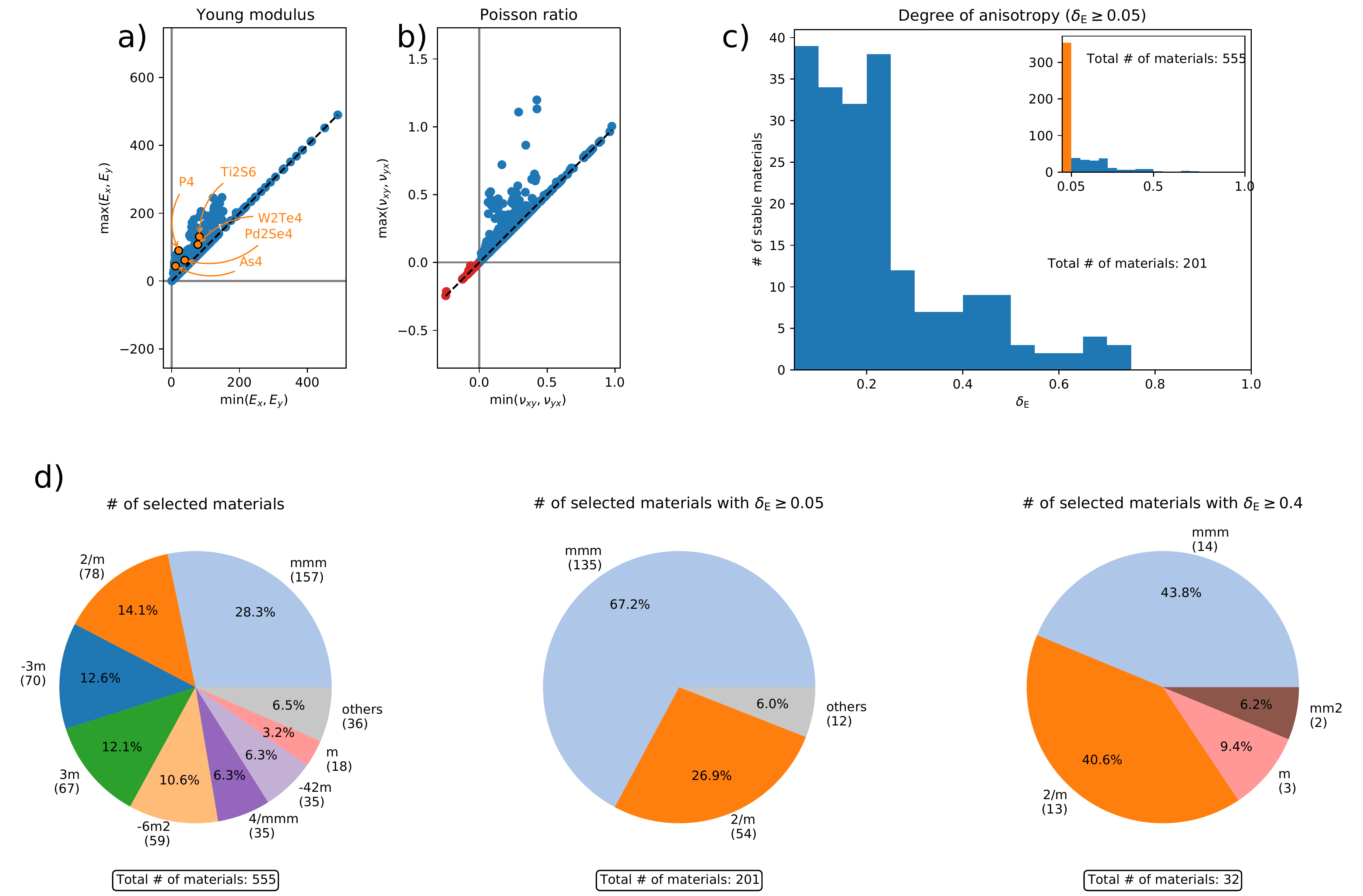}
    \caption{
        \label{fig:elastic_1}
        Elastic properties of the 555 orthotropic materials in C2DB. Panels (a) and (b) show a comparison between Young modulus and Poisson ratios along opposite directions, respectively. In panel (a), some known anisotropic materials are highlighted in orange, while in panel (b) we use red markers for auxetic materials. Panel (c) shows the distribution of the anisotropic degree $\delta_E$ for the 201 materials having $\delta_E \geq 0.05$, with the inset showing the full distribution of all 555 materials for comparison. Panel (d) shows the point group distributions of orthotropic materials for three different threshold values of $\delta_E$ (namely 0, 0.05, and 0.4 from left to right).
    }
\end{figure*}

In Fig.\ \ref{fig:elastic_1}a we show an overview of the direction-dependent Young modulus for all orthotropic and stable materials in C2DB. We plot $\mathrm{max}(E_x, E_y)$ against $\mathrm{min}(E_x, E_y)$, which means that all data point lying outside the diagonal represent a material with anisotropic elastic properties.
Well known anisotropic structures such as WTe$_2$, PdSe$_2$, TiS$_3$, P$_4$ and As$_4$ are all identified by this method, while hundreds of unexplored anisotropic materials are predicted as well.
In Fig.\ \ref{fig:elastic_1}b we use a similar method to show the anisotropy of the Poisson ratio. While this does not add much information with respect to panel a --- since $E_x/E_y = \nu_{yx}/\nu_{xy}$, as one can easily infer from Eqs.\ \ref{eq:def_Young_Poisson} --- we notice that Poisson ratios can also take negative values, differently from the Young modulus.
In such a case, a material stretched (or compressed) along the $x$ direction will also expand (shrink) along the perpendicular $y$ direction, a quite counterintuitive property called \textit{auxetic} behavior \cite{Akinwande17, Jiang16}. We will investigate such cases in more detail in the following.

To describe elastic anisotropy in a more quantitative manner, we define an elastic anisotropy degree (or anisotropy parameter) for each material as
\begin{equation}
    \delta_E = \frac{|E_x - E_y|}{E_x + E_y} .
\end{equation}
Such a parameter will be always bounded between 0 and 1, with $\delta_E = 0$ signifying a perfectly isotropic materials while $\delta \approx 1$ for an extremely anisotropic medium.

In Fig.\ \ref{fig:elastic_1}c we show the distribution of the elastic anisotropy degree for all materials having $\delta_E \geq 0.05$ (corresponding to a difference of at least 10\% between $x$ and $y$ Young modulus), with the inset showing the full distribution including materials with $\delta_E < 0.05$.
We notice that more than one third of the selected materials (201 out of 555) show an elastic anisotropy exceeding this threshold value, while 162 of them exceed the value $\delta_E = 0.1$ (signifying a difference of roughly 20\% or more between $E_x$ and $E_y$) and 32 of them show a highly anisotropic elastic behaviour with $\delta_E \geq 0.4$.

The distribution of point groups corresponding to different threshold values for $\delta_E$ is shown as a series of pie charts in Fig.\ \ref{fig:elastic_1}d.
On the left, we plot the distribution of point groups for all 555 selected materials.
A comparison with Fig.\ \ref{fig:overview} shows that our choice of selecting only orthotropic materials tends to favor orthorhombic structures (especially group \textit{mmm}) with respect to trigonal ones, while the proportions between remaining point groups remain basically unaffected.
However, when selecting all materials with at least 10\% of difference between $E_x$ and $E_y$ ($\delta_E \geq 0.05$, in the middle), the proportions change drastically, with all trigonal and hexagonal groups suppressed in favor of orthorhombic and monoclinic structures. This shows that symmetric crystal structures such as the ones of TMDCs in the H and T phase, graphene and hBN are generally isotropic, with little difference in the elastic properties along $x$ and $y$ directions. On the other hand, TMDCs in the distorted T' phase (such as WTe$_2$), pentagonal structures (PdSe$_2$) and puckered layers (phosphorene) stand out for their markedly anisotropic elastic properties due to their asymmetric crystal lattice.

\begin{figure*}
    \includegraphics[width=\linewidth]{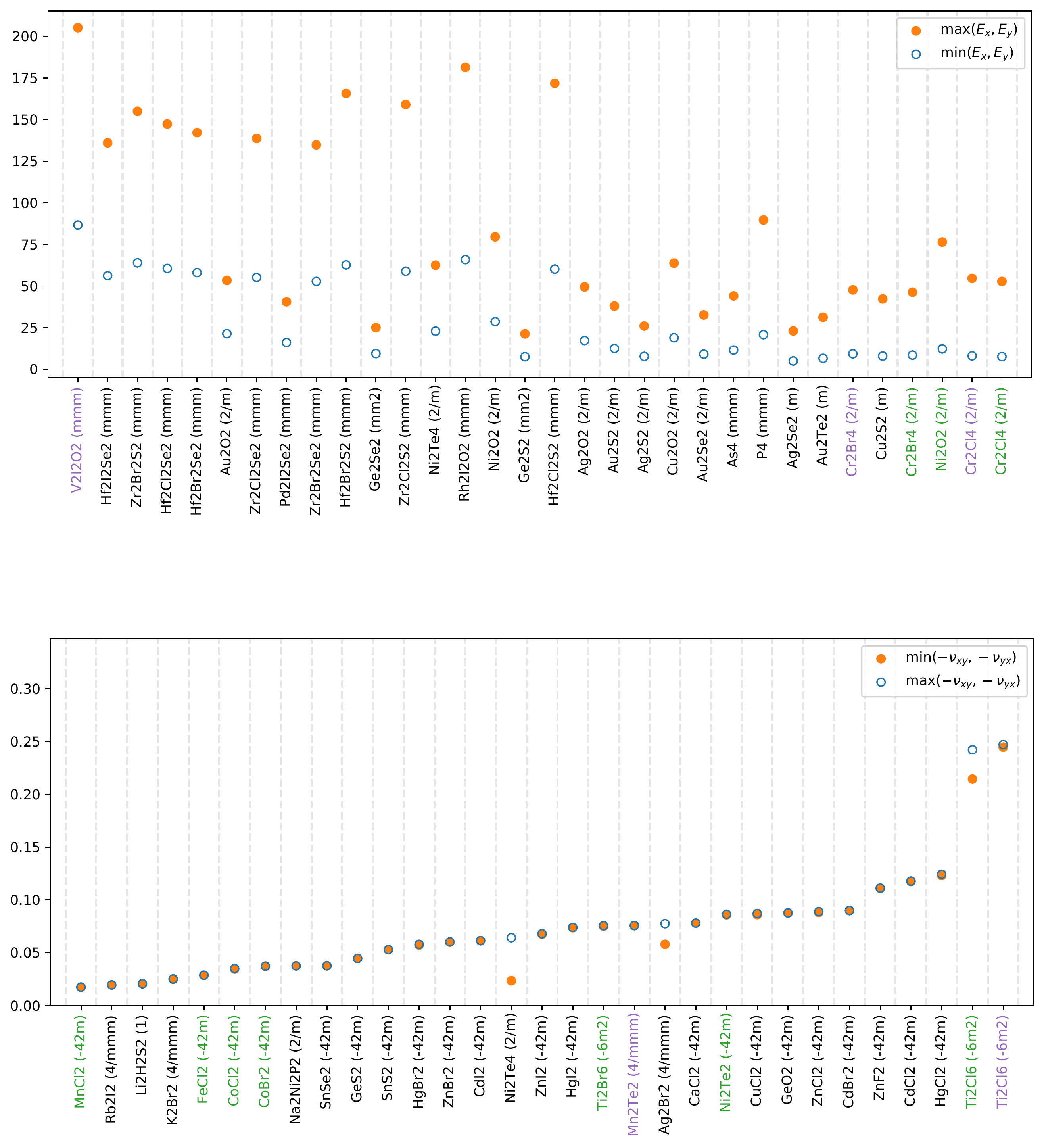}
    \caption{
        \label{fig:elastic_2}
        Top: Materials with highly anisotropic Young modulus $\delta_\mathrm{E} \geq 0.4$, sorted from lowest to highest $\delta_\mathrm{E}$.
        Bottom: Materials with negative Poisson ratio, sorted from lowest to highest value of $\mathrm{max}(-\nu_{xy}, -\nu_{yx})$.
        Green (purple) labels indicates ferromagnetic (antiferromagnetic) materials.
        }
\end{figure*}

When restricting to highly anisotropic materials having $\delta_E \geq 0.4$, monoclinic and orthorhombic structures share exactly 50\% of the total each. The Young moduli of these 32 structures are plotted in the top panel of \ref{fig:elastic_2}, sorted from lowest to highest value of $\delta_E$.
Besides known structures such as phosphorene (P$_4$) and puckered arsenene (As$_4$), we find many new stable structures with exceptionally high elastic anisotropy.
Four out of the first six materials are compounds of the form CrX$_2$ (with X a halogen element) in both the AFM and FM magnetic state, which also stand out for their markedly anisotropic magnetic behavior as described previously.
These are, however, not the most stable structures with the same constituent elements, since they all have a competing phase of the form CrX$_3$ with a more favorable formation energy (one of them is shown in Fig.\ \ref{fig:overview}). However, this is not the case for the monoclinic structures AuSe and AuTe, which represent the most stable phase of their respective elements. One of them (AuSe, which is shown in Fig.\ \ref{fig:overview} as well) has also been identified as an easily exfoliable materials by independent work of Mounet \textit{et al.} \cite{Mounet2017}, making this one of the most appealing material for anisotropic elastic applications found in this work.
We also note that puckered compounds GeS and GeSe seem also to be easily exfoliable from their respective three-dimensional parent structures, which is again confirmed in the literature \cite{Mounet2017}.

Finally, it is worth mentioning the presence of several entries in the structural prototype \textit{ABC-59-ab} (especially Hf- and Zr-based compounds), whose relevance has already been discussed in the previous section. We note that Ref.\ \onlinecite{Mounet2017} lists HfNBr, ZrNBr and ZrNI as easily exfoliable layered materials. We find a relatively low elastic anisotropy degree $\delta_E = 0.07$ for ZrNI, but we suggest that materials with much higher elastic anisotropy such as HfBrX and ZrBrX (with X = S, Se) should in principle be available by susbtitution of Nitrogen with an element from the halogen group.

Let us now come back to the subset of materials showing a negative in-plane Poisson ratio, that is the ones highlighted with red markers in Fig.\ \ref{fig:elastic_1}b.
The auxetic effect is not necessarily associated to anisotropy as both Poisson ratios $\nu_{xy}$ and $\nu_{yx}$ can take negative values without necessarily being different from each other. Indeed, such an effect does not originate from the material having a different elastic response along orthogonal axes, but rather from the presence of special re-entrant structures or rigid blocks linked by flexible hinges in the crystalline structure, that can compress or extend in counter-intuitive fashions.
Nevertheless, our framework allows for the systematic search of novel 2D materials with negative Poisson ratio, which itself is an active field of research \cite{Jiang16}. Moreover, Poisson ratios of anisotropic materials can take arbitrarily large values (positive or negative), differently from ordinary isotropic media \cite{Ting05}.

In the bottom panel of Fig.\ \ref{fig:elastic_2} we plot the Poisson ratios of the 31 stable materials in C2DB showing auxetic behavior, sorting from lowest to highest value of $\mathrm{max}(-\nu_{xy}, -\nu_{yx})$.
The largest negative Poisson ratio is found for TiCl$_3$ in the hexagonal crystal structure (shown in Fig.\ \ref{fig:overview}), with both AFM and FM magnetic configurations. Once again, this is not the most stable phase for such a compound, which reaches the lowest energy configuration when arranged in a trigonal phase, in the same crystal prototype as the ferromagnetic insulator CrI$_3$ \cite{Huang2017_FM}.

There are several interesting candidates among the materials with tetragonal structure. In particular, materials with stoichiometry AB$_2$ in point group \textit{-42m}, such as the case of ZnCl$_2$ shown in Fig.\ \ref{fig:overview}, represent a large majority of stable auxetic materials in C2DB. 
Notable examples are metal di-halides involving Co, Mn, or Fe as the metallic element. Such materials are all exfoliable from a 3D parent compound with trigonal point group \cite{Mounet2017}, but their tetragonal phases generally have total energies that are comparable or even lower than the trigonal monolayer phase (which is also present in C2DB).
A second notable example is given by group 12 di-halides involving Zn, Cd, and Hg, for which the tetragonal auxetic structure turns out to be the most stable phase. Interestingly, both HgI$_2$ and ZnCl$_2$ are reported as easily exfoliable materials by Mounet \textit{et al.}\cite{Mounet2017}, making these two materials very appealing candidates for novel auxetic 2D materials.
We also note that MnTe, AgBr, and GeO$_2$ all seem to have total energies very close to the convex hull, and thus also belong to the set of  predicted stable auxetic monolayers.

Let us note that a significant majority of known auxetic 2D structures display negative Poisson ratio in the out-of-plane direction \cite{Liu19, Kong18, Gomes15, Jiang14, Du16}, while only very few materials were previously predicted to exhibit in-plane auxetic response \cite{Yu17, Qin20}. Ref.\ \onlinecite{Yu17} reports negative Poisson ratio for monolayers of groups 6–7 transition-metal dichalcogenides (MX$_2$ with M=Mo, W, Tc, Re and X=S, Se, Te) in the 1T-phase. We do find a negative $\nu_{xy}$ in C2DB for all of them, but they have low dynamical and thermodynamical stability, and therefore are not identified by our analysis.

\subsection{Effective masses}

Monolayer 2D materials with a finite band gap can display large anisotropies in the effective masses along two orthogonal directions. This makes them appealing for highly directional-dependent transport, with applications in anisotropic field-effect transistors, polarization-sensitive detectors and non-volatile memory devices among others \cite{Jin15, Liu15_ReS2_FET, Zhang16_ReS2, Liu20_ReSe2, Wang19_anisotropic_resistance, Wang15_phospho}.

In C2DB, effective masses for conduction and valence bands are calculated for all materials having a finite band gap greater than 0.01 eV at the PBE level. We define the effective mass, $m$, from the curvature, $a$, at the band maximum (minimum) for valence (conduction) bands as $m = 1/2a$. To determine the curvature we start from a self-consistent ground state calculation performed at a $k$-point density of 12 \AA$^{-1}$. From these $k$-points a preliminary band extremum is found and a second, non-self-consistent calculation is performed with higher density of $k$-points centered around the preliminary extremum. Then, from these values a final extremum is determined and the energies for a number of $k$-points spaced very closely around the extremum are calculated non-self-consistently. The $k$-points used for the first refinement step are by default chosen to lie in a sphere around the extremum with a radius of 250 meV (for a mass of 1) and the same number of $k$-points as the original calculation (but at least 19). The last refinement uses a 1 meV sphere and 9 points. The points calculated in the final refinement step are used to determine the curvature. We first do a fit to a second order polynomial to determine a preliminary extremum. Then we perform a fit to a third order polynomial and find the new extremum, unless the optimization algorithm diverges (as may happen for third order polynomials) in which case we revert to the original fit. We have found that the third order polynomial fit does provide an improvement to the description of the band extremum and in some cases is necessary, e.g. in the presence of parabolic bands crossing as in Rashba splitting. From the fit we find the curvature $a$ at the extremum and the mass is calculated as $m = 1/2a$.

To measure the presence of anisotropic effects in the effective masses, we define the parameters
\begin{subequations}
\begin{align}
    \delta_\mathrm{me} & = \frac{|m^\mathrm{(e)}_x - m^\mathrm{(e)}_y|}{m^\mathrm{(e)}_x + m^\mathrm{(e)}_y} ,\\
    \delta_\mathrm{mh} & = \frac{|m^\mathrm{(h)}_x - m^\mathrm{(h)}_y|}{m^\mathrm{(h)}_x + m^\mathrm{(h)}_y} ,
\end{align}
\end{subequations}
where:
\begin{itemize}
    \item $m^\mathrm{(e)}_i$ is the effective electron mass calculated along the $i$ direction around the conduction band minimum;
    \item $m^\mathrm{(h)}_i$ is the effective hole mass calculated along the $i$ direction around the valence band maximum.
\end{itemize}
Unfortunately, getting a very accurate value for the effective masses in a fully automated fashion turns out to be a quite challenging task, with some fits being not accurate enough, or picking a wrong sign for the electron or hole mass in the case of a particularly heavy effective mass.
We therefore remove all materials having $m^\mathrm{(e/h)}_i \geq 20 m_{\mathrm{e}}$, with $m_{\mathrm{e}}$ the free electron mass, and materials with extremely high ratio $m^\mathrm{(e/h)}_i / m^\mathrm{(e/h)}_j \geq 20$. We stress that these threshold values are arbitrary. They have been primarily chosen so that we discard all wrong results, while also keeping the highly anisotropic materials with accurate results into the analysis as much as possible. 

\begin{figure*}
    \includegraphics[width=\linewidth]{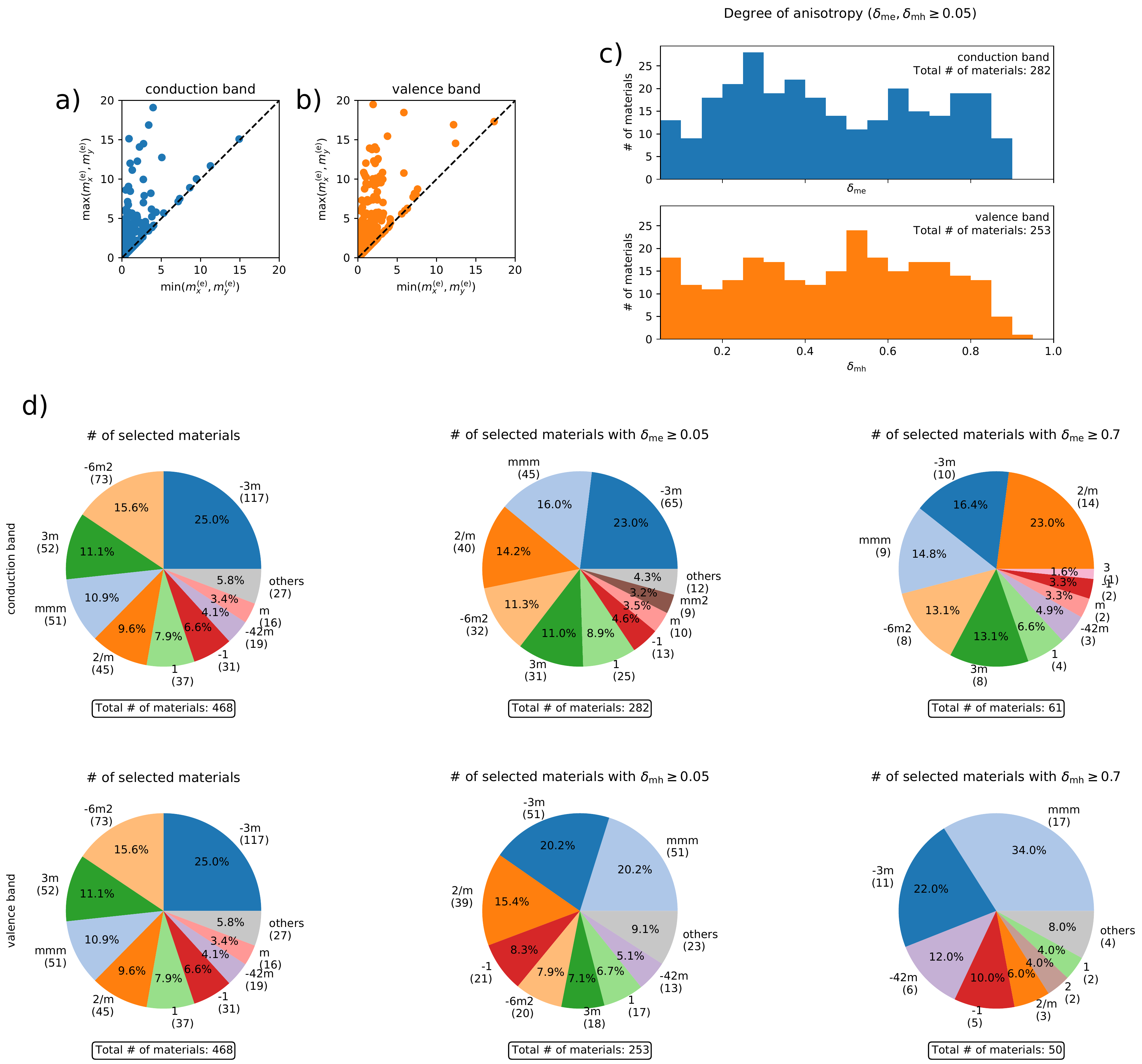}
    \caption{
        \label{fig:emasses_1}
        Distribution of the effective masses in C2DB. In panels a) and b) we show the effective mass along the heaviest direction against the effective mass along the lightest direction for conduction and valence bands respectively. Effective masses are plotted in units of the free electron mass $m_{\mathrm{e}}$. Panel c) shows a distribution of the anisotropic degrees $\delta_\mathrm{me}$ ($\delta_\mathrm{mh}$) for the such cases where $\delta_\mathrm{me}$ ($\delta_\mathrm{mh}$) is greater than 0.05. Panel d) shows, from left to right, the distribution of $\delta_\mathrm{me}$ ($\delta_\mathrm{mh}$) in terms of point group symmetry for three different threshold values of 0, 0.05, and 0.7 respectively.
        }
\end{figure*}

The C2DB database contains 574 dynamically and thermodynamically stable materials with a PBE band gap greater than 0.01 eV, of which 106 fall outside the range of validity described above. 
This leaves us with a total of 468 materials, whose effective electron and hole masses are shown as a scatter plot in Fig.\ \ref{fig:emasses_1}a and \ref{fig:emasses_1}b.
We find a rather large set of materials with anisotropic effective masses, as one can immediately notice from the large number of points falling outside the main diagonal.
Indeed, as shown in Fig.\ \ref{fig:emasses_1}c, 60\% of the selected materials (282 out of 468) show a difference of at least 10\% between electron effective masses, while 54\% of them (253 out of 468) have at least 10\% of difference between hole effective masses. Moreover, a quite large fraction of materials have extremely high values of $\delta_\mathrm{me}$ or $\delta_\mathrm{mh}$ as compared with the case of elastic anisotropy in Fig.\ \ref{fig:elastic_1}c.

\begin{figure*}
    \includegraphics[width=\linewidth]{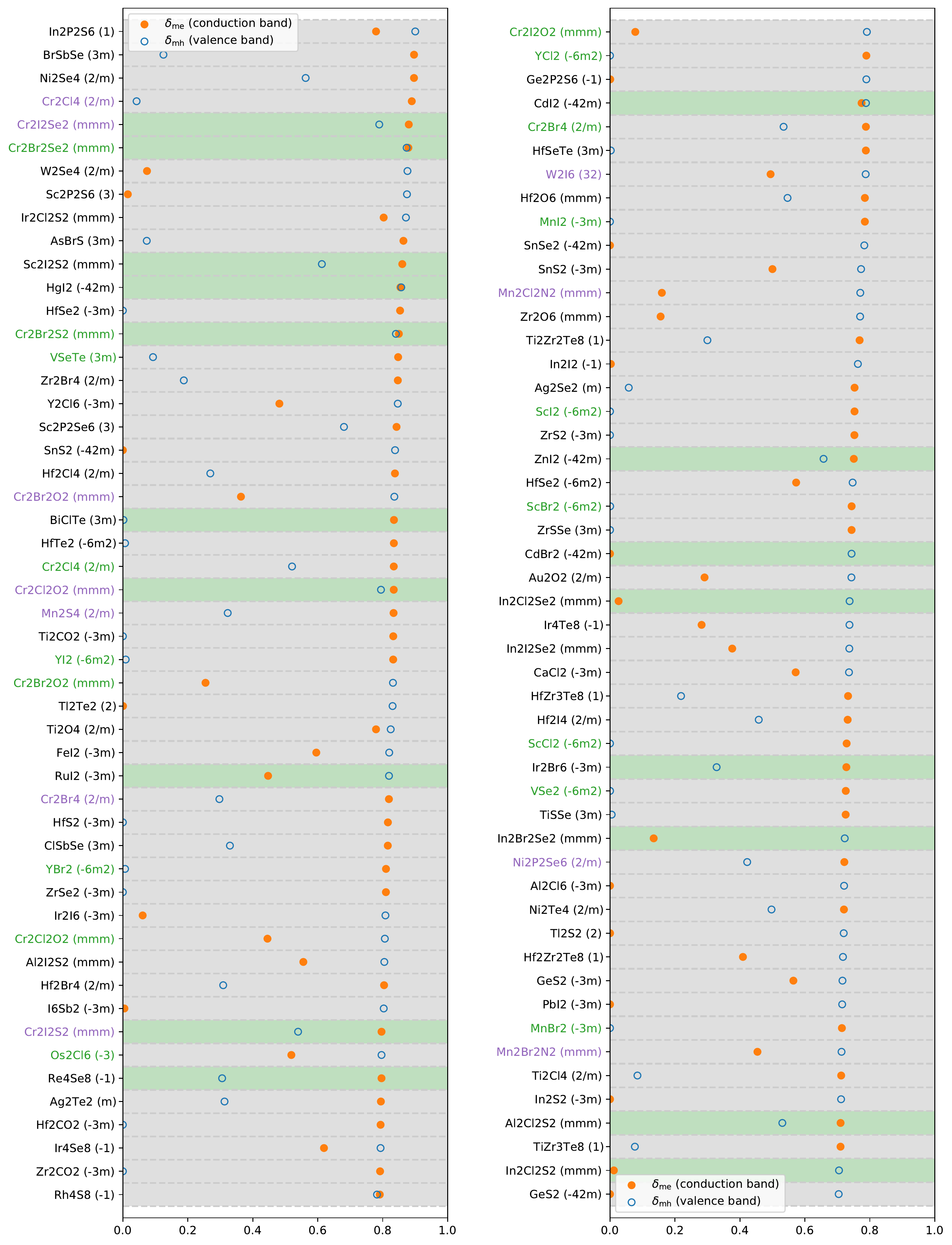}
    \caption{
        \label{fig:emasses_2}
        Materials with highly anisotropic effective masses $\delta_\mathrm{me} \geq 0.7$ or $\delta_\mathrm{mh} \geq 0.7$, sorted from highest (top left) to lowest (bottom right) value of $\mathrm{max}(\delta_\mathrm{me}, \delta_\mathrm{mh})$.
        Green (purple) labels indicates ferromagnetic (antiferromagnetic) materials. A green (light grey) background indicates the presence of a direct (indirect) band gap at the PBE level.
        }
\end{figure*}

When considering the distribution of point groups for materials with effective masses anisotropy, a quite different behavior with respect to previous cases emerges.
First, as shown in the left column of Fig.\ \ref{fig:emasses_2}, let us notice that the restriction to semiconductors with a band gap of at least 0.01 eV removes many structures in the orthorhombic (\textit{mmm}) and monoclinic (\textit{2/m}) groups, while favoring structures with trigonal (\textit{-3m, 3m}) and hexagonal (\textit{-62m}) symmetry with respect to the general case shown in Fig.\ \ref{fig:overview}.
More importantly, we notice that these structures are not filtered out even when we select materials with increasingly high effective masses anisotropy ($\delta_\mathrm{me (mh)} \geq 0.05$ in the middle, $\delta_\mathrm{me (mh)} \geq 0.7$ on the right).
This means that, despite their structural symmetry, materials such as TMDCs in the T and H phase and Janus structures display a quite strong anisotropy in the effective masses.
One should bear in mind that we only calculate the curvatures of valence and conduction bands in one particular valley. While these are not bound to symmetries of the crystal, the overall transport properties (such as, for instance, the mobility) are determined by adding up contributions from all valleys, which in the end cancels out any anisotropic effect and restores the Neumann principle. However, it's worth noting that transport properties of a single anisotropic valley should in principle be accessible in experiments with valley-selection techniques such as circularly polarized optical excitation and gating \cite{Cao2012_ValleyHall, Mak14_ValleyHall, Lee2016_ValleyHall}

For the case of electron effective masses, we find that 61 stable materials have a rather high anisotropy degree $\delta_\mathrm{me} \geq 0.7$. While this group is dominated by monoclinic structures in the \textit{2/m} point group (mostly TMCD in the distorted T' phase) we find a rather large set of triclinic structures which sum up to roughly one third of the total, and a significant 13\% of share for hexagonal structures.
The situation is different for the case of hole effective masses, where orthorhombic structures represent a 34\% of the 50 materials with $\delta_\mathrm{mh} \geq 0.7$. However, we still find that 22\% of structures are in a trigonal point group as well.

We have reported all the 101 structures with $\delta_\mathrm{me}$ or $\delta_\mathrm{mh}$ greater that 0.7 in Fig.\ \ref{fig:emasses_2}, with 10 of them having both parameters above the threshold value. 
Among this group, we notice a recurrent presence of Hafnium, Bismuth and Antimony-based Janus structures, and Chromium based compounds (especially Chromium halides, which also bear elastic and magnetic anisotropy).
Most importantly, we find few materials that have already been exfoliated down to monolayer thickness in experiments, including four Hafnium and Zirconium-based TMDC in the stable T phase and the already mentioned FM compound CrBrS.
We also point out the presence of monolayers SbI$_3$, BiClTe, SbITe, and other ternary compounds in the crystal prototype \textit{ABC-59-ab} (point group \textit{mmm}) such as CrBrO and CrClO, which all seem to be easily exfoliable from the layered 3D parent bulk structure \cite{Mounet2017}.
Finally, we note that the direct or indirect character of the band gap does not seem to be relevant for attaining high anisotropy in the effective masses. Indeed, we find 18 out of 101 highly anisotropic materials with direct gap (that is 18\%), which is not dramatically different from the total fraction of semiconductors and insulators with direct gap among the 468 materials considered in this analysis (29 \%).

A list of experimentally available or easily exfoliable materials with highly anisotropic effective masses is presented in Table \ref{tab:emasses_candidates}, where we also report three experimentally available materials and one easily exfoliable material having $0.5 \leq \delta_\mathrm{me/mh} \leq 0.7$ (namely TiS$_3$, SnSe$_2$, GaTe, AuSe).
We also include additional structures that can be obtained from this subset by replacing a constituent element with atoms from the same group. This a relevant case for Janus monolayer, which can be obtained from already available structures by stripping off an outer layer of chalcogen atoms and substituting them with an element from the same family \cite{Lu17_Janus}. A similar method is likely to be applicable to halogen and chalcogen atoms in ternary orthorhombic compounds.
For each entry of Table \ref{tab:emasses_candidates}, we have double-checked the accuracy of the parabolic fit.

\begin{table}
    \centering
    %\begin{tabular}{ccccll}
    \begin{tabular}{c|c|c|c|l|l}
        material & pointgroup & $\delta_\mathrm{me}$ & $\delta_\mathrm{mh}$ & Ref.\ (exp.) & \makecell{exfol.\ \\ from bulk} \\
        \hline
        CrBrS (\textcolor{tab_purple}{AFM}) & mmm   & 0.89  & 0.94  & Ref.\ \onlinecite{Lee2020_arxiv_CrSBr}  & yes  \\  %Br2Cr2S2-d0a8f0957736
        W$_2$Se$_4$     & 2/m   & 0.07  & 0.88  & Ref.\ \onlinecite{Chen18_W2Se4}       &       \\  % W2Se4-800574ca4834
        CrBrS (\textcolor{tab_green}{FM})   & mmm   & 0.85  & 0.84  & Ref.\ \onlinecite{Lee2020_arxiv_CrSBr}  & yes  \\  % Br2Cr2S2-806bd2cbdf65
        HfSe$_2$        & -3m   & 0.85  & 0.00  & Ref.\ \onlinecite{Aretouli15_HfSe2}   & yes   \\  % HfSe2-7a708c5759cf
        Ti$_2$CO$_2$    & -3m   & 0.83  & 0.00  & Ref.\ \onlinecite{Melchior18_TiCO2}   &       \\  % CO2Ti2-c5bb3a35a784
        HfS$_2$         & -3m   & 0.81  & 0.00  & Ref.\ \onlinecite{Xu15_HfS2}          & yes   \\  % HfS2-d48ddb34811e
        ZrSe$_2$        & -3m   & 0.81  & 0.00  & Ref.\ \onlinecite{Manas16_ZrSe2}      & yes   \\  % ZrSe2-7dc2a0a57b42
        Re$_4$Se$_8$   & -1    & 0.80  & 0.30  & Ref.\ \onlinecite{Zhang16_ReS2}       & yes   \\  % Re4Se8-c3bc24117d8e
        SnS$_2$         & -3m   & 0.50  & 0.78  & Ref.\ \onlinecite{Sun12_SnS2}         & yes   \\  % SnS2-42a44e4e7298
        ZrS$_2$         & -3m   & 0.75  & 0.00  & Ref.\ \onlinecite{Zhang15_ZrS2}       & yes   \\  % ZrS2-8913f4e54692
        PbI$_2$         & -3m   & 0.00  & 0.72  & Ref.\ \onlinecite{Zheng19_PbI2}       & yes   \\  % PbI2-82db29775962
        Ti$_2$S$_6$     & 2/m   & 0.60  & 0.52  & Ref.\ \onlinecite{Island14}           & yes   \\  % Ti2S6-5529ea42c5a9
        SnSe$_2$        & -3m   & 0.55  & 0.00  & Ref.\ \onlinecite{Park16_SnSe2}       & yes   \\  % SnSe2-2c6d4c024ca0
        Ga$_2$Te$_2$    & -6m2  & 0.51  & 0.37  & Ref.\ \onlinecite{PozoZamudio15}      & yes   \\  % Ga2Te2-55c23ca88a05
        CrBrO (\textcolor{tab_purple}{AFM}) & mmm   & 0.37  & 0.84  &   & yes\cite{Haastrup18, Mounet2017}  \\  % Br2Cr2O2-3ab713cd571f \\
        CrClO (\textcolor{tab_purple}{AFM}) & mmm   & 0.83  & 0.79  &   & yes\cite{Haastrup18, Mounet2017}  \\  % Cl2Cr2O2-49392dbd4e5d \\
        CrBrO (\textcolor{tab_green}{FM})   & mmm   & 0.25  & 0.83  &   & yes\cite{Haastrup18, Mounet2017}  \\  % Br2Cr2O2-70429132c5ff \\
        CrClO (\textcolor{tab_green}{FM})   & mmm   & 0.45  & 0.81  &   & yes\cite{Haastrup18, Mounet2017}  \\  % Cl2Cr2O2-f0c029f3be24 \\
        I$_6$Sb$_2$       & -3m   & 0.01  & 0.80  &   & yes\cite{Haastrup18}              \\  % Sb2I6-d7ab89c33511
        Au$_2$Se$_2$      & 2/m   & 0.65  & 0.26  &   & yes\cite{Haastrup18, Mounet2017}  \\  % Au2Se2-3a5cf30b34d6
        \hline
        % BiClTe  & 3m    & 0.96  & 0.02  &   & yes\cite{Mounet2017}              \\  % BiClTe-968a6902b7f5
        % ISbTe   & 3m    & 0.96  & 0.72  &   & yes\cite{Mounet2017}              \\  % ISbTe-0f02957b17cf
        
        \rule{0pt}{4ex}
        material & pointgroup & $\delta_\mathrm{me}$ & $\delta_\mathrm{mh}$ & \multicolumn{2}{l}{\makecell{obtainable via \\ substitution}} \\
        \hline
        CrBrSe (\textcolor{tab_purple}{AFM})    & mmm   & 0.94  & 0.92  & \multicolumn{2}{l}{from CrBrS\cite{Mounet2017}} \\ % Br2Cr2Se2-1e7e654fbccd
        CrIS (\textcolor{tab_green}{FM})        & mmm   & 0.93  & 0.73  & \multicolumn{2}{l}{from CrBrS\cite{Mounet2017}} \\ % Cr2I2S2-a68f6e0481d3
        BrSbSe  & 3m    & 0.90  & 0.13  & \multicolumn{2}{l}{from ISbTe \cite{Mounet2017}}     \\  % BrSbSe-89b15ddef41d
        CrISe (\textcolor{tab_purple}{AFM})     & mmm   & 0.89  & 0.80  & \multicolumn{2}{l}{from CrBrS\cite{Mounet2017}} \\ % Cr2I2Se2-aabf9ecc5e0f
        CrBrSe (\textcolor{tab_green}{FM})      & mmm   & 0.88  & 0.88  & \multicolumn{2}{l}{from CrBrS\cite{Mounet2017}} \\ % Br2Cr2Se2-6d18e7a36c5a
        CrIS (\textcolor{tab_purple}{AFM})      & mmm   & 0.79  & 0.53  & \multicolumn{2}{l}{from CrBrS\cite{Mounet2017}} \\ % Cr2I2S2-7c12ce545360
        CrIO (\textcolor{tab_green}{FM})        & mmm   & 0.08  & 0.79  & \multicolumn{2}{l}{from CrBrO\cite{Mounet2017}, CrClO\cite{Mounet2017}} \\ % Cr2I2O2-30c4e34d1e6e
        HfSeTe  & 3m    & 0.79  & 0.01  & \multicolumn{2}{l}{from HfS$_2$\cite{Xu15_HfS2}, HfSe$_2$\cite{Aretouli15_HfSe2}} \\  % HfSeTe-305c779b8752
        % HfSTe   & 3m    & 0.78  & 0.00  & \multicolumn{2}{l}{from HfS$_2$\cite{Xu15_HfS2}} \\  % HfSTe-2602a918955b
        ZrSSe   & 3m    & 0.75  & 0.00  & \multicolumn{2}{l}{from ZrS$_2$\cite{Zhang15_ZrS2}, ZrSe$_2$\cite{Manas16_ZrSe2}} \\  % SSeZr-1a9901838600
        CrIO (\textcolor{tab_purple}{AFM})      & mmm   & 0.15  & 0.58  & \multicolumn{2}{l}{from CrBrO\cite{Mounet2017}, CrClO\cite{Mounet2017}} \\ % Cr2I2O2-19903e43bd98
        \hline
        % BrSSb   & 3m    & 0.96  & 0.13  & \multicolumn{2}{l}{from ISbTe}    \\  % BrSSb-4da5c6be60db
        % BiBrTe  & 3m    & 0.94  & 0.36  & \multicolumn{2}{l}{from BiTeI \cite{Fulop18_Janus}}   \\  % BiBrTe-304bc6a92d82
        % BiBrTe  & 3m    & 0.92  & 0.00  & \multicolumn{2}{l}{from BiTeI \cite{Fulop18_Janus}}   \\  % BiBrTe-f4f45fcade85
        % ClSbSe  & 3m    & 0.82  & 0.33  & \multicolumn{2}{l}{from ISbTe}    \\  % ClSbSe-f705a30af945

    \end{tabular}
    \caption{List of experimentally available or easily exfoliable materials with high anisotropy in the effective electron or hole masses. The list includes materials that can be obtained from available or exfoliable compounds by chemical substitution of one or more atomic species.}
    \label{tab:emasses_candidates}
\end{table}

\subsection{Polarizability}

The polarizability of a material relates the induced electric dipole moment density to an applied electric field to linear order.\cite{LeRu2008}. For 2D materials, this relation takes the form:
\begin{equation}
    P^{2D}_i(\vec{q},\omega)=\sum_j\alpha^{2D}_{ij}(\vec{q},\omega)E_j(\vec{q},\omega)
\end{equation}
where $P^{2D}$ is the induced polarization in the material averaged over the area of the unit cell, $E(\vec{q},\omega)$ is the the applied electric field, and $\alpha^{2D}$ is the polarizability.\cite{Haastrup18}.

In general, the polarizability can be split into a contribution from the electrons, $\alpha^{e}_{ij}(\vec{q},\omega)$, and a contribution from the lattice, $\alpha^{lat}_{ij}(\vec{q},\omega)$. 
Since the characteristic response time of the electrons is much faster than that of the lattice, the relevance of the two contributions depends on the timescale of the considered problem. For optical processes involving electromagnetic waves with frequency well above the characteristic phonon frequency of the lattice, only the electronic polarizability is relevant and we can write $\alpha_{ij}(\vec{q},\omega)\approx\alpha^{e}_{ij}(\vec{q},\omega)$. On the other hand, for processes involving infrared light, the lattice response must be considered as well and can in some case even dominate the electronic response. 

The polarizability determines the degree of dielectric screening in a material and as such it sets the strength of Coulomb interaction between charged particles\cite{Huser2013,TianSantos2019}. It thereby governs several of the unique properties that made 2D materials famous over the last decade \cite{Novoselov04,CastroNeto2009,Mak10_MoS2} including excitons, plasmons, and band gap renormalization effects\cite{thygesen2017calculating}.
In this context, the in-plane anisotropy of 2D materials has attracted significant interest since the synthesis of few-layer black phosporous (P$_4$) in 2014 \cite{Li14_BP,LiuYe2014,Xia14_phospho}. For example, the anisotropic optical absorption (essentially the imaginary part of the electronic polarizability) makes the material act as a linear polarizer\cite{TranYang2014}, which finds applications in diverse fields such as liquid-crystal displays, medical applications, or optical quantum computers \cite{Knill2001,Zeng2009}. In addition, other fundamental properties, such as the electron-phonon coupling and electron-hole interactions, are influenced by an anisotropic polarizability resulting in formation of quasiparticles, e.g. polarons, excitons, trions, with unconventional shapes and dispersion relations.\cite{Dresselhaus2016,LiTaniguchi2016,TranYang2014,xu2016extraordinarily,yang2015optical,deilmann2018unraveling,gjerding2020efficient}.

In the C2DB the electronic polarizability is calculated within the random phase approximation (RPA) \cite{RPA,RPA_2} using PBE wave functions and eigenvalues, see Ref \onlinecite{Haastrup18} for further details. To keep the discussion general we focus here on the polarizability in the static ($\omega=0$) and long wavelength ($q=0$) limits. As a measure of the degree of anisotropy we adopt the $\delta$ parameter defined above and define
\begin{equation}
    \delta_{\alpha^p}=\frac{|\alpha^p_x-\alpha^p_y|}{|\alpha^p_x|+|\alpha^p_y|} ,
\end{equation}
with $p = \{e, lat\}$ for the electronic and lattice polarizability respectively.

\begin{figure*}
    \includegraphics[width=\linewidth]{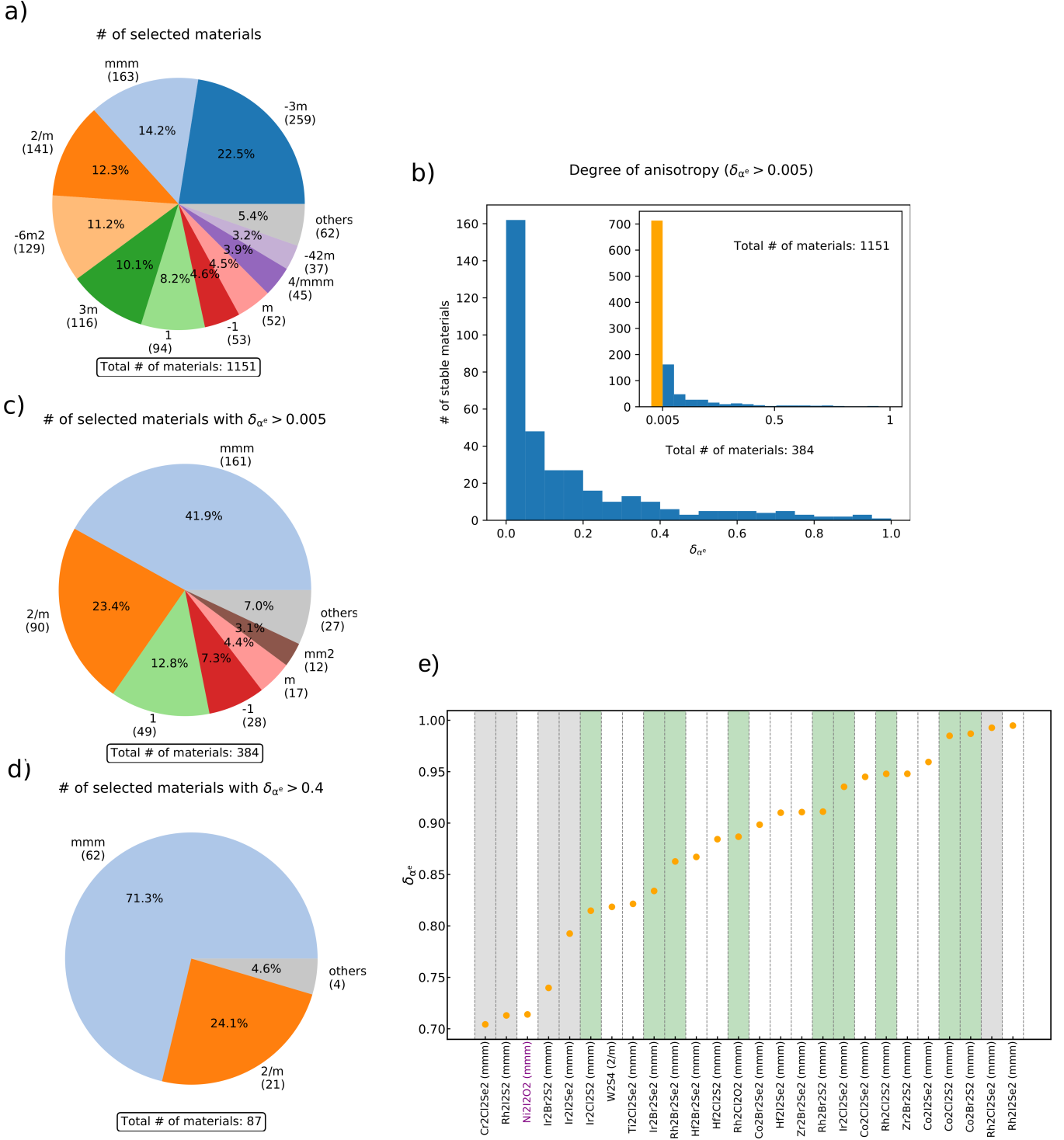}
    \caption{
        \label{fig:polari}
        a) Distribution of materials predicted to be stable in the C2DB database, according to their point group and their in-plane optical polarizability anisotropy. The percentage of materials represented by each point group and their number of occurrences in the database is explicitly shown, except for those point groups representing less than 3$\%$ of materials. In b) and c), classification of materials in the database with a low ($\delta_{\alpha^e}$>0.005) in-plane optical polarizability anisotropy threshold; in a histogram according to their degree of anisotropy in b), and in a pie chart according to their point group in c). The main chart in b) shows the distribution of the materials above the threshold, while in the inset an orange bin represents all materials below the threshold. In d) and e) the materials with a high anisotropy are classified. In d) a threshold of $\delta_{\alpha^e}$>0.4 is set. In e) the materials over a threshold of $\delta_{\alpha^e}$>0.7 are sorted by their anisotropy, and their magnetic state is indicated by the font color (purple for AFM).Green, grey and white background colors indicate the presence of a direct and indirect band gap and zero band gap according to HSE calculations
        }
\end{figure*}

In figure \ref{fig:polari} panels (c) and (d) we show the statistical distribution of the materials with electronic polarizability anisotropy above 0.005 and 0.4, respectively. For reference, the distribution of all the materials for which the polarizability has been calculated is shown in panel (a), and the materials are classified according to point group symmetry as usual. As the threshold is increased we see the same trend as for the magnetic, elastic, and effective masses anisotropies, namely, the orthorhomic \textit{mmm} point group followed by the monoclinic \textit{2/m} becomes increasingly dominant. Both the trigonal and tetragonal phases disappear from the distribution already for $\delta_{\alpha^{e}}$ >0.005. The othorhombic \textit{mmm} group is particularly ubiquitous among the materials with high $\delta_{\alpha^{e}}$, surpassing 70 $\%$ of the remaining materials already at a moderate threshold of $\delta_{\alpha^{e}}$ > 0.4.
In figure \ref{fig:polari}e we show the materials with the largest anisotropies found in the range $\delta_{\alpha^{e}}$ > 0.7. We note that our analysis correctly identifies the known in-plane anisotropic compounds such as P$_4$ \cite{Li14_BP} (phosphorene), As$_4$ \cite{Kamal15_As} (arsenene), MoS$_2$ (in the T' phase) and WTe$_2$ \cite{Tang17_WTe2} among others.
%applic: https://pubs.acs.org/doi/10.1021/acs.jpcc.7b07939

The materials with the highest $\delta_{\alpha^e}$ are ternary compounds and therefore more challenging to realize experimentally than the more common binary 2D materials, a notable exception being the recently synthesised Cs$_2$Br$_2$S$_2$ \cite{Lee2020_arxiv_CrSBr}, mentioned above, which has in addition a high electronic polarizability anisotropy. Regarding binary compounds, several of the materials with large $\delta_{\alpha^{e}}$ have been predicted to be exfoliable from known parent bulk materials \cite{Mounet2017}. Among our anisotropic materials that match the stoichiometry of the materials listed in \onlinecite{Mounet2017} as easily exfoliable, the most promising of these are listed in Table \ref{tab:pole}.

\begin{table}[htb]
   \centering
    \begin{tabular}{c|c|c|c|c|c}
         & Sym.&$E_{\mathrm{hull}}$(meV) & \makecell{Lowest $E_{\mathrm{hull}}$ \\ monolayer?} &$\delta_{\alpha^{e}}$\\
         \hline
         Cs$_2$Br$_2$S$_2$ \cite{Lee2020_arxiv_CrSBr}& Pmmm&0.0&Yes&0.41\\
         %\hline
         W$_2$Se$_4 \cite{Chen18_W2Se4}$& P$2_1$/m&91.7&No&0.47\\
         %\hline
         Zr$_2$I$_4$&P$2_1$/m&0.0&Yes&0.51\\
         %\hline
         Mo$_2$Se$_4$&P$2_1$/m&109.4&No&0.54\\
         %\hline
         Zr$_2$Cl$_4$&P$2_1$/m&31.9&No&0.60\\
         %\hline
         Ti$_2$Cl$_4$&P$2_1$/m&0.0&Yes&0.65\\
         %\hline
         W$_2$S$_4$&P$2_1$/m&177.9&No&0.82
    \end{tabular}
    \caption{Monolayers with the highest in-plane electronic polarizability anisotropy ($\delta_{\alpha^{e}}$) in the C2DB database whose stoichiometry matches that of the entries predicted to be easily exfoliable from a known layered bulk material in Ref. \onlinecite{Mounet2017}. Their space group symmetry, energy above convex hull, magnetic state, and in-plane electronic polarizability anisotropies are listed.}
    \label{tab:pole}
\end{table}

We highlight the experimentally known ternary compound Cs$_2$Br$_2$S$_2$ among the easily exfoliable materials from Table \ref{tab:pole}, presenting a high optical polarizability anisotropy of $\delta_{\alpha^{e}}$>0.41. Among the compounds not known yet experimentally we highlight two in the table, that have $\delta_{\alpha^{e}}$>0.65 and $\delta_{\alpha^{e}}$>0.5, respectively: Ti$_2$Cl$_4$ and Zr$_2$I$_4$. We stress that $\delta_{\alpha^{e}}=0.65$, for instance, implies that the polarizability in one direction of the plane is 4.7 times larger than in the other direction. Consequently, these materials are very promising candidates for anisotropic optical applications such as light polarizers.

%In spite of being a local property of atoms in a crystal, the electronic polarizability is an important variable ruling the dielectric behavior of 2D materials . A main reason is that, unlike in the bulk, the dielectric constant does not unequivocally define the dielectric properties of two-dimensional (2D) materials, due to the locality of their electrostatic screening \cite{LiSamori2016}. Therefore we can consider the anisotropy of the electronic polarization as a key property behind several of the outstanding electronic properties of 2D materials that have been discovered in the recent years. As an example of these properties, a 2D material may present 

\begin{figure*}
    \includegraphics[width=0.6\textwidth]{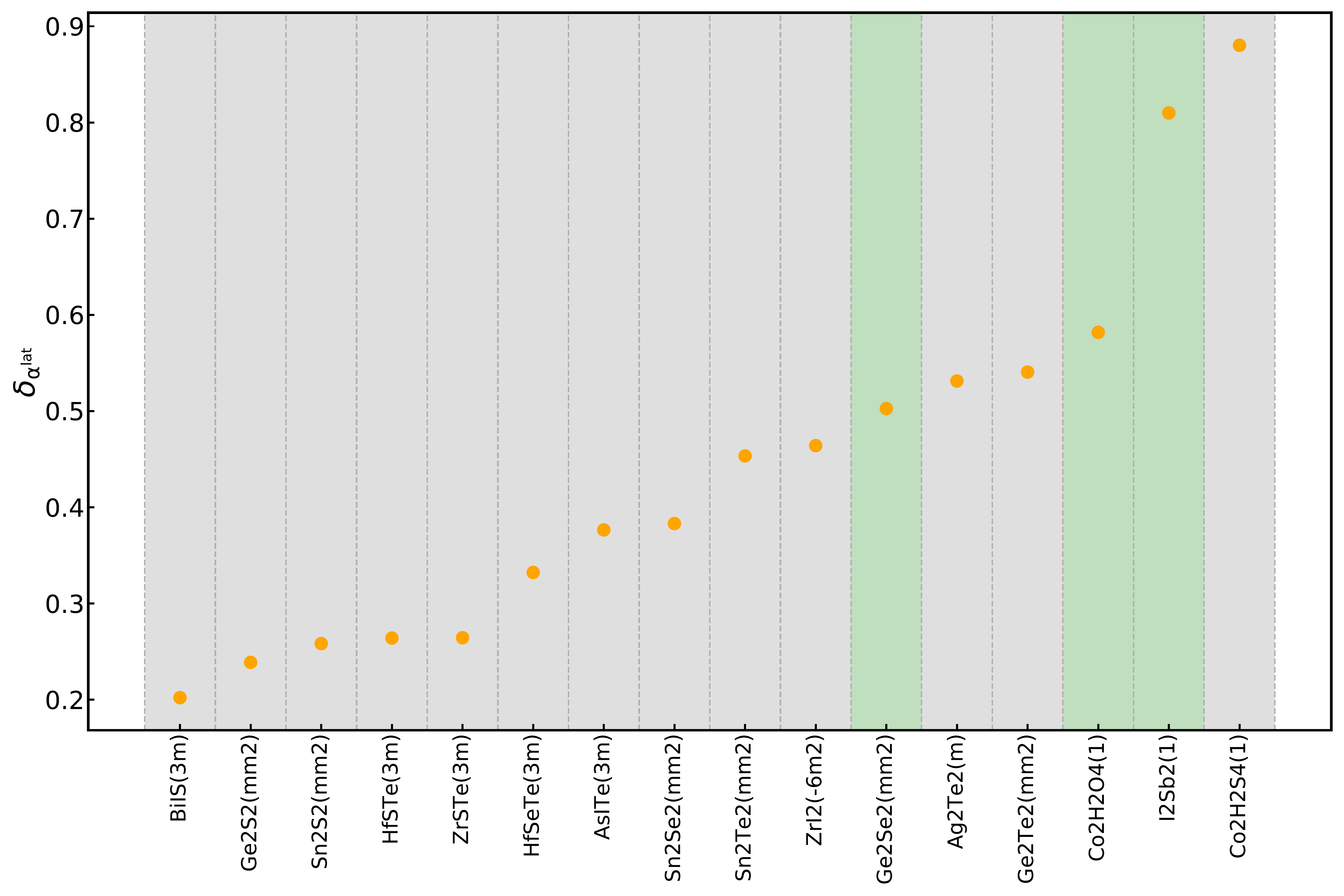}
    \caption{
        \label{fig:polat}
        Materials in the C2DB database with high infrared polarizability anisotropy ($\delta_{\alpha^{lat}}$>0.2) predicted to be stable.
        Green, grey and white background colors indicate the presence of a direct and indirect band gap and zero band gap according to HSE calculations.
        }
\end{figure*}

The lattice or infrared polarizability $\alpha^{lat}$ is calculated in the C2DB database only for materials meeting the requirements of high stability and band gap > 10 meV, which represent about 15\% of the materials in the database. Hence we will limit our analysis to extracting the most promising individual materials, since a statistical analysis would not be representative of the real distribution of the materials in the database. In Figure \ref{fig:polat} we show all 16 materials with $\delta_{\alpha^{lat}}$>0.2. As it is the case with the rest of properties, we find several materials with a significant anisotropy. For instance, the monolayer Sn$_2$Te$_2$ has a $\delta_{\alpha^{lat}}$ value 0.45, is at the bottom of the convex hull combining Sn and Te among monolayers and is considered to be easily exfoliable \cite{Mounet2017}. This makes it a very interesting material for further experimental and theoretical exploration. We find other materials whose stoichiometry matches easily exfoliable entries in Ref.\ \onlinecite{Mounet2017} among those with $\delta_{\alpha^{lat}}$>0.25, and are listed in Table \ref{tab:polat}. Taking into account that these promising materials are selected among only a little fraction of the entire C2DB database, we anticipate that there are a large amount of promising infrared anisotropic materials in the database yet to be discovered.

\begin{table}[htb]
   \centering
    \begin{tabular}{c|c|c|c|c|c}
         & Sym.&$E_{\mathrm{hull}}$(meV) & \makecell{Lowest $E_{\mathrm{hull}}$ \\ monolayer?} &$\delta_{\alpha^{lat}}$\\
         \hline
         Sn$_2$S$_2$& Pmn$2_1$&42.6&No&0.26\\
         %\hline
         Sn$_2$Se$_2$&Pmn$2_1$&42.9&No&0.38\\
         %\hline
         Sn$_2$Te$_2$&Pmn$2_1$&62.9&Yes&0.45\\
         %\hline
         ZrI$_2$&P$\bar{6}$m2&27.0&No&0.46\\
         %\hline
         Ge$_2$Se$_2$&Pmn$2_1$&24.9&No&0.50
    \end{tabular}
    \caption{Stable materials with the highest in-plane infrared polarizability anisotropy in the C2DB database whose stoichiometry matches that of entries in the easily exfoliable 2D materials list in Ref \onlinecite{Mounet2017}. Their space group symmetry, energy over the hull minimum, magnetic state and in-plane infrared polarizability anisotropies are given.}
    \label{tab:polat}
\end{table}

\section{Conclusions}
\label{sec:conclusions}

In this work, we have analyzed the presence of anisotropic behavior among more than one thousand 2D materials predicted to be stable in the C2DB database \cite{Haastrup18}. Specifically, we have identified materials with in-plane magnetic anisotropy, anisotropic Young's modulus and/or negative Poisson ratio, anisotropic effective masses, and anisotropic polarizabilities. 

Consistent with the Neumann principle, we have found that there are two main features in the C2DB database that favour anisotropy, namely (i) a lower symmetry and (ii) a larger number of constituent elements. In addition, our analysis satisfactorily captures the specific symmetry requirements of each anisotropy type: elastic and polarizability anisotropies, derived from second order tensors, are forbidden for trigonal, tetragonal or hexagonal compounds; the magnetic anisotropic materials do not include hexagonal and tetragonal groups, and the effective mass anisotropy is allowed for all symmetry groups in the database.
Several of the materials identified in this study outperform the known 2D materials in terms of anisotropic figures of merit and are predicted to be stable and/or exfoliable from known parent bulk crystals\cite{Mounet2017}, providing useful guidelines for future experimental investigations. 

The most prominent material class resulting from our analysis is the ternary orthorhombic compound prototype \textit{ABC-59-ab}\cite{Haastrup18}. This material class combines three different atomic species in a low symmetry structure, often resulting in strongly anisotropic properties. To the best of our knowledge, one of such materials (namely CrSBr) has been isolated in monolayer form very recently \cite{Lee2020_arxiv_CrSBr}, and further experimental efforts in this direction could hopefully be motivated by our work.

We find several binary monolayers with interesting anisotropic behaviors that are predicted to be stable and in some cases even predicted as easily exfoliable. A transition metal (in particular Ni, V, Cr, Os) combined with a halide in a low symmetry structure appears to be the best recipe for obtaining in-plane magnetic and elastic anisotropy. For instance, VI$_2$ is an exfoliable and very stable compound with an in-plane magnetic anisotropy that competes with the highest out-of-plane anisotropies known to date. Moreover, there are multiple compounds with large predicted anisotropies, that match the stoichiometry of exfoliable materials and with similar total energies. Such materials could be stabilized under the right experimental conditions. Among them, we highlight anti-ferromagnetic Ni$_2$I$_4$, which has an exceptional in-plane anisotropy exceeding 20 meV/unit cell, that makes it a candidate for realization of high temperature in-plane 2D antiferromagnetism. Likewise, the chromium di-halides stand out for their markedly anisotropic behavior in the elastic response and effective electron and hole masses.
Among the non-magnetic materials, we identify AuX (X=S, Se, Te) as a new class of potentially stable 2D materials with high anisotropy in several physical properties, with AuSe being reported as easily exfoliable in the literature.

On the other hand, transport properties of a single valley deserve to be mentioned separately, as they do not seem to be bound to the symmetries of the crystal lattice and could be experimentally accessed by means of circularly polarized light.
Several TMDCs and Janus structures have highly symmetric crystal struture with trigonal symmetry but strong effective mass anisotropy, and we identify HfX$_2$, ZrX$_2$, and SnX$_2$ (X = S, Se) as the most interesting monolayers for anisotropic transport applications already available in experiments, together with the above mentioned CrSBr. Moreover, new undiscovered structures with very low inter-layer binding energy such as SbI3, AuSe, and ternary magnetic compounds CrOBr and CrOCl display strong effective masses anisotropies.

Regarding the electronic polarizability, we also find a large amount of anisotropic materials, nearly all being ternary compounds of orthorhombic symmetry. In addition, some binary compounds, mostly involving a transition metal and a halide or a chalcogen, that are predicted to be easily exfoliable are identified and listed in the text. Finally, we also identified materials with interesting infrared polarizability anisotropy values among a smaller set of candidates in the C2DB. The most promising prospects for experimental realization are listed in the text.

Among the materials with negative Poisson ratios (so-called auxetic materials) identified in our study, we highlight HgI$_2$ and ZnCl$_2$, which are both predicted as easily exfoliable\cite{Mounet2017}, and MnTe, AgBr, and GeO$_2$, which are predicted to be stable in their monolayer form.

\begin{acknowledgments}

We thank Alireza Taghizadeh for useful discussions and suggestions.
The research leading to these results has received funding from the European Union's Horizon 2020 research and innovation program under the Marie Sk\l{}odowska-Curie grant agreement No.\ 754462. (EuroTechPostdoc).
KST acknowledges funding from the European Research Council (ERC) under the European Union's Horizon 2020 research and innovation programme (Grant No.\ 773122, LIMA).
The Center for Nanostructured Graphene is sponsored by the Danish National Research Foundation, Project DNRF103.

\end{acknowledgments}

\section*{Data availability}

The data that support the findings of this study are openly available online at the C2DB website \cite{C2DB}.

\section*{References}

\bibliography{biblio_anisotropy}

%merlin.mbs aipnum4-1.bst 2010-07-25 4.21a (PWD, AO, DPC) hacked
%Control: key (0)
%Control: author (8) initials jnrlst
%Control: editor formatted (1) identically to author
%Control: production of article title (0) allowed
%Control: page (1) range
%Control: year (1) truncated
%Control: production of eprint (0) enabled
\providecommand{\noopsort}[1]{}\providecommand{\singleletter}[1]{#1}%
\begin{thebibliography}{121}%
\makeatletter
\providecommand \@ifxundefined [1]{%
 \@ifx{#1\undefined}
}%
\providecommand \@ifnum [1]{%
 \ifnum #1\expandafter \@firstoftwo
 \else \expandafter \@secondoftwo
 \fi
}%
\providecommand \@ifx [1]{%
 \ifx #1\expandafter \@firstoftwo
 \else \expandafter \@secondoftwo
 \fi
}%
\providecommand \natexlab [1]{#1}%
\providecommand \enquote  [1]{``#1''}%
\providecommand \bibnamefont  [1]{#1}%
\providecommand \bibfnamefont [1]{#1}%
\providecommand \citenamefont [1]{#1}%
\providecommand \href@noop [0]{\@secondoftwo}%
\providecommand \href [0]{\begingroup \@sanitize@url \@href}%
\providecommand \@href[1]{\@@startlink{#1}\@@href}%
\providecommand \@@href[1]{\endgroup#1\@@endlink}%
\providecommand \@sanitize@url [0]{\catcode `\\12\catcode `\$12\catcode
  `\&12\catcode `\#12\catcode `\^12\catcode `\_12\catcode `\%12\relax}%
\providecommand \@@startlink[1]{}%
\providecommand \@@endlink[0]{}%
\providecommand \url  [0]{\begingroup\@sanitize@url \@url }%
\providecommand \@url [1]{\endgroup\@href {#1}{\urlprefix }}%
\providecommand \urlprefix  [0]{URL }%
\providecommand \Eprint [0]{\href }%
\providecommand \doibase [0]{http://dx.doi.org/}%
\providecommand \selectlanguage [0]{\@gobble}%
\providecommand \bibinfo  [0]{\@secondoftwo}%
\providecommand \bibfield  [0]{\@secondoftwo}%
\providecommand \translation [1]{[#1]}%
\providecommand \BibitemOpen [0]{}%
\providecommand \bibitemStop [0]{}%
\providecommand \bibitemNoStop [0]{.\EOS\space}%
\providecommand \EOS [0]{\spacefactor3000\relax}%
\providecommand \BibitemShut  [1]{\csname bibitem#1\endcsname}%
\let\auto@bib@innerbib\@empty
%</preamble>
\bibitem [{\citenamefont {{Gjerding}}\ \emph {et~al.}(2017)\citenamefont
  {{Gjerding}}, \citenamefont {{Petersen}}, \citenamefont {{Pedersen}},
  \citenamefont {{Mortensen}},\ and\ \citenamefont
  {{Thygesen}}}]{Gjerding17_hyperbolic}%
  \BibitemOpen
  \bibfield  {author} {\bibinfo {author} {\bibfnamefont {M.~N.}\ \bibnamefont
  {{Gjerding}}}, \bibinfo {author} {\bibfnamefont {R.}~\bibnamefont
  {{Petersen}}}, \bibinfo {author} {\bibfnamefont {T.~G.}\ \bibnamefont
  {{Pedersen}}}, \bibinfo {author} {\bibfnamefont {N.~A.}\ \bibnamefont
  {{Mortensen}}}, \ and\ \bibinfo {author} {\bibfnamefont {K.~S.}\ \bibnamefont
  {{Thygesen}}},\ }\bibfield  {title} {\enquote {\bibinfo {title} {{Layered van
  der Waals crystals with hyperbolic light dispersion}},}\ }\href {\doibase
  10.1038/s41467-017-00412-y} {\bibfield  {journal} {\bibinfo  {journal}
  {Nature Communications}\ }\textbf {\bibinfo {volume} {8}},\ \bibinfo {eid}
  {320} (\bibinfo {year} {2017})}\BibitemShut {NoStop}%
\bibitem [{\citenamefont {Jacob}, \citenamefont {Alekseyev},\ and\
  \citenamefont {Narimanov}(2006)}]{jacob2006optical}%
  \BibitemOpen
  \bibfield  {author} {\bibinfo {author} {\bibfnamefont {Z.}~\bibnamefont
  {Jacob}}, \bibinfo {author} {\bibfnamefont {L.~V.}\ \bibnamefont
  {Alekseyev}}, \ and\ \bibinfo {author} {\bibfnamefont {E.}~\bibnamefont
  {Narimanov}},\ }\bibfield  {title} {\enquote {\bibinfo {title} {Optical
  hyperlens: far-field imaging beyond the diffraction limit},}\ }\href
  {\doibase 10.1364/OE.14.008247} {\bibfield  {journal} {\bibinfo  {journal}
  {Optics express}\ }\textbf {\bibinfo {volume} {14}},\ \bibinfo {pages}
  {8247--8256} (\bibinfo {year} {2006})}\BibitemShut {NoStop}%
\bibitem [{\citenamefont {Lu}\ \emph {et~al.}(2014)\citenamefont {Lu},
  \citenamefont {Kan}, \citenamefont {Fullerton},\ and\ \citenamefont
  {Liu}}]{lu2014enhancing}%
  \BibitemOpen
  \bibfield  {author} {\bibinfo {author} {\bibfnamefont {D.}~\bibnamefont
  {Lu}}, \bibinfo {author} {\bibfnamefont {J.~J.}\ \bibnamefont {Kan}},
  \bibinfo {author} {\bibfnamefont {E.~E.}\ \bibnamefont {Fullerton}}, \ and\
  \bibinfo {author} {\bibfnamefont {Z.}~\bibnamefont {Liu}},\ }\bibfield
  {title} {\enquote {\bibinfo {title} {Enhancing spontaneous emission rates of
  molecules using nanopatterned multilayer hyperbolic metamaterials},}\ }\href
  {\doibase 10.1038/nnano.2013.276} {\bibfield  {journal} {\bibinfo  {journal}
  {Nature nanotechnology}\ }\textbf {\bibinfo {volume} {9}},\ \bibinfo {pages}
  {48--53} (\bibinfo {year} {2014})}\BibitemShut {NoStop}%
\bibitem [{\citenamefont {Novoselov}\ \emph {et~al.}(2004)\citenamefont
  {Novoselov}, \citenamefont {Geim}, \citenamefont {Morozov}, \citenamefont
  {Jiang}, \citenamefont {Zhang}, \citenamefont {Dubonos}, \citenamefont
  {Grigorieva},\ and\ \citenamefont {Firsov}}]{Novoselov04}%
  \BibitemOpen
  \bibfield  {author} {\bibinfo {author} {\bibfnamefont {K.~S.}\ \bibnamefont
  {Novoselov}}, \bibinfo {author} {\bibfnamefont {A.~K.}\ \bibnamefont {Geim}},
  \bibinfo {author} {\bibfnamefont {S.~V.}\ \bibnamefont {Morozov}}, \bibinfo
  {author} {\bibfnamefont {D.}~\bibnamefont {Jiang}}, \bibinfo {author}
  {\bibfnamefont {Y.}~\bibnamefont {Zhang}}, \bibinfo {author} {\bibfnamefont
  {S.~V.}\ \bibnamefont {Dubonos}}, \bibinfo {author} {\bibfnamefont {I.~V.}\
  \bibnamefont {Grigorieva}}, \ and\ \bibinfo {author} {\bibfnamefont {A.~A.}\
  \bibnamefont {Firsov}},\ }\bibfield  {title} {\enquote {\bibinfo {title}
  {{Electric Field Effect in Atomically Thin Carbon Films}},}\ }\href {\doibase
  10.1126/science.1102896} {\bibfield  {journal} {\bibinfo  {journal}
  {Science}\ }\textbf {\bibinfo {volume} {306}},\ \bibinfo {pages} {666--669}
  (\bibinfo {year} {2004})}\BibitemShut {NoStop}%
\bibitem [{\citenamefont {Ci}\ \emph {et~al.}(2010)\citenamefont {Ci},
  \citenamefont {Song}, \citenamefont {Jin}, \citenamefont {Jariwala},
  \citenamefont {Wu}, \citenamefont {Li}, \citenamefont {Srivastava},
  \citenamefont {Wang}, \citenamefont {Storr}, \citenamefont {Balicas} \emph
  {et~al.}}]{Ci10_BN}%
  \BibitemOpen
  \bibfield  {author} {\bibinfo {author} {\bibfnamefont {L.}~\bibnamefont
  {Ci}}, \bibinfo {author} {\bibfnamefont {L.}~\bibnamefont {Song}}, \bibinfo
  {author} {\bibfnamefont {C.}~\bibnamefont {Jin}}, \bibinfo {author}
  {\bibfnamefont {D.}~\bibnamefont {Jariwala}}, \bibinfo {author}
  {\bibfnamefont {D.}~\bibnamefont {Wu}}, \bibinfo {author} {\bibfnamefont
  {Y.}~\bibnamefont {Li}}, \bibinfo {author} {\bibfnamefont {A.}~\bibnamefont
  {Srivastava}}, \bibinfo {author} {\bibfnamefont {Z.}~\bibnamefont {Wang}},
  \bibinfo {author} {\bibfnamefont {K.}~\bibnamefont {Storr}}, \bibinfo
  {author} {\bibfnamefont {L.}~\bibnamefont {Balicas}},  \emph {et~al.},\
  }\bibfield  {title} {\enquote {\bibinfo {title} {Atomic layers of hybridized
  boron nitride and graphene domains},}\ }\href {\doibase 10.1038/nmat2711}
  {\bibfield  {journal} {\bibinfo  {journal} {Nature materials}\ }\textbf
  {\bibinfo {volume} {9}},\ \bibinfo {pages} {430--435} (\bibinfo {year}
  {2010})}\BibitemShut {NoStop}%
\bibitem [{\citenamefont {Mak}\ \emph {et~al.}(2010)\citenamefont {Mak},
  \citenamefont {Lee}, \citenamefont {Hone}, \citenamefont {Shan},\ and\
  \citenamefont {Heinz}}]{Mak10_MoS2}%
  \BibitemOpen
  \bibfield  {author} {\bibinfo {author} {\bibfnamefont {K.~F.}\ \bibnamefont
  {Mak}}, \bibinfo {author} {\bibfnamefont {C.}~\bibnamefont {Lee}}, \bibinfo
  {author} {\bibfnamefont {J.}~\bibnamefont {Hone}}, \bibinfo {author}
  {\bibfnamefont {J.}~\bibnamefont {Shan}}, \ and\ \bibinfo {author}
  {\bibfnamefont {T.~F.}\ \bibnamefont {Heinz}},\ }\bibfield  {title} {\enquote
  {\bibinfo {title} {{Atomically Thin ${\mathrm{MoS}}_{2}$: A New Direct-Gap
  Semiconductor}},}\ }\href {\doibase 10.1103/PhysRevLett.105.136805}
  {\bibfield  {journal} {\bibinfo  {journal} {Phys. Rev. Lett.}\ }\textbf
  {\bibinfo {volume} {105}},\ \bibinfo {pages} {136805} (\bibinfo {year}
  {2010})}\BibitemShut {NoStop}%
\bibitem [{\citenamefont {Li}\ \emph {et~al.}(2014)\citenamefont {Li},
  \citenamefont {Yu}, \citenamefont {Ye}, \citenamefont {Ge}, \citenamefont
  {Ou}, \citenamefont {Wu}, \citenamefont {Feng}, \citenamefont {Chen},\ and\
  \citenamefont {Zhang}}]{Li14_BP}%
  \BibitemOpen
  \bibfield  {author} {\bibinfo {author} {\bibfnamefont {L.}~\bibnamefont
  {Li}}, \bibinfo {author} {\bibfnamefont {Y.}~\bibnamefont {Yu}}, \bibinfo
  {author} {\bibfnamefont {G.~J.}\ \bibnamefont {Ye}}, \bibinfo {author}
  {\bibfnamefont {Q.}~\bibnamefont {Ge}}, \bibinfo {author} {\bibfnamefont
  {X.}~\bibnamefont {Ou}}, \bibinfo {author} {\bibfnamefont {H.}~\bibnamefont
  {Wu}}, \bibinfo {author} {\bibfnamefont {D.}~\bibnamefont {Feng}}, \bibinfo
  {author} {\bibfnamefont {X.~H.}\ \bibnamefont {Chen}}, \ and\ \bibinfo
  {author} {\bibfnamefont {Y.}~\bibnamefont {Zhang}},\ }\bibfield  {title}
  {\enquote {\bibinfo {title} {Black phosphorus field-effect transistors},}\
  }\href {\doibase 10.1038/nnano.2014.35} {\bibfield  {journal} {\bibinfo
  {journal} {Nature nanotechnology}\ }\textbf {\bibinfo {volume} {9}},\
  \bibinfo {pages} {372} (\bibinfo {year} {2014})}\BibitemShut {NoStop}%
\bibitem [{\citenamefont {Fei}\ and\ \citenamefont
  {Yang}(2014)}]{Fei14_phospho}%
  \BibitemOpen
  \bibfield  {author} {\bibinfo {author} {\bibfnamefont {R.}~\bibnamefont
  {Fei}}\ and\ \bibinfo {author} {\bibfnamefont {L.}~\bibnamefont {Yang}},\
  }\bibfield  {title} {\enquote {\bibinfo {title} {{Strain-Engineering the
  Anisotropic Electrical Conductance of Few-Layer Black Phosphorus}},}\ }\href
  {\doibase 10.1021/nl500935z} {\bibfield  {journal} {\bibinfo  {journal} {Nano
  Letters}\ }\textbf {\bibinfo {volume} {14}},\ \bibinfo {pages} {2884--2889}
  (\bibinfo {year} {2014})}\BibitemShut {NoStop}%
\bibitem [{\citenamefont {Xia}, \citenamefont {Wang},\ and\ \citenamefont
  {Jia}(2014)}]{Xia14_phospho}%
  \BibitemOpen
  \bibfield  {author} {\bibinfo {author} {\bibfnamefont {F.}~\bibnamefont
  {Xia}}, \bibinfo {author} {\bibfnamefont {H.}~\bibnamefont {Wang}}, \ and\
  \bibinfo {author} {\bibfnamefont {Y.}~\bibnamefont {Jia}},\ }\bibfield
  {title} {\enquote {\bibinfo {title} {Rediscovering black phosphorus as an
  anisotropic layered material for optoelectronics and electronics},}\ }\href
  {\doibase 10.1038/ncomms5458} {\bibfield  {journal} {\bibinfo  {journal}
  {Nat. Commun.}\ }\textbf {\bibinfo {volume} {5}},\ \bibinfo {pages} {4458}
  (\bibinfo {year} {2014})}\BibitemShut {NoStop}%
\bibitem [{\citenamefont {Wei}\ and\ \citenamefont {Peng}(2014)}]{Wei14}%
  \BibitemOpen
  \bibfield  {author} {\bibinfo {author} {\bibfnamefont {Q.}~\bibnamefont
  {Wei}}\ and\ \bibinfo {author} {\bibfnamefont {X.}~\bibnamefont {Peng}},\
  }\bibfield  {title} {\enquote {\bibinfo {title} {Superior mechanical
  flexibility of phosphorene and few-layer black phosphorus},}\ }\href
  {\doibase 10.1063/1.4885215} {\bibfield  {journal} {\bibinfo  {journal}
  {Applied Physics Letters}\ }\textbf {\bibinfo {volume} {104}},\ \bibinfo
  {pages} {251915} (\bibinfo {year} {2014})}\BibitemShut {NoStop}%
\bibitem [{\citenamefont {Low}\ \emph {et~al.}(2014)\citenamefont {Low},
  \citenamefont {Rold\'an}, \citenamefont {Wang}, \citenamefont {Xia},
  \citenamefont {Avouris}, \citenamefont {Moreno},\ and\ \citenamefont
  {Guinea}}]{Low14}%
  \BibitemOpen
  \bibfield  {author} {\bibinfo {author} {\bibfnamefont {T.}~\bibnamefont
  {Low}}, \bibinfo {author} {\bibfnamefont {R.}~\bibnamefont {Rold\'an}},
  \bibinfo {author} {\bibfnamefont {H.}~\bibnamefont {Wang}}, \bibinfo {author}
  {\bibfnamefont {F.}~\bibnamefont {Xia}}, \bibinfo {author} {\bibfnamefont
  {P.}~\bibnamefont {Avouris}}, \bibinfo {author} {\bibfnamefont {L.~M.}\
  \bibnamefont {Moreno}}, \ and\ \bibinfo {author} {\bibfnamefont
  {F.}~\bibnamefont {Guinea}},\ }\bibfield  {title} {\enquote {\bibinfo {title}
  {{Plasmons and Screening in Monolayer and Multilayer Black Phosphorus}},}\
  }\href {\doibase 10.1103/PhysRevLett.113.106802} {\bibfield  {journal}
  {\bibinfo  {journal} {Phys. Rev. Lett.}\ }\textbf {\bibinfo {volume} {113}},\
  \bibinfo {pages} {106802} (\bibinfo {year} {2014})}\BibitemShut {NoStop}%
\bibitem [{\citenamefont {Wang}\ \emph {et~al.}(2015)\citenamefont {Wang},
  \citenamefont {Kutana}, \citenamefont {Zou},\ and\ \citenamefont
  {Yakobson}}]{Wang15_phospho}%
  \BibitemOpen
  \bibfield  {author} {\bibinfo {author} {\bibfnamefont {L.}~\bibnamefont
  {Wang}}, \bibinfo {author} {\bibfnamefont {A.}~\bibnamefont {Kutana}},
  \bibinfo {author} {\bibfnamefont {X.}~\bibnamefont {Zou}}, \ and\ \bibinfo
  {author} {\bibfnamefont {B.~I.}\ \bibnamefont {Yakobson}},\ }\bibfield
  {title} {\enquote {\bibinfo {title} {Electro-mechanical anisotropy of
  phosphorene},}\ }\href {\doibase 10.1039/C5NR00355E} {\bibfield  {journal}
  {\bibinfo  {journal} {Nanoscale}\ }\textbf {\bibinfo {volume} {7}},\ \bibinfo
  {pages} {9746--9751} (\bibinfo {year} {2015})}\BibitemShut {NoStop}%
\bibitem [{\citenamefont {Jain}\ and\ \citenamefont
  {McGaughey}(2015)}]{Jain15_phospho}%
  \BibitemOpen
  \bibfield  {author} {\bibinfo {author} {\bibfnamefont {A.}~\bibnamefont
  {Jain}}\ and\ \bibinfo {author} {\bibfnamefont {A.}~\bibnamefont
  {McGaughey}},\ }\bibfield  {title} {\enquote {\bibinfo {title} {Strongly
  anisotropic in-plane thermal transport in single-layer black phosphorene},}\
  }\href {\doibase 10.1038/srep08501} {\bibfield  {journal} {\bibinfo
  {journal} {Sci Rep}\ }\textbf {\bibinfo {volume} {5}},\ \bibinfo {pages}
  {8501} (\bibinfo {year} {2015})}\BibitemShut {NoStop}%
\bibitem [{\citenamefont {{Wang}}\ \emph {et~al.}(2015)\citenamefont {{Wang}},
  \citenamefont {{Jones}}, \citenamefont {{Seyler}}, \citenamefont {{Tran}},
  \citenamefont {{Jia}}, \citenamefont {{Zhao}}, \citenamefont {{Wang}},
  \citenamefont {{Yang}}, \citenamefont {{Xu}},\ and\ \citenamefont
  {{Xia}}}]{Wang15}%
  \BibitemOpen
  \bibfield  {author} {\bibinfo {author} {\bibfnamefont {X.}~\bibnamefont
  {{Wang}}}, \bibinfo {author} {\bibfnamefont {A.~M.}\ \bibnamefont {{Jones}}},
  \bibinfo {author} {\bibfnamefont {K.~L.}\ \bibnamefont {{Seyler}}}, \bibinfo
  {author} {\bibfnamefont {V.}~\bibnamefont {{Tran}}}, \bibinfo {author}
  {\bibfnamefont {Y.}~\bibnamefont {{Jia}}}, \bibinfo {author} {\bibfnamefont
  {H.}~\bibnamefont {{Zhao}}}, \bibinfo {author} {\bibfnamefont
  {H.}~\bibnamefont {{Wang}}}, \bibinfo {author} {\bibfnamefont
  {L.}~\bibnamefont {{Yang}}}, \bibinfo {author} {\bibfnamefont
  {X.}~\bibnamefont {{Xu}}}, \ and\ \bibinfo {author} {\bibfnamefont
  {F.}~\bibnamefont {{Xia}}},\ }\bibfield  {title} {\enquote {\bibinfo {title}
  {{Highly anisotropic and robust excitons in monolayer black phosphorus}},}\
  }\href {\doibase 10.1038/nnano.2015.71} {\bibfield  {journal} {\bibinfo
  {journal} {Nature Nanotechnology}\ }\textbf {\bibinfo {volume} {10}},\
  \bibinfo {pages} {517--521} (\bibinfo {year} {2015})}\BibitemShut {NoStop}%
\bibitem [{\citenamefont {Tian}\ \emph {et~al.}(2016)\citenamefont {Tian},
  \citenamefont {Guo}, \citenamefont {Xie}, \citenamefont {Zhao}, \citenamefont
  {Li}, \citenamefont {Cha}, \citenamefont {Xia},\ and\ \citenamefont
  {Wang}}]{Tian16_BP_synapses}%
  \BibitemOpen
  \bibfield  {author} {\bibinfo {author} {\bibfnamefont {H.}~\bibnamefont
  {Tian}}, \bibinfo {author} {\bibfnamefont {Q.}~\bibnamefont {Guo}}, \bibinfo
  {author} {\bibfnamefont {Y.}~\bibnamefont {Xie}}, \bibinfo {author}
  {\bibfnamefont {H.}~\bibnamefont {Zhao}}, \bibinfo {author} {\bibfnamefont
  {C.}~\bibnamefont {Li}}, \bibinfo {author} {\bibfnamefont {J.~J.}\
  \bibnamefont {Cha}}, \bibinfo {author} {\bibfnamefont {F.}~\bibnamefont
  {Xia}}, \ and\ \bibinfo {author} {\bibfnamefont {H.}~\bibnamefont {Wang}},\
  }\bibfield  {title} {\enquote {\bibinfo {title} {{Anisotropic Black
  Phosphorus Synaptic Device for Neuromorphic Applications}},}\ }\href
  {\doibase 10.1002/adma.201600166} {\bibfield  {journal} {\bibinfo  {journal}
  {Advanced Materials}\ }\textbf {\bibinfo {volume} {28}},\ \bibinfo {pages}
  {4991--4997} (\bibinfo {year} {2016})}\BibitemShut {NoStop}%
\bibitem [{\citenamefont {Yang}\ \emph {et~al.}(2017)\citenamefont {Yang},
  \citenamefont {Jussila}, \citenamefont {Autere}, \citenamefont {Komsa},
  \citenamefont {Ye}, \citenamefont {Chen}, \citenamefont {Hasan},\ and\
  \citenamefont {Sun}}]{Yang17_birefringence}%
  \BibitemOpen
  \bibfield  {author} {\bibinfo {author} {\bibfnamefont {H.}~\bibnamefont
  {Yang}}, \bibinfo {author} {\bibfnamefont {H.}~\bibnamefont {Jussila}},
  \bibinfo {author} {\bibfnamefont {A.}~\bibnamefont {Autere}}, \bibinfo
  {author} {\bibfnamefont {H.-P.}\ \bibnamefont {Komsa}}, \bibinfo {author}
  {\bibfnamefont {G.}~\bibnamefont {Ye}}, \bibinfo {author} {\bibfnamefont
  {X.}~\bibnamefont {Chen}}, \bibinfo {author} {\bibfnamefont {T.}~\bibnamefont
  {Hasan}}, \ and\ \bibinfo {author} {\bibfnamefont {Z.}~\bibnamefont {Sun}},\
  }\bibfield  {title} {\enquote {\bibinfo {title} {{Optical Waveplates Based on
  Birefringence of Anisotropic Two-Dimensional Layered Materials}},}\ }\href
  {\doibase 10.1021/acsphotonics.7b00507} {\bibfield  {journal} {\bibinfo
  {journal} {ACS Photonics}\ }\textbf {\bibinfo {volume} {4}},\ \bibinfo
  {pages} {3023--3030} (\bibinfo {year} {2017})}\BibitemShut {NoStop}%
\bibitem [{\citenamefont {Wang}\ \emph {et~al.}(2020)\citenamefont {Wang},
  \citenamefont {Zhang}, \citenamefont {Huang}, \citenamefont {Xie},\ and\
  \citenamefont {Yan}}]{Wang20_optical_anisotropy}%
  \BibitemOpen
  \bibfield  {author} {\bibinfo {author} {\bibfnamefont {C.}~\bibnamefont
  {Wang}}, \bibinfo {author} {\bibfnamefont {G.}~\bibnamefont {Zhang}},
  \bibinfo {author} {\bibfnamefont {S.}~\bibnamefont {Huang}}, \bibinfo
  {author} {\bibfnamefont {Y.}~\bibnamefont {Xie}}, \ and\ \bibinfo {author}
  {\bibfnamefont {H.}~\bibnamefont {Yan}},\ }\bibfield  {title} {\enquote
  {\bibinfo {title} {{The Optical Properties and Plasmonics of Anisotropic 2D
  Materials}},}\ }\href {\doibase 10.1002/adom.201900996} {\bibfield  {journal}
  {\bibinfo  {journal} {Advanced Optical Materials}\ }\textbf {\bibinfo
  {volume} {8}},\ \bibinfo {pages} {1900996} (\bibinfo {year}
  {2020})}\BibitemShut {NoStop}%
\bibitem [{\citenamefont {Ma}\ \emph {et~al.}(2016)\citenamefont {Ma},
  \citenamefont {Chen}, \citenamefont {Han},\ and\ \citenamefont {Li}}]{Ma16}%
  \BibitemOpen
  \bibfield  {author} {\bibinfo {author} {\bibfnamefont {J.}~\bibnamefont
  {Ma}}, \bibinfo {author} {\bibfnamefont {Y.}~\bibnamefont {Chen}}, \bibinfo
  {author} {\bibfnamefont {Z.}~\bibnamefont {Han}}, \ and\ \bibinfo {author}
  {\bibfnamefont {W.}~\bibnamefont {Li}},\ }\bibfield  {title} {\enquote
  {\bibinfo {title} {{Strong anisotropic thermal conductivity of monolayer
  WTe$_2$}},}\ }\href {\doibase 10.1088/2053-1583/3/4/045010} {\bibfield
  {journal} {\bibinfo  {journal} {2D Materials}\ }\textbf {\bibinfo {volume}
  {3}},\ \bibinfo {pages} {045010} (\bibinfo {year} {2016})}\BibitemShut
  {NoStop}%
\bibitem [{\citenamefont {Torun}\ \emph {et~al.}(2016)\citenamefont {Torun},
  \citenamefont {Sahin}, \citenamefont {Cahangirov}, \citenamefont {Rubio},\
  and\ \citenamefont {Peeters}}]{Torun16}%
  \BibitemOpen
  \bibfield  {author} {\bibinfo {author} {\bibfnamefont {E.}~\bibnamefont
  {Torun}}, \bibinfo {author} {\bibfnamefont {H.}~\bibnamefont {Sahin}},
  \bibinfo {author} {\bibfnamefont {S.}~\bibnamefont {Cahangirov}}, \bibinfo
  {author} {\bibfnamefont {A.}~\bibnamefont {Rubio}}, \ and\ \bibinfo {author}
  {\bibfnamefont {F.~M.}\ \bibnamefont {Peeters}},\ }\bibfield  {title}
  {\enquote {\bibinfo {title} {{Anisotropic electronic, mechanical, and optical
  properties of monolayer WTe2}},}\ }\href {\doibase 10.1063/1.4942162}
  {\bibfield  {journal} {\bibinfo  {journal} {Journal of Applied Physics}\
  }\textbf {\bibinfo {volume} {119}},\ \bibinfo {pages} {074307} (\bibinfo
  {year} {2016})}\BibitemShut {NoStop}%
\bibitem [{\citenamefont {Zhang}\ \emph
  {et~al.}(2019{\natexlab{a}})\citenamefont {Zhang}, \citenamefont {Wang},
  \citenamefont {Dong}, \citenamefont {Wang}, \citenamefont {Fu},\ and\
  \citenamefont {Wang}}]{Zhang19}%
  \BibitemOpen
  \bibfield  {author} {\bibinfo {author} {\bibfnamefont {Y.-J.}\ \bibnamefont
  {Zhang}}, \bibinfo {author} {\bibfnamefont {R.-N.}\ \bibnamefont {Wang}},
  \bibinfo {author} {\bibfnamefont {G.-Y.}\ \bibnamefont {Dong}}, \bibinfo
  {author} {\bibfnamefont {S.-F.}\ \bibnamefont {Wang}}, \bibinfo {author}
  {\bibfnamefont {G.-S.}\ \bibnamefont {Fu}}, \ and\ \bibinfo {author}
  {\bibfnamefont {J.-L.}\ \bibnamefont {Wang}},\ }\bibfield  {title} {\enquote
  {\bibinfo {title} {{Mechanical properties of 1T-, 1T'-, and 1H-MX2 monolayers
  and their 1H/1T'-MX2 (M = Mo, W and X = S, Se, Te) heterostructures}},}\
  }\href {\doibase 10.1063/1.5128849} {\bibfield  {journal} {\bibinfo
  {journal} {AIP Advances}\ }\textbf {\bibinfo {volume} {9}},\ \bibinfo {pages}
  {125208} (\bibinfo {year} {2019}{\natexlab{a}})}\BibitemShut {NoStop}%
\bibitem [{\citenamefont {Zhang}\ \emph
  {et~al.}(2019{\natexlab{b}})\citenamefont {Zhang}, \citenamefont {Zhang},
  \citenamefont {Chen}, \citenamefont {Shen}, \citenamefont {An}, \citenamefont
  {Hu}, \citenamefont {Dong}, \citenamefont {Liu},\ and\ \citenamefont
  {Zhu}}]{Zhang19_WTe2}%
  \BibitemOpen
  \bibfield  {author} {\bibinfo {author} {\bibfnamefont {Q.}~\bibnamefont
  {Zhang}}, \bibinfo {author} {\bibfnamefont {R.}~\bibnamefont {Zhang}},
  \bibinfo {author} {\bibfnamefont {J.}~\bibnamefont {Chen}}, \bibinfo {author}
  {\bibfnamefont {W.}~\bibnamefont {Shen}}, \bibinfo {author} {\bibfnamefont
  {C.}~\bibnamefont {An}}, \bibinfo {author} {\bibfnamefont {X.}~\bibnamefont
  {Hu}}, \bibinfo {author} {\bibfnamefont {M.}~\bibnamefont {Dong}}, \bibinfo
  {author} {\bibfnamefont {J.}~\bibnamefont {Liu}}, \ and\ \bibinfo {author}
  {\bibfnamefont {L.}~\bibnamefont {Zhu}},\ }\bibfield  {title} {\enquote
  {\bibinfo {title} {{Remarkable electronic and optical anisotropy of layered
  1T'-WTe2 2D materials}},}\ }\href {\doibase 10.3762/bjnano.10.170} {\bibfield
   {journal} {\bibinfo  {journal} {Beilstein J. Nanotechnol.}\ }\textbf
  {\bibinfo {volume} {10}},\ \bibinfo {pages} {1745–1753} (\bibinfo {year}
  {2019}{\natexlab{b}})}\BibitemShut {NoStop}%
\bibitem [{\citenamefont {{Wang}}\ \emph {et~al.}(2020)\citenamefont {{Wang}},
  \citenamefont {{Huang}}, \citenamefont {{Xing}}, \citenamefont {{Xie}},
  \citenamefont {{Song}}, \citenamefont {{Wang}},\ and\ \citenamefont
  {{Yan}}}]{Wang20_WTe2_hyperbolic}%
  \BibitemOpen
  \bibfield  {author} {\bibinfo {author} {\bibfnamefont {C.}~\bibnamefont
  {{Wang}}}, \bibinfo {author} {\bibfnamefont {S.}~\bibnamefont {{Huang}}},
  \bibinfo {author} {\bibfnamefont {Q.}~\bibnamefont {{Xing}}}, \bibinfo
  {author} {\bibfnamefont {Y.}~\bibnamefont {{Xie}}}, \bibinfo {author}
  {\bibfnamefont {C.}~\bibnamefont {{Song}}}, \bibinfo {author} {\bibfnamefont
  {F.}~\bibnamefont {{Wang}}}, \ and\ \bibinfo {author} {\bibfnamefont
  {H.}~\bibnamefont {{Yan}}},\ }\bibfield  {title} {\enquote {\bibinfo {title}
  {{Van der Waals thin films of WTe$_{2}$ for natural hyperbolic plasmonic
  surfaces}},}\ }\href {\doibase 10.1038/s41467-020-15001-9} {\bibfield
  {journal} {\bibinfo  {journal} {Nature Communications}\ }\textbf {\bibinfo
  {volume} {11}},\ \bibinfo {eid} {1158} (\bibinfo {year} {2020})}\BibitemShut
  {NoStop}%
\bibitem [{\citenamefont {Island}\ \emph {et~al.}(2014)\citenamefont {Island},
  \citenamefont {Buscema}, \citenamefont {Barawi}, \citenamefont {Clamagirand},
  \citenamefont {Ares}, \citenamefont {Sánchez}, \citenamefont {Ferrer},
  \citenamefont {Steele}, \citenamefont {van~der Zant},\ and\ \citenamefont
  {Castellanos-Gomez}}]{Island14}%
  \BibitemOpen
  \bibfield  {author} {\bibinfo {author} {\bibfnamefont {J.~O.}\ \bibnamefont
  {Island}}, \bibinfo {author} {\bibfnamefont {M.}~\bibnamefont {Buscema}},
  \bibinfo {author} {\bibfnamefont {M.}~\bibnamefont {Barawi}}, \bibinfo
  {author} {\bibfnamefont {J.~M.}\ \bibnamefont {Clamagirand}}, \bibinfo
  {author} {\bibfnamefont {J.~R.}\ \bibnamefont {Ares}}, \bibinfo {author}
  {\bibfnamefont {C.}~\bibnamefont {Sánchez}}, \bibinfo {author}
  {\bibfnamefont {I.~J.}\ \bibnamefont {Ferrer}}, \bibinfo {author}
  {\bibfnamefont {G.~A.}\ \bibnamefont {Steele}}, \bibinfo {author}
  {\bibfnamefont {H.~S.~J.}\ \bibnamefont {van~der Zant}}, \ and\ \bibinfo
  {author} {\bibfnamefont {A.}~\bibnamefont {Castellanos-Gomez}},\ }\bibfield
  {title} {\enquote {\bibinfo {title} {{Ultrahigh Photoresponse of Few-Layer
  TiS3 Nanoribbon Transistors}},}\ }\href {\doibase 10.1002/adom.201400043}
  {\bibfield  {journal} {\bibinfo  {journal} {Advanced Optical Materials}\
  }\textbf {\bibinfo {volume} {2}},\ \bibinfo {pages} {641--645} (\bibinfo
  {year} {2014})}\BibitemShut {NoStop}%
\bibitem [{\citenamefont {Island}\ \emph {et~al.}(2015)\citenamefont {Island},
  \citenamefont {Barawi}, \citenamefont {Biele}, \citenamefont {Almazán},
  \citenamefont {Clamagirand}, \citenamefont {Ares}, \citenamefont {Sánchez},
  \citenamefont {van~der Zant}, \citenamefont {Álvarez}, \citenamefont
  {D'Agosta}, \citenamefont {Ferrer},\ and\ \citenamefont
  {Castellanos-Gomez}}]{Island15}%
  \BibitemOpen
  \bibfield  {author} {\bibinfo {author} {\bibfnamefont {J.~O.}\ \bibnamefont
  {Island}}, \bibinfo {author} {\bibfnamefont {M.}~\bibnamefont {Barawi}},
  \bibinfo {author} {\bibfnamefont {R.}~\bibnamefont {Biele}}, \bibinfo
  {author} {\bibfnamefont {A.}~\bibnamefont {Almazán}}, \bibinfo {author}
  {\bibfnamefont {J.~M.}\ \bibnamefont {Clamagirand}}, \bibinfo {author}
  {\bibfnamefont {J.~R.}\ \bibnamefont {Ares}}, \bibinfo {author}
  {\bibfnamefont {C.}~\bibnamefont {Sánchez}}, \bibinfo {author}
  {\bibfnamefont {H.~S.~J.}\ \bibnamefont {van~der Zant}}, \bibinfo {author}
  {\bibfnamefont {J.~V.}\ \bibnamefont {Álvarez}}, \bibinfo {author}
  {\bibfnamefont {R.}~\bibnamefont {D'Agosta}}, \bibinfo {author}
  {\bibfnamefont {I.~J.}\ \bibnamefont {Ferrer}}, \ and\ \bibinfo {author}
  {\bibfnamefont {A.}~\bibnamefont {Castellanos-Gomez}},\ }\bibfield  {title}
  {\enquote {\bibinfo {title} {{TiS3 Transistors with Tailored Morphology and
  Electrical Properties}},}\ }\href {\doibase 10.1002/adma.201405632}
  {\bibfield  {journal} {\bibinfo  {journal} {Advanced Materials}\ }\textbf
  {\bibinfo {volume} {27}},\ \bibinfo {pages} {2595--2601} (\bibinfo {year}
  {2015})}\BibitemShut {NoStop}%
\bibitem [{\citenamefont {Kang}, \citenamefont {Sahin},\ and\ \citenamefont
  {Peeters}(2015)}]{Kang15}%
  \BibitemOpen
  \bibfield  {author} {\bibinfo {author} {\bibfnamefont {J.}~\bibnamefont
  {Kang}}, \bibinfo {author} {\bibfnamefont {H.}~\bibnamefont {Sahin}}, \ and\
  \bibinfo {author} {\bibfnamefont {F.~M.}\ \bibnamefont {Peeters}},\
  }\bibfield  {title} {\enquote {\bibinfo {title} {Mechanical properties of
  monolayer sulphides: a comparative study between mos2, hfs2 and tis3},}\
  }\href {\doibase 10.1039/C5CP04576B} {\bibfield  {journal} {\bibinfo
  {journal} {Phys. Chem. Chem. Phys.}\ }\textbf {\bibinfo {volume} {17}},\
  \bibinfo {pages} {27742--27749} (\bibinfo {year} {2015})}\BibitemShut
  {NoStop}%
\bibitem [{\citenamefont {Jin}, \citenamefont {Li},\ and\ \citenamefont
  {Yang}(2015)}]{Jin15}%
  \BibitemOpen
  \bibfield  {author} {\bibinfo {author} {\bibfnamefont {Y.}~\bibnamefont
  {Jin}}, \bibinfo {author} {\bibfnamefont {X.}~\bibnamefont {Li}}, \ and\
  \bibinfo {author} {\bibfnamefont {J.}~\bibnamefont {Yang}},\ }\bibfield
  {title} {\enquote {\bibinfo {title} {{Single layer of MX3 (M = Ti{,} Zr; X =
  S{,} Se{,} Te): a new platform for nano-electronics and optics}},}\ }\href
  {\doibase 10.1039/C5CP02813B} {\bibfield  {journal} {\bibinfo  {journal}
  {Phys. Chem. Chem. Phys.}\ }\textbf {\bibinfo {volume} {17}},\ \bibinfo
  {pages} {18665--18669} (\bibinfo {year} {2015})}\BibitemShut {NoStop}%
\bibitem [{\citenamefont {Silva-Guill{\'{e}}n}\ \emph
  {et~al.}(2017)\citenamefont {Silva-Guill{\'{e}}n}, \citenamefont {Canadell},
  \citenamefont {Ordej{\'{o}}n}, \citenamefont {Guinea},\ and\ \citenamefont
  {Rold{\'{a}}n}}]{SilvaGuillen17}%
  \BibitemOpen
  \bibfield  {author} {\bibinfo {author} {\bibfnamefont {J.~A.}\ \bibnamefont
  {Silva-Guill{\'{e}}n}}, \bibinfo {author} {\bibfnamefont {E.}~\bibnamefont
  {Canadell}}, \bibinfo {author} {\bibfnamefont {P.}~\bibnamefont
  {Ordej{\'{o}}n}}, \bibinfo {author} {\bibfnamefont {F.}~\bibnamefont
  {Guinea}}, \ and\ \bibinfo {author} {\bibfnamefont {R.}~\bibnamefont
  {Rold{\'{a}}n}},\ }\bibfield  {title} {\enquote {\bibinfo {title}
  {{Anisotropic features in the electronic structure of the two-dimensional
  transition metal trichalcogenide {TiS}$_3$: electron doping and plasmons}},}\
  }\href {\doibase 10.1088/2053-1583/aa6b92} {\bibfield  {journal} {\bibinfo
  {journal} {2D Materials}\ }\textbf {\bibinfo {volume} {4}},\ \bibinfo {pages}
  {025085} (\bibinfo {year} {2017})}\BibitemShut {NoStop}%
\bibitem [{\citenamefont {Khatibi}\ \emph {et~al.}(2019)\citenamefont
  {Khatibi}, \citenamefont {Godiksen}, \citenamefont {Basuvalingam},
  \citenamefont {Pellegrino}, \citenamefont {Bol}, \citenamefont {Shokri},\
  and\ \citenamefont {Curto}}]{Khatibi19}%
  \BibitemOpen
  \bibfield  {author} {\bibinfo {author} {\bibfnamefont {A.}~\bibnamefont
  {Khatibi}}, \bibinfo {author} {\bibfnamefont {R.~H.}\ \bibnamefont
  {Godiksen}}, \bibinfo {author} {\bibfnamefont {S.~B.}\ \bibnamefont
  {Basuvalingam}}, \bibinfo {author} {\bibfnamefont {D.}~\bibnamefont
  {Pellegrino}}, \bibinfo {author} {\bibfnamefont {A.~A.}\ \bibnamefont {Bol}},
  \bibinfo {author} {\bibfnamefont {B.}~\bibnamefont {Shokri}}, \ and\ \bibinfo
  {author} {\bibfnamefont {A.~G.}\ \bibnamefont {Curto}},\ }\bibfield  {title}
  {\enquote {\bibinfo {title} {{Anisotropic infrared light emission from
  quasi-1D layered {TiS}$_3$}},}\ }\href {\doibase 10.1088/2053-1583/ab57ef}
  {\bibfield  {journal} {\bibinfo  {journal} {2D Materials}\ }\textbf {\bibinfo
  {volume} {7}},\ \bibinfo {pages} {015022} (\bibinfo {year}
  {2019})}\BibitemShut {NoStop}%
\bibitem [{\citenamefont {{Liu}}\ \emph {et~al.}(2015)\citenamefont {{Liu}},
  \citenamefont {{Fu}}, \citenamefont {{Wang}}, \citenamefont {{Feng}},
  \citenamefont {{Liu}}, \citenamefont {{Wan}}, \citenamefont {{Zhou}},
  \citenamefont {{Wang}}, \citenamefont {{Shao}}, \citenamefont {{Ho}},
  \citenamefont {{Huang}}, \citenamefont {{Cao}}, \citenamefont {{Wang}},
  \citenamefont {{Li}}, \citenamefont {{Zeng}}, \citenamefont {{Song}},
  \citenamefont {{Wang}}, \citenamefont {{Shi}}, \citenamefont {{Yuan}},
  \citenamefont {{Hwang}}, \citenamefont {{Cui}}, \citenamefont {{Miao}},\ and\
  \citenamefont {{Xing}}}]{Liu15_ReS2_FET}%
  \BibitemOpen
  \bibfield  {author} {\bibinfo {author} {\bibfnamefont {E.}~\bibnamefont
  {{Liu}}}, \bibinfo {author} {\bibfnamefont {Y.}~\bibnamefont {{Fu}}},
  \bibinfo {author} {\bibfnamefont {Y.}~\bibnamefont {{Wang}}}, \bibinfo
  {author} {\bibfnamefont {Y.}~\bibnamefont {{Feng}}}, \bibinfo {author}
  {\bibfnamefont {H.}~\bibnamefont {{Liu}}}, \bibinfo {author} {\bibfnamefont
  {X.}~\bibnamefont {{Wan}}}, \bibinfo {author} {\bibfnamefont
  {W.}~\bibnamefont {{Zhou}}}, \bibinfo {author} {\bibfnamefont
  {B.}~\bibnamefont {{Wang}}}, \bibinfo {author} {\bibfnamefont
  {L.}~\bibnamefont {{Shao}}}, \bibinfo {author} {\bibfnamefont {C.-H.}\
  \bibnamefont {{Ho}}}, \bibinfo {author} {\bibfnamefont {Y.-S.}\ \bibnamefont
  {{Huang}}}, \bibinfo {author} {\bibfnamefont {Z.}~\bibnamefont {{Cao}}},
  \bibinfo {author} {\bibfnamefont {L.}~\bibnamefont {{Wang}}}, \bibinfo
  {author} {\bibfnamefont {A.}~\bibnamefont {{Li}}}, \bibinfo {author}
  {\bibfnamefont {J.}~\bibnamefont {{Zeng}}}, \bibinfo {author} {\bibfnamefont
  {F.}~\bibnamefont {{Song}}}, \bibinfo {author} {\bibfnamefont
  {X.}~\bibnamefont {{Wang}}}, \bibinfo {author} {\bibfnamefont
  {Y.}~\bibnamefont {{Shi}}}, \bibinfo {author} {\bibfnamefont
  {H.}~\bibnamefont {{Yuan}}}, \bibinfo {author} {\bibfnamefont {H.~Y.}\
  \bibnamefont {{Hwang}}}, \bibinfo {author} {\bibfnamefont {Y.}~\bibnamefont
  {{Cui}}}, \bibinfo {author} {\bibfnamefont {F.}~\bibnamefont {{Miao}}}, \
  and\ \bibinfo {author} {\bibfnamefont {D.}~\bibnamefont {{Xing}}},\
  }\bibfield  {title} {\enquote {\bibinfo {title} {{Integrated digital
  inverters based on two-dimensional anisotropic ReS2 field-effect
  transistors}},}\ }\href {\doibase 10.1038/ncomms7991} {\bibfield  {journal}
  {\bibinfo  {journal} {Nat Commun}\ }\textbf {\bibinfo {volume} {6}},\
  \bibinfo {pages} {6991} (\bibinfo {year} {2015})}\BibitemShut {NoStop}%
\bibitem [{\citenamefont {Zhang}\ \emph {et~al.}(2016)\citenamefont {Zhang},
  \citenamefont {Wang}, \citenamefont {Li}, \citenamefont {Wang}, \citenamefont
  {Song}, \citenamefont {Huang}, \citenamefont {Chen}, \citenamefont {Yang},
  \citenamefont {Zhang}, \citenamefont {Lu}, \citenamefont {Wang},
  \citenamefont {Liu}, \citenamefont {Fang}, \citenamefont {Zhou},
  \citenamefont {Yan}, \citenamefont {Zou}, \citenamefont {Wan}, \citenamefont
  {Zhou}, \citenamefont {Hu},\ and\ \citenamefont {Xiu}}]{Zhang16_ReS2}%
  \BibitemOpen
  \bibfield  {author} {\bibinfo {author} {\bibfnamefont {E.}~\bibnamefont
  {Zhang}}, \bibinfo {author} {\bibfnamefont {P.}~\bibnamefont {Wang}},
  \bibinfo {author} {\bibfnamefont {Z.}~\bibnamefont {Li}}, \bibinfo {author}
  {\bibfnamefont {H.}~\bibnamefont {Wang}}, \bibinfo {author} {\bibfnamefont
  {C.}~\bibnamefont {Song}}, \bibinfo {author} {\bibfnamefont {C.}~\bibnamefont
  {Huang}}, \bibinfo {author} {\bibfnamefont {Z.-G.}\ \bibnamefont {Chen}},
  \bibinfo {author} {\bibfnamefont {L.}~\bibnamefont {Yang}}, \bibinfo {author}
  {\bibfnamefont {K.}~\bibnamefont {Zhang}}, \bibinfo {author} {\bibfnamefont
  {S.}~\bibnamefont {Lu}}, \bibinfo {author} {\bibfnamefont {W.}~\bibnamefont
  {Wang}}, \bibinfo {author} {\bibfnamefont {S.}~\bibnamefont {Liu}}, \bibinfo
  {author} {\bibfnamefont {H.}~\bibnamefont {Fang}}, \bibinfo {author}
  {\bibfnamefont {X.}~\bibnamefont {Zhou}}, \bibinfo {author} {\bibfnamefont
  {H.}~\bibnamefont {Yan}}, \bibinfo {author} {\bibfnamefont {J.}~\bibnamefont
  {Zou}}, \bibinfo {author} {\bibfnamefont {X.}~\bibnamefont {Wan}}, \bibinfo
  {author} {\bibfnamefont {P.}~\bibnamefont {Zhou}}, \bibinfo {author}
  {\bibfnamefont {W.}~\bibnamefont {Hu}}, \ and\ \bibinfo {author}
  {\bibfnamefont {F.}~\bibnamefont {Xiu}},\ }\bibfield  {title} {\enquote
  {\bibinfo {title} {{Tunable Ambipolar Polarization-Sensitive Photodetectors
  Based on High-Anisotropy ReSe2 Nanosheets}},}\ }\href {\doibase
  10.1021/acsnano.6b04165} {\bibfield  {journal} {\bibinfo  {journal} {ACS
  Nano}\ }\textbf {\bibinfo {volume} {10}},\ \bibinfo {pages} {8067--8077}
  (\bibinfo {year} {2016})}\BibitemShut {NoStop}%
\bibitem [{\citenamefont {Echeverry}\ and\ \citenamefont
  {Gerber}(2018)}]{Echeverry18}%
  \BibitemOpen
  \bibfield  {author} {\bibinfo {author} {\bibfnamefont {J.~P.}\ \bibnamefont
  {Echeverry}}\ and\ \bibinfo {author} {\bibfnamefont {I.~C.}\ \bibnamefont
  {Gerber}},\ }\bibfield  {title} {\enquote {\bibinfo {title} {{Theoretical
  investigations of the anisotropic optical properties of distorted 1T ReS$_2$
  and ReSe$_2$ monolayers, bilayers, and in the bulk limit}},}\ }\href
  {\doibase 10.1103/PhysRevB.97.075123} {\bibfield  {journal} {\bibinfo
  {journal} {Phys. Rev. B}\ }\textbf {\bibinfo {volume} {97}},\ \bibinfo
  {pages} {075123} (\bibinfo {year} {2018})}\BibitemShut {NoStop}%
\bibitem [{\citenamefont {Liu}\ \emph {et~al.}(2020)\citenamefont {Liu},
  \citenamefont {Yuan}, \citenamefont {Wang}, \citenamefont {Deacon},
  \citenamefont {Yoo}, \citenamefont {Sun},\ and\ \citenamefont
  {Ishibashi}}]{Liu20_ReSe2}%
  \BibitemOpen
  \bibfield  {author} {\bibinfo {author} {\bibfnamefont {X.}~\bibnamefont
  {Liu}}, \bibinfo {author} {\bibfnamefont {Y.}~\bibnamefont {Yuan}}, \bibinfo
  {author} {\bibfnamefont {Z.}~\bibnamefont {Wang}}, \bibinfo {author}
  {\bibfnamefont {R.~S.}\ \bibnamefont {Deacon}}, \bibinfo {author}
  {\bibfnamefont {W.~J.}\ \bibnamefont {Yoo}}, \bibinfo {author} {\bibfnamefont
  {J.}~\bibnamefont {Sun}}, \ and\ \bibinfo {author} {\bibfnamefont
  {K.}~\bibnamefont {Ishibashi}},\ }\bibfield  {title} {\enquote {\bibinfo
  {title} {{Directly Probing Effective-Mass Anisotropy of Two-Dimensional
  ${\mathrm{Re}\mathrm{Se}}_{2}$ in Schottky Tunnel Transistors}},}\ }\href
  {\doibase 10.1103/PhysRevApplied.13.044056} {\bibfield  {journal} {\bibinfo
  {journal} {Phys. Rev. Applied}\ }\textbf {\bibinfo {volume} {13}},\ \bibinfo
  {pages} {044056} (\bibinfo {year} {2020})}\BibitemShut {NoStop}%
\bibitem [{\citenamefont {{Wang}}\ \emph {et~al.}(2019)\citenamefont {{Wang}},
  \citenamefont {{Chen}}, \citenamefont {{Zhu}}, \citenamefont {{Wang}},
  \citenamefont {{Dong}}, \citenamefont {{Sun}}, \citenamefont {{Zhang}},
  \citenamefont {{Cao}}, \citenamefont {{Li}}, \citenamefont {{Huang}},
  \citenamefont {{Zhang}}, \citenamefont {{Liu}}, \citenamefont {{Sun}},
  \citenamefont {{Ye}}, \citenamefont {{Song}}, \citenamefont {{Wang}},
  \citenamefont {{Han}}, \citenamefont {{Yang}}, \citenamefont {{Guo}},
  \citenamefont {{Qin}}, \citenamefont {{Xiao}}, \citenamefont {{Zhang}},
  \citenamefont {{Chen}}, \citenamefont {{Han}},\ and\ \citenamefont
  {{Zhang}}}]{Wang19_anisotropic_resistance}%
  \BibitemOpen
  \bibfield  {author} {\bibinfo {author} {\bibfnamefont {H.}~\bibnamefont
  {{Wang}}}, \bibinfo {author} {\bibfnamefont {M.-L.}\ \bibnamefont {{Chen}}},
  \bibinfo {author} {\bibfnamefont {M.}~\bibnamefont {{Zhu}}}, \bibinfo
  {author} {\bibfnamefont {Y.}~\bibnamefont {{Wang}}}, \bibinfo {author}
  {\bibfnamefont {B.}~\bibnamefont {{Dong}}}, \bibinfo {author} {\bibfnamefont
  {X.}~\bibnamefont {{Sun}}}, \bibinfo {author} {\bibfnamefont
  {X.}~\bibnamefont {{Zhang}}}, \bibinfo {author} {\bibfnamefont
  {S.}~\bibnamefont {{Cao}}}, \bibinfo {author} {\bibfnamefont
  {X.}~\bibnamefont {{Li}}}, \bibinfo {author} {\bibfnamefont {J.}~\bibnamefont
  {{Huang}}}, \bibinfo {author} {\bibfnamefont {L.}~\bibnamefont {{Zhang}}},
  \bibinfo {author} {\bibfnamefont {W.}~\bibnamefont {{Liu}}}, \bibinfo
  {author} {\bibfnamefont {D.}~\bibnamefont {{Sun}}}, \bibinfo {author}
  {\bibfnamefont {Y.}~\bibnamefont {{Ye}}}, \bibinfo {author} {\bibfnamefont
  {K.}~\bibnamefont {{Song}}}, \bibinfo {author} {\bibfnamefont
  {J.}~\bibnamefont {{Wang}}}, \bibinfo {author} {\bibfnamefont
  {Y.}~\bibnamefont {{Han}}}, \bibinfo {author} {\bibfnamefont
  {T.}~\bibnamefont {{Yang}}}, \bibinfo {author} {\bibfnamefont
  {H.}~\bibnamefont {{Guo}}}, \bibinfo {author} {\bibfnamefont
  {C.}~\bibnamefont {{Qin}}}, \bibinfo {author} {\bibfnamefont
  {L.}~\bibnamefont {{Xiao}}}, \bibinfo {author} {\bibfnamefont
  {J.}~\bibnamefont {{Zhang}}}, \bibinfo {author} {\bibfnamefont
  {J.}~\bibnamefont {{Chen}}}, \bibinfo {author} {\bibfnamefont
  {Z.}~\bibnamefont {{Han}}}, \ and\ \bibinfo {author} {\bibfnamefont
  {Z.}~\bibnamefont {{Zhang}}},\ }\bibfield  {title} {\enquote {\bibinfo
  {title} {{Gate tunable giant anisotropic resistance in ultra-thin GaTe}},}\
  }\href {\doibase 10.1038/s41467-019-10256-3} {\bibfield  {journal} {\bibinfo
  {journal} {Nat Commun}\ }\textbf {\bibinfo {volume} {10}},\ \bibinfo {pages}
  {2302} (\bibinfo {year} {2019})}\BibitemShut {NoStop}%
\bibitem [{\citenamefont {Oyedele}\ \emph {et~al.}(2017)\citenamefont
  {Oyedele}, \citenamefont {Yang}, \citenamefont {Liang}, \citenamefont
  {Puretzky}, \citenamefont {Wang}, \citenamefont {Zhang}, \citenamefont {Yu},
  \citenamefont {Pudasaini}, \citenamefont {Ghosh}, \citenamefont {Liu},
  \citenamefont {Rouleau}, \citenamefont {Sumpter}, \citenamefont {Chisholm},
  \citenamefont {Zhou}, \citenamefont {Rack}, \citenamefont {Geohegan},\ and\
  \citenamefont {Xiao}}]{Oyedele17}%
  \BibitemOpen
  \bibfield  {author} {\bibinfo {author} {\bibfnamefont {A.~D.}\ \bibnamefont
  {Oyedele}}, \bibinfo {author} {\bibfnamefont {S.}~\bibnamefont {Yang}},
  \bibinfo {author} {\bibfnamefont {L.}~\bibnamefont {Liang}}, \bibinfo
  {author} {\bibfnamefont {A.~A.}\ \bibnamefont {Puretzky}}, \bibinfo {author}
  {\bibfnamefont {K.}~\bibnamefont {Wang}}, \bibinfo {author} {\bibfnamefont
  {J.}~\bibnamefont {Zhang}}, \bibinfo {author} {\bibfnamefont
  {P.}~\bibnamefont {Yu}}, \bibinfo {author} {\bibfnamefont {P.~R.}\
  \bibnamefont {Pudasaini}}, \bibinfo {author} {\bibfnamefont {A.~W.}\
  \bibnamefont {Ghosh}}, \bibinfo {author} {\bibfnamefont {Z.}~\bibnamefont
  {Liu}}, \bibinfo {author} {\bibfnamefont {C.~M.}\ \bibnamefont {Rouleau}},
  \bibinfo {author} {\bibfnamefont {B.~G.}\ \bibnamefont {Sumpter}}, \bibinfo
  {author} {\bibfnamefont {M.~F.}\ \bibnamefont {Chisholm}}, \bibinfo {author}
  {\bibfnamefont {W.}~\bibnamefont {Zhou}}, \bibinfo {author} {\bibfnamefont
  {P.~D.}\ \bibnamefont {Rack}}, \bibinfo {author} {\bibfnamefont {D.~B.}\
  \bibnamefont {Geohegan}}, \ and\ \bibinfo {author} {\bibfnamefont
  {K.}~\bibnamefont {Xiao}},\ }\bibfield  {title} {\enquote {\bibinfo {title}
  {{PdSe2: Pentagonal Two-Dimensional Layers with High Air Stability for
  Electronics}},}\ }\href {\doibase 10.1021/jacs.7b04865} {\bibfield  {journal}
  {\bibinfo  {journal} {Journal of the American Chemical Society}\ }\textbf
  {\bibinfo {volume} {139}},\ \bibinfo {pages} {14090--14097} (\bibinfo {year}
  {2017})}\BibitemShut {NoStop}%
\bibitem [{\citenamefont {Lu}\ \emph {et~al.}(2020)\citenamefont {Lu},
  \citenamefont {Chen}, \citenamefont {Cheng}, \citenamefont {Chuu},
  \citenamefont {Lu}, \citenamefont {Chen}, \citenamefont {Lu}, \citenamefont
  {Chuang}, \citenamefont {Wei}, \citenamefont {Chueh}, \citenamefont {Jian},
  \citenamefont {Li}, \citenamefont {Chang}, \citenamefont {Li},\ and\
  \citenamefont {Chang}}]{Lu20_anisotropy_PdSe2}%
  \BibitemOpen
  \bibfield  {author} {\bibinfo {author} {\bibfnamefont {L.-S.}\ \bibnamefont
  {Lu}}, \bibinfo {author} {\bibfnamefont {G.-H.}\ \bibnamefont {Chen}},
  \bibinfo {author} {\bibfnamefont {H.-Y.}\ \bibnamefont {Cheng}}, \bibinfo
  {author} {\bibfnamefont {C.-P.}\ \bibnamefont {Chuu}}, \bibinfo {author}
  {\bibfnamefont {K.-C.}\ \bibnamefont {Lu}}, \bibinfo {author} {\bibfnamefont
  {C.-H.}\ \bibnamefont {Chen}}, \bibinfo {author} {\bibfnamefont {M.-Y.}\
  \bibnamefont {Lu}}, \bibinfo {author} {\bibfnamefont {T.-H.}\ \bibnamefont
  {Chuang}}, \bibinfo {author} {\bibfnamefont {D.-H.}\ \bibnamefont {Wei}},
  \bibinfo {author} {\bibfnamefont {W.-C.}\ \bibnamefont {Chueh}}, \bibinfo
  {author} {\bibfnamefont {W.-B.}\ \bibnamefont {Jian}}, \bibinfo {author}
  {\bibfnamefont {M.-Y.}\ \bibnamefont {Li}}, \bibinfo {author} {\bibfnamefont
  {Y.-M.}\ \bibnamefont {Chang}}, \bibinfo {author} {\bibfnamefont {L.-J.}\
  \bibnamefont {Li}}, \ and\ \bibinfo {author} {\bibfnamefont {W.-H.}\
  \bibnamefont {Chang}},\ }\bibfield  {title} {\enquote {\bibinfo {title}
  {{Layer-Dependent and In-Plane Anisotropic Properties of Low-Temperature
  Synthesized Few-Layer PdSe2 Single Crystals}},}\ }\href {\doibase
  10.1021/acsnano.0c01139} {\bibfield  {journal} {\bibinfo  {journal} {ACS
  Nano}\ }\textbf {\bibinfo {volume} {14}},\ \bibinfo {pages} {4963--4972}
  (\bibinfo {year} {2020})}\BibitemShut {NoStop}%
\bibitem [{\citenamefont {Haastrup}\ \emph {et~al.}(2018)\citenamefont
  {Haastrup}, \citenamefont {Strange}, \citenamefont {Pandey}, \citenamefont
  {Deilmann}, \citenamefont {Schmidt}, \citenamefont {Hinsche}, \citenamefont
  {Gjerding}, \citenamefont {Torelli}, \citenamefont {Larsen}, \citenamefont
  {Riis-Jensen}, \citenamefont {Gath}, \citenamefont {Jacobsen}, \citenamefont
  {Mortensen}, \citenamefont {Olsen},\ and\ \citenamefont
  {Thygesen}}]{Haastrup18}%
  \BibitemOpen
  \bibfield  {author} {\bibinfo {author} {\bibfnamefont {S.}~\bibnamefont
  {Haastrup}}, \bibinfo {author} {\bibfnamefont {M.}~\bibnamefont {Strange}},
  \bibinfo {author} {\bibfnamefont {M.}~\bibnamefont {Pandey}}, \bibinfo
  {author} {\bibfnamefont {T.}~\bibnamefont {Deilmann}}, \bibinfo {author}
  {\bibfnamefont {P.~S.}\ \bibnamefont {Schmidt}}, \bibinfo {author}
  {\bibfnamefont {N.~F.}\ \bibnamefont {Hinsche}}, \bibinfo {author}
  {\bibfnamefont {M.~N.}\ \bibnamefont {Gjerding}}, \bibinfo {author}
  {\bibfnamefont {D.}~\bibnamefont {Torelli}}, \bibinfo {author} {\bibfnamefont
  {P.~M.}\ \bibnamefont {Larsen}}, \bibinfo {author} {\bibfnamefont {A.~C.}\
  \bibnamefont {Riis-Jensen}}, \bibinfo {author} {\bibfnamefont
  {J.}~\bibnamefont {Gath}}, \bibinfo {author} {\bibfnamefont {K.~W.}\
  \bibnamefont {Jacobsen}}, \bibinfo {author} {\bibfnamefont {J.~J.}\
  \bibnamefont {Mortensen}}, \bibinfo {author} {\bibfnamefont {T.}~\bibnamefont
  {Olsen}}, \ and\ \bibinfo {author} {\bibfnamefont {K.~S.}\ \bibnamefont
  {Thygesen}},\ }\bibfield  {title} {\enquote {\bibinfo {title} {{The
  Computational 2D Materials Database: high-throughput modeling and discovery
  of atomically thin crystals}},}\ }\href {\doibase 10.1088/2053-1583/aacfc1}
  {\bibfield  {journal} {\bibinfo  {journal} {2D Mater.}\ }\textbf {\bibinfo
  {volume} {5}},\ \bibinfo {pages} {042002} (\bibinfo {year}
  {2018})}\BibitemShut {NoStop}%
\bibitem [{\citenamefont {Mounet}\ \emph {et~al.}(2018)\citenamefont {Mounet},
  \citenamefont {Gibertini}, \citenamefont {Schwaller}, \citenamefont {Campi},
  \citenamefont {Merkys}, \citenamefont {Marrazzo}, \citenamefont {Sohier},
  \citenamefont {Castelli}, \citenamefont {Cepellotti}, \citenamefont {Pizzi},\
  and\ \citenamefont {Marzari}}]{Mounet2017}%
  \BibitemOpen
  \bibfield  {author} {\bibinfo {author} {\bibfnamefont {N.}~\bibnamefont
  {Mounet}}, \bibinfo {author} {\bibfnamefont {M.}~\bibnamefont {Gibertini}},
  \bibinfo {author} {\bibfnamefont {P.}~\bibnamefont {Schwaller}}, \bibinfo
  {author} {\bibfnamefont {D.}~\bibnamefont {Campi}}, \bibinfo {author}
  {\bibfnamefont {A.}~\bibnamefont {Merkys}}, \bibinfo {author} {\bibfnamefont
  {A.}~\bibnamefont {Marrazzo}}, \bibinfo {author} {\bibfnamefont
  {T.}~\bibnamefont {Sohier}}, \bibinfo {author} {\bibfnamefont {I.~E.}\
  \bibnamefont {Castelli}}, \bibinfo {author} {\bibfnamefont {A.}~\bibnamefont
  {Cepellotti}}, \bibinfo {author} {\bibfnamefont {G.}~\bibnamefont {Pizzi}}, \
  and\ \bibinfo {author} {\bibfnamefont {N.}~\bibnamefont {Marzari}},\
  }\bibfield  {title} {\enquote {\bibinfo {title} {{Two-dimensional materials
  from high-throughput computational exfoliation of experimentally known
  compounds}},}\ }\href {\doibase 10.1038/s41565-017-0035-5} {\bibfield
  {journal} {\bibinfo  {journal} {Nature Nanotechnology}\ }\textbf {\bibinfo
  {volume} {13}},\ \bibinfo {pages} {246–252} (\bibinfo {year}
  {2018})}\BibitemShut {NoStop}%
\bibitem [{\citenamefont {Ashton}\ \emph {et~al.}(2017)\citenamefont {Ashton},
  \citenamefont {Paul}, \citenamefont {Sinnott},\ and\ \citenamefont
  {Hennig}}]{Ashton17}%
  \BibitemOpen
  \bibfield  {author} {\bibinfo {author} {\bibfnamefont {M.}~\bibnamefont
  {Ashton}}, \bibinfo {author} {\bibfnamefont {J.}~\bibnamefont {Paul}},
  \bibinfo {author} {\bibfnamefont {S.~B.}\ \bibnamefont {Sinnott}}, \ and\
  \bibinfo {author} {\bibfnamefont {R.~G.}\ \bibnamefont {Hennig}},\ }\bibfield
   {title} {\enquote {\bibinfo {title} {{Topology-Scaling Identification of
  Layered Solids and Stable Exfoliated 2D Materials}},}\ }\href {\doibase
  10.1103/PhysRevLett.118.106101} {\bibfield  {journal} {\bibinfo  {journal}
  {Phys. Rev. Lett.}\ }\textbf {\bibinfo {volume} {118}},\ \bibinfo {pages}
  {106101} (\bibinfo {year} {2017})}\BibitemShut {NoStop}%
\bibitem [{\citenamefont {{Choudhary}}\ \emph {et~al.}(2017)\citenamefont
  {{Choudhary}}, \citenamefont {{Kalish}}, \citenamefont {{Beams}},\ and\
  \citenamefont {{Tavazza}}}]{Choudhary17}%
  \BibitemOpen
  \bibfield  {author} {\bibinfo {author} {\bibfnamefont {K.}~\bibnamefont
  {{Choudhary}}}, \bibinfo {author} {\bibfnamefont {I.}~\bibnamefont
  {{Kalish}}}, \bibinfo {author} {\bibfnamefont {R.}~\bibnamefont {{Beams}}}, \
  and\ \bibinfo {author} {\bibfnamefont {F.}~\bibnamefont {{Tavazza}}},\
  }\bibfield  {title} {\enquote {\bibinfo {title} {{High-throughput
  Identification and Characterization of Two-dimensional Materials using
  Density functional theory}},}\ }\href {\doibase 10.1038/s41598-017-05402-0}
  {\bibfield  {journal} {\bibinfo  {journal} {Scientific Reports}\ }\textbf
  {\bibinfo {volume} {7}},\ \bibinfo {eid} {5179} (\bibinfo {year}
  {2017})}\BibitemShut {NoStop}%
\bibitem [{\citenamefont {Cheon}\ \emph {et~al.}(2017)\citenamefont {Cheon},
  \citenamefont {Duerloo}, \citenamefont {Sendek}, \citenamefont {Porter},
  \citenamefont {Chen},\ and\ \citenamefont {Reed}}]{Cheon17}%
  \BibitemOpen
  \bibfield  {author} {\bibinfo {author} {\bibfnamefont {G.}~\bibnamefont
  {Cheon}}, \bibinfo {author} {\bibfnamefont {K.-A.~N.}\ \bibnamefont
  {Duerloo}}, \bibinfo {author} {\bibfnamefont {A.~D.}\ \bibnamefont {Sendek}},
  \bibinfo {author} {\bibfnamefont {C.}~\bibnamefont {Porter}}, \bibinfo
  {author} {\bibfnamefont {Y.}~\bibnamefont {Chen}}, \ and\ \bibinfo {author}
  {\bibfnamefont {E.~J.}\ \bibnamefont {Reed}},\ }\bibfield  {title} {\enquote
  {\bibinfo {title} {Data mining for new two- and one-dimensional weakly bonded
  solids and lattice-commensurate heterostructures},}\ }\href {\doibase
  10.1021/acs.nanolett.6b05229} {\bibfield  {journal} {\bibinfo  {journal}
  {Nano Letters}\ }\textbf {\bibinfo {volume} {17}},\ \bibinfo {pages}
  {1915--1923} (\bibinfo {year} {2017})}\BibitemShut {NoStop}%
\bibitem [{\citenamefont {{Zhou}}\ \emph {et~al.}(2019)\citenamefont {{Zhou}},
  \citenamefont {{Shen}}, \citenamefont {{Costa}}, \citenamefont {{Persson}},
  \citenamefont {{Ong}}, \citenamefont {{Huck}}, \citenamefont {{Lu}},
  \citenamefont {{Ma}}, \citenamefont {{Chen}}, \citenamefont {{Tang}},\ and\
  \citenamefont {{Feng}}}]{Zhou2019_2DMatPedia}%
  \BibitemOpen
  \bibfield  {author} {\bibinfo {author} {\bibfnamefont {J.}~\bibnamefont
  {{Zhou}}}, \bibinfo {author} {\bibfnamefont {L.}~\bibnamefont {{Shen}}},
  \bibinfo {author} {\bibfnamefont {M.~D.}\ \bibnamefont {{Costa}}}, \bibinfo
  {author} {\bibfnamefont {K.~A.}\ \bibnamefont {{Persson}}}, \bibinfo {author}
  {\bibfnamefont {S.~P.}\ \bibnamefont {{Ong}}}, \bibinfo {author}
  {\bibfnamefont {P.}~\bibnamefont {{Huck}}}, \bibinfo {author} {\bibfnamefont
  {Y.}~\bibnamefont {{Lu}}}, \bibinfo {author} {\bibfnamefont {X.}~\bibnamefont
  {{Ma}}}, \bibinfo {author} {\bibfnamefont {Y.}~\bibnamefont {{Chen}}},
  \bibinfo {author} {\bibfnamefont {H.}~\bibnamefont {{Tang}}}, \ and\ \bibinfo
  {author} {\bibfnamefont {Y.~P.}\ \bibnamefont {{Feng}}},\ }\bibfield  {title}
  {\enquote {\bibinfo {title} {{2DMatPedia, an open computational database of
  two-dimensional materials from top-down and bottom-up approaches}},}\ }\href
  {\doibase 10.1038/s41597-019-0097-3} {\bibfield  {journal} {\bibinfo
  {journal} {Scientific Data}\ }\textbf {\bibinfo {volume} {6}},\ \bibinfo
  {eid} {86} (\bibinfo {year} {2019})}\BibitemShut {NoStop}%
\bibitem [{\citenamefont {Enkovaara}\ \emph {et~al.}(2010)\citenamefont
  {Enkovaara}, \citenamefont {Rostgaard}, \citenamefont {Mortensen},
  \citenamefont {Chen}, \citenamefont {Du{\l}ak}, \citenamefont {Ferrighi},
  \citenamefont {Gavnholt}, \citenamefont {Glinsvad}, \citenamefont {Haikola},
  \citenamefont {Hansen}, \citenamefont {Kristoffersen}, \citenamefont
  {Kuisma}, \citenamefont {Larsen}, \citenamefont {Lehtovaara}, \citenamefont
  {Ljungberg}, \citenamefont {Lopez-Acevedo}, \citenamefont {Moses},
  \citenamefont {Ojanen}, \citenamefont {Olsen}, \citenamefont {Petzold},
  \citenamefont {Romero}, \citenamefont {Stausholm-M{\o}ller}, \citenamefont
  {Strange}, \citenamefont {Tritsaris}, \citenamefont {Vanin}, \citenamefont
  {Walter}, \citenamefont {Hammer}, \citenamefont {H\"akkinen}, \citenamefont
  {Madsen}, \citenamefont {Nieminen}, \citenamefont {N{\o}rskov}, \citenamefont
  {Puska}, \citenamefont {Rantala}, \citenamefont {Schi{\o}tz}, \citenamefont
  {Thygesen},\ and\ \citenamefont {Jacobsen}}]{Enkovaara10}%
  \BibitemOpen
  \bibfield  {author} {\bibinfo {author} {\bibfnamefont {J.}~\bibnamefont
  {Enkovaara}}, \bibinfo {author} {\bibfnamefont {C.}~\bibnamefont
  {Rostgaard}}, \bibinfo {author} {\bibfnamefont {J.~J.}\ \bibnamefont
  {Mortensen}}, \bibinfo {author} {\bibfnamefont {J.}~\bibnamefont {Chen}},
  \bibinfo {author} {\bibfnamefont {M.}~\bibnamefont {Du{\l}ak}}, \bibinfo
  {author} {\bibfnamefont {L.}~\bibnamefont {Ferrighi}}, \bibinfo {author}
  {\bibfnamefont {J.}~\bibnamefont {Gavnholt}}, \bibinfo {author}
  {\bibfnamefont {C.}~\bibnamefont {Glinsvad}}, \bibinfo {author}
  {\bibfnamefont {V.}~\bibnamefont {Haikola}}, \bibinfo {author} {\bibfnamefont
  {H.~A.}\ \bibnamefont {Hansen}}, \bibinfo {author} {\bibfnamefont {H.~H.}\
  \bibnamefont {Kristoffersen}}, \bibinfo {author} {\bibfnamefont
  {M.}~\bibnamefont {Kuisma}}, \bibinfo {author} {\bibfnamefont {A.~H.}\
  \bibnamefont {Larsen}}, \bibinfo {author} {\bibfnamefont {L.}~\bibnamefont
  {Lehtovaara}}, \bibinfo {author} {\bibfnamefont {M.}~\bibnamefont
  {Ljungberg}}, \bibinfo {author} {\bibfnamefont {O.}~\bibnamefont
  {Lopez-Acevedo}}, \bibinfo {author} {\bibfnamefont {P.~G.}\ \bibnamefont
  {Moses}}, \bibinfo {author} {\bibfnamefont {J.}~\bibnamefont {Ojanen}},
  \bibinfo {author} {\bibfnamefont {T.}~\bibnamefont {Olsen}}, \bibinfo
  {author} {\bibfnamefont {V.}~\bibnamefont {Petzold}}, \bibinfo {author}
  {\bibfnamefont {N.~A.}\ \bibnamefont {Romero}}, \bibinfo {author}
  {\bibfnamefont {J.}~\bibnamefont {Stausholm-M{\o}ller}}, \bibinfo {author}
  {\bibfnamefont {M.}~\bibnamefont {Strange}}, \bibinfo {author} {\bibfnamefont
  {G.~A.}\ \bibnamefont {Tritsaris}}, \bibinfo {author} {\bibfnamefont
  {M.}~\bibnamefont {Vanin}}, \bibinfo {author} {\bibfnamefont
  {M.}~\bibnamefont {Walter}}, \bibinfo {author} {\bibfnamefont
  {B.}~\bibnamefont {Hammer}}, \bibinfo {author} {\bibfnamefont
  {H.}~\bibnamefont {H\"akkinen}}, \bibinfo {author} {\bibfnamefont {G.~K.~H.}\
  \bibnamefont {Madsen}}, \bibinfo {author} {\bibfnamefont {R.~M.}\
  \bibnamefont {Nieminen}}, \bibinfo {author} {\bibfnamefont {J.~K.}\
  \bibnamefont {N{\o}rskov}}, \bibinfo {author} {\bibfnamefont
  {M.}~\bibnamefont {Puska}}, \bibinfo {author} {\bibfnamefont {T.~T.}\
  \bibnamefont {Rantala}}, \bibinfo {author} {\bibfnamefont {J.}~\bibnamefont
  {Schi{\o}tz}}, \bibinfo {author} {\bibfnamefont {K.~S.}\ \bibnamefont
  {Thygesen}}, \ and\ \bibinfo {author} {\bibfnamefont {K.~W.}\ \bibnamefont
  {Jacobsen}},\ }\bibfield  {title} {\enquote {\bibinfo {title} {{Electronic
  structure calculations with GPAW: a real-space implementation of the
  projector augmented-wave method}},}\ }\href {\doibase
  10.1088/0953-8984/22/25/253202} {\bibfield  {journal} {\bibinfo  {journal}
  {J. Phys. Condens. Matter}\ }\textbf {\bibinfo {volume} {22}},\ \bibinfo
  {pages} {253202} (\bibinfo {year} {2010})}\BibitemShut {NoStop}%
\bibitem [{\citenamefont {Larsen}\ \emph {et~al.}(2017)\citenamefont {Larsen},
  \citenamefont {Mortensen}, \citenamefont {Blomqvist}, \citenamefont
  {Castelli}, \citenamefont {Christensen}, \citenamefont {Du{\l}ak},
  \citenamefont {Friis}, \citenamefont {Groves}, \citenamefont {Hammer},
  \citenamefont {Hargus}, \citenamefont {Hermes}, \citenamefont {Jennings},
  \citenamefont {Jensen}, \citenamefont {Kermode}, \citenamefont {Kitchin},
  \citenamefont {Kolsbjerg}, \citenamefont {Kubal}, \citenamefont {Kaasbjerg},
  \citenamefont {Lysgaard}, \citenamefont {Maronsson}, \citenamefont {Maxson},
  \citenamefont {Olsen}, \citenamefont {Pastewka}, \citenamefont {Peterson},
  \citenamefont {Rostgaard}, \citenamefont {Schi{\o}tz}, \citenamefont
  {Schütt}, \citenamefont {Strange}, \citenamefont {Thygesen}, \citenamefont
  {Vegge}, \citenamefont {Vilhelmsen}, \citenamefont {Walter}, \citenamefont
  {Zeng},\ and\ \citenamefont {Jacobsen}}]{HjorthLarsen17}%
  \BibitemOpen
  \bibfield  {author} {\bibinfo {author} {\bibfnamefont {A.~H.}\ \bibnamefont
  {Larsen}}, \bibinfo {author} {\bibfnamefont {J.~J.}\ \bibnamefont
  {Mortensen}}, \bibinfo {author} {\bibfnamefont {J.}~\bibnamefont
  {Blomqvist}}, \bibinfo {author} {\bibfnamefont {I.~E.}\ \bibnamefont
  {Castelli}}, \bibinfo {author} {\bibfnamefont {R.}~\bibnamefont
  {Christensen}}, \bibinfo {author} {\bibfnamefont {M.}~\bibnamefont
  {Du{\l}ak}}, \bibinfo {author} {\bibfnamefont {J.}~\bibnamefont {Friis}},
  \bibinfo {author} {\bibfnamefont {M.~N.}\ \bibnamefont {Groves}}, \bibinfo
  {author} {\bibfnamefont {B.}~\bibnamefont {Hammer}}, \bibinfo {author}
  {\bibfnamefont {C.}~\bibnamefont {Hargus}}, \bibinfo {author} {\bibfnamefont
  {E.~D.}\ \bibnamefont {Hermes}}, \bibinfo {author} {\bibfnamefont {P.~C.}\
  \bibnamefont {Jennings}}, \bibinfo {author} {\bibfnamefont {P.~B.}\
  \bibnamefont {Jensen}}, \bibinfo {author} {\bibfnamefont {J.}~\bibnamefont
  {Kermode}}, \bibinfo {author} {\bibfnamefont {J.~R.}\ \bibnamefont
  {Kitchin}}, \bibinfo {author} {\bibfnamefont {E.~L.}\ \bibnamefont
  {Kolsbjerg}}, \bibinfo {author} {\bibfnamefont {J.}~\bibnamefont {Kubal}},
  \bibinfo {author} {\bibfnamefont {K.}~\bibnamefont {Kaasbjerg}}, \bibinfo
  {author} {\bibfnamefont {S.}~\bibnamefont {Lysgaard}}, \bibinfo {author}
  {\bibfnamefont {J.~B.}\ \bibnamefont {Maronsson}}, \bibinfo {author}
  {\bibfnamefont {T.}~\bibnamefont {Maxson}}, \bibinfo {author} {\bibfnamefont
  {T.}~\bibnamefont {Olsen}}, \bibinfo {author} {\bibfnamefont
  {L.}~\bibnamefont {Pastewka}}, \bibinfo {author} {\bibfnamefont
  {A.}~\bibnamefont {Peterson}}, \bibinfo {author} {\bibfnamefont
  {C.}~\bibnamefont {Rostgaard}}, \bibinfo {author} {\bibfnamefont
  {J.}~\bibnamefont {Schi{\o}tz}}, \bibinfo {author} {\bibfnamefont
  {O.}~\bibnamefont {Schütt}}, \bibinfo {author} {\bibfnamefont
  {M.}~\bibnamefont {Strange}}, \bibinfo {author} {\bibfnamefont {K.~S.}\
  \bibnamefont {Thygesen}}, \bibinfo {author} {\bibfnamefont {T.}~\bibnamefont
  {Vegge}}, \bibinfo {author} {\bibfnamefont {L.}~\bibnamefont {Vilhelmsen}},
  \bibinfo {author} {\bibfnamefont {M.}~\bibnamefont {Walter}}, \bibinfo
  {author} {\bibfnamefont {Z.}~\bibnamefont {Zeng}}, \ and\ \bibinfo {author}
  {\bibfnamefont {K.~W.}\ \bibnamefont {Jacobsen}},\ }\bibfield  {title}
  {\enquote {\bibinfo {title} {{The atomic simulation environment{\textemdash}a
  Python library for working with atoms}},}\ }\href {\doibase
  10.1088/1361-648x/aa680e} {\bibfield  {journal} {\bibinfo  {journal} {J.
  Phys. Condens. Matter}\ }\textbf {\bibinfo {volume} {29}},\ \bibinfo {pages}
  {273002} (\bibinfo {year} {2017})}\BibitemShut {NoStop}%
\bibitem [{ASR()}]{ASR_docs}%
  \BibitemOpen
  \href@noop {} {\enquote {\bibinfo {title} {{Atomic Simulation Recipes
  (ASR)}},}\ }\bibinfo {howpublished}
  {\url{https://asr.readthedocs.io/en/latest/}}\BibitemShut {NoStop}%
\bibitem [{\citenamefont {Mortensen}, \citenamefont {Gjerding},\ and\
  \citenamefont {Thygesen}(2020)}]{Mortensen20}%
  \BibitemOpen
  \bibfield  {author} {\bibinfo {author} {\bibfnamefont {J.~J.}\ \bibnamefont
  {Mortensen}}, \bibinfo {author} {\bibfnamefont {M.}~\bibnamefont {Gjerding}},
  \ and\ \bibinfo {author} {\bibfnamefont {K.~S.}\ \bibnamefont {Thygesen}},\
  }\bibfield  {title} {\enquote {\bibinfo {title} {{MyQueue: Task and workflow
  scheduling system}},}\ }\href {\doibase 10.21105/joss.01844} {\bibfield
  {journal} {\bibinfo  {journal} {Journal of Open Source Software}\ }\textbf
  {\bibinfo {volume} {5}},\ \bibinfo {pages} {1844} (\bibinfo {year}
  {2020})}\BibitemShut {NoStop}%
\bibitem [{\citenamefont {Perdew}, \citenamefont {Burke},\ and\ \citenamefont
  {Ernzerhof}(1996)}]{Perdew96}%
  \BibitemOpen
  \bibfield  {author} {\bibinfo {author} {\bibfnamefont {J.~P.}\ \bibnamefont
  {Perdew}}, \bibinfo {author} {\bibfnamefont {K.}~\bibnamefont {Burke}}, \
  and\ \bibinfo {author} {\bibfnamefont {M.}~\bibnamefont {Ernzerhof}},\
  }\bibfield  {title} {\enquote {\bibinfo {title} {{Generalized Gradient
  Approximation Made Simple}},}\ }\href {\doibase 10.1103/PhysRevLett.77.3865}
  {\bibfield  {journal} {\bibinfo  {journal} {Phys. Rev. Lett.}\ }\textbf
  {\bibinfo {volume} {77}},\ \bibinfo {pages} {3865--3868} (\bibinfo {year}
  {1996})}\BibitemShut {NoStop}%
\bibitem [{\citenamefont {Saal}\ \emph {et~al.}(2013)\citenamefont {Saal},
  \citenamefont {Kirklin}, \citenamefont {Aykol}, \citenamefont {Meredig},\
  and\ \citenamefont {Wolverton}}]{saal2013materials}%
  \BibitemOpen
  \bibfield  {author} {\bibinfo {author} {\bibfnamefont {J.~E.}\ \bibnamefont
  {Saal}}, \bibinfo {author} {\bibfnamefont {S.}~\bibnamefont {Kirklin}},
  \bibinfo {author} {\bibfnamefont {M.}~\bibnamefont {Aykol}}, \bibinfo
  {author} {\bibfnamefont {B.}~\bibnamefont {Meredig}}, \ and\ \bibinfo
  {author} {\bibfnamefont {C.}~\bibnamefont {Wolverton}},\ }\bibfield  {title}
  {\enquote {\bibinfo {title} {{Materials design and discovery with
  high-throughput density functional theory: the open quantum materials
  database (OQMD)}},}\ }\href {\doibase 10.1007/s11837-013-0755-4} {\bibfield
  {journal} {\bibinfo  {journal} {Jom}\ }\textbf {\bibinfo {volume} {65}},\
  \bibinfo {pages} {1501--1509} (\bibinfo {year} {2013})}\BibitemShut {NoStop}%
\bibitem [{\citenamefont {Neumann}\ and\ \citenamefont
  {Meyer}(1885)}]{Neumann1885}%
  \BibitemOpen
  \bibfield  {author} {\bibinfo {author} {\bibfnamefont {F.}~\bibnamefont
  {Neumann}}\ and\ \bibinfo {author} {\bibfnamefont {O.}~\bibnamefont
  {Meyer}},\ }\href@noop {} {\emph {\bibinfo {title} {{Vorlesungen {\"u}ber die
  Theorie der Elasticit{\"a}t der festen K{\"o}rper und des
  Licht{\"a}thers}}}},\ Vorlesungen {\"u}ber mathematische Physik\ (\bibinfo
  {publisher} {Teubner},\ \bibinfo {year} {1885})\BibitemShut {NoStop}%
\bibitem [{\citenamefont {{Togo}}\ and\ \citenamefont
  {{Tanaka}}(2018)}]{Togo18_Spglib}%
  \BibitemOpen
  \bibfield  {author} {\bibinfo {author} {\bibfnamefont {A.}~\bibnamefont
  {{Togo}}}\ and\ \bibinfo {author} {\bibfnamefont {I.}~\bibnamefont
  {{Tanaka}}},\ }\bibfield  {title} {\enquote {\bibinfo {title}
  {{$\texttt{Spglib}$: a software library for crystal symmetry search}},}\
  }\href@noop {} {\bibfield  {journal} {\bibinfo  {journal} {arXiv e-prints}\
  ,\ \bibinfo {eid} {arXiv:1808.01590}} (\bibinfo {year} {2018})},\ \Eprint
  {http://arxiv.org/abs/1808.01590} {arXiv:1808.01590 [cond-mat.mtrl-sci]}
  \BibitemShut {NoStop}%
\bibitem [{Note1()}]{Note1}%
  \BibitemOpen
  \bibinfo {note} {Note that we will substitute the common overline notation
  with a dash (that is, we will use $-n$ instead of $\protect \overline
  n$)}\BibitemShut {NoStop}%
\bibitem [{\citenamefont {Xu}\ \emph {et~al.}(2015)\citenamefont {Xu},
  \citenamefont {Wang}, \citenamefont {Wang}, \citenamefont {Huang},
  \citenamefont {Wang}, \citenamefont {Yin}, \citenamefont {Jiang},\ and\
  \citenamefont {He}}]{Xu15_HfS2}%
  \BibitemOpen
  \bibfield  {author} {\bibinfo {author} {\bibfnamefont {K.}~\bibnamefont
  {Xu}}, \bibinfo {author} {\bibfnamefont {Z.}~\bibnamefont {Wang}}, \bibinfo
  {author} {\bibfnamefont {F.}~\bibnamefont {Wang}}, \bibinfo {author}
  {\bibfnamefont {Y.}~\bibnamefont {Huang}}, \bibinfo {author} {\bibfnamefont
  {F.}~\bibnamefont {Wang}}, \bibinfo {author} {\bibfnamefont {L.}~\bibnamefont
  {Yin}}, \bibinfo {author} {\bibfnamefont {C.}~\bibnamefont {Jiang}}, \ and\
  \bibinfo {author} {\bibfnamefont {J.}~\bibnamefont {He}},\ }\bibfield
  {title} {\enquote {\bibinfo {title} {{Ultrasensitive Phototransistors Based
  on Few-Layered HfS2}},}\ }\href {\doibase 10.1002/adma.201503864} {\bibfield
  {journal} {\bibinfo  {journal} {Advanced Materials}\ }\textbf {\bibinfo
  {volume} {27}},\ \bibinfo {pages} {7881--7887} (\bibinfo {year}
  {2015})}\BibitemShut {NoStop}%
\bibitem [{\citenamefont {{Zhu}}\ \emph {et~al.}(2015)\citenamefont {{Zhu}},
  \citenamefont {{Chen}}, \citenamefont {{Xu}}, \citenamefont {{Gao}},
  \citenamefont {{Guan}}, \citenamefont {{Liu}}, \citenamefont {{Qian}},
  \citenamefont {{Zhang}},\ and\ \citenamefont {{Jia}}}]{Zhu15_stanene}%
  \BibitemOpen
  \bibfield  {author} {\bibinfo {author} {\bibfnamefont {F.-F.}\ \bibnamefont
  {{Zhu}}}, \bibinfo {author} {\bibfnamefont {W.-J.}\ \bibnamefont {{Chen}}},
  \bibinfo {author} {\bibfnamefont {Y.}~\bibnamefont {{Xu}}}, \bibinfo {author}
  {\bibfnamefont {C.-L.}\ \bibnamefont {{Gao}}}, \bibinfo {author}
  {\bibfnamefont {D.-D.}\ \bibnamefont {{Guan}}}, \bibinfo {author}
  {\bibfnamefont {C.-H.}\ \bibnamefont {{Liu}}}, \bibinfo {author}
  {\bibfnamefont {D.}~\bibnamefont {{Qian}}}, \bibinfo {author} {\bibfnamefont
  {S.-C.}\ \bibnamefont {{Zhang}}}, \ and\ \bibinfo {author} {\bibfnamefont
  {J.-F.}\ \bibnamefont {{Jia}}},\ }\bibfield  {title} {\enquote {\bibinfo
  {title} {{Epitaxial growth of two-dimensional stanene}},}\ }\href {\doibase
  10.1038/nmat4384} {\bibfield  {journal} {\bibinfo  {journal} {Nature
  Materials}\ }\textbf {\bibinfo {volume} {14}},\ \bibinfo {pages} {1020--1025}
  (\bibinfo {year} {2015})}\BibitemShut {NoStop}%
\bibitem [{\citenamefont {{Elias}}\ \emph {et~al.}(2009)\citenamefont
  {{Elias}}, \citenamefont {{Nair}}, \citenamefont {{Mohiuddin}}, \citenamefont
  {{Morozov}}, \citenamefont {{Blake}}, \citenamefont {{Halsall}},
  \citenamefont {{Ferrari}}, \citenamefont {{Boukhvalov}}, \citenamefont
  {{Katsnelson}}, \citenamefont {{Geim}},\ and\ \citenamefont
  {{Novoselov}}}]{Elias09}%
  \BibitemOpen
  \bibfield  {author} {\bibinfo {author} {\bibfnamefont {D.~C.}\ \bibnamefont
  {{Elias}}}, \bibinfo {author} {\bibfnamefont {R.~R.}\ \bibnamefont {{Nair}}},
  \bibinfo {author} {\bibfnamefont {T.~M.~G.}\ \bibnamefont {{Mohiuddin}}},
  \bibinfo {author} {\bibfnamefont {S.~V.}\ \bibnamefont {{Morozov}}}, \bibinfo
  {author} {\bibfnamefont {P.}~\bibnamefont {{Blake}}}, \bibinfo {author}
  {\bibfnamefont {M.~P.}\ \bibnamefont {{Halsall}}}, \bibinfo {author}
  {\bibfnamefont {A.~C.}\ \bibnamefont {{Ferrari}}}, \bibinfo {author}
  {\bibfnamefont {D.~W.}\ \bibnamefont {{Boukhvalov}}}, \bibinfo {author}
  {\bibfnamefont {M.~I.}\ \bibnamefont {{Katsnelson}}}, \bibinfo {author}
  {\bibfnamefont {A.~K.}\ \bibnamefont {{Geim}}}, \ and\ \bibinfo {author}
  {\bibfnamefont {K.~S.}\ \bibnamefont {{Novoselov}}},\ }\bibfield  {title}
  {\enquote {\bibinfo {title} {{Control of Graphene{\textquoteright}s
  Properties by Reversible Hydrogenation: Evidence for Graphane}},}\ }\href
  {\doibase 10.1126/science.1167130} {\bibfield  {journal} {\bibinfo  {journal}
  {Science}\ }\textbf {\bibinfo {volume} {323}},\ \bibinfo {pages} {610}
  (\bibinfo {year} {2009})}\BibitemShut {NoStop}%
\bibitem [{\citenamefont {Lu}\ \emph {et~al.}(2017)\citenamefont {Lu},
  \citenamefont {Zhu}, \citenamefont {Xiao}, \citenamefont {Chuu},
  \citenamefont {Han}, \citenamefont {Chiu}, \citenamefont {Cheng},
  \citenamefont {Yang}, \citenamefont {Wei}, \citenamefont {Yang} \emph
  {et~al.}}]{Lu17_Janus}%
  \BibitemOpen
  \bibfield  {author} {\bibinfo {author} {\bibfnamefont {A.-Y.}\ \bibnamefont
  {Lu}}, \bibinfo {author} {\bibfnamefont {H.}~\bibnamefont {Zhu}}, \bibinfo
  {author} {\bibfnamefont {J.}~\bibnamefont {Xiao}}, \bibinfo {author}
  {\bibfnamefont {C.-P.}\ \bibnamefont {Chuu}}, \bibinfo {author}
  {\bibfnamefont {Y.}~\bibnamefont {Han}}, \bibinfo {author} {\bibfnamefont
  {M.-H.}\ \bibnamefont {Chiu}}, \bibinfo {author} {\bibfnamefont {C.-C.}\
  \bibnamefont {Cheng}}, \bibinfo {author} {\bibfnamefont {C.-W.}\ \bibnamefont
  {Yang}}, \bibinfo {author} {\bibfnamefont {K.-H.}\ \bibnamefont {Wei}},
  \bibinfo {author} {\bibfnamefont {Y.}~\bibnamefont {Yang}},  \emph {et~al.},\
  }\bibfield  {title} {\enquote {\bibinfo {title} {Janus monolayers of
  transition metal dichalcogenides},}\ }\href {\doibase 10.1038/nnano.2017.100}
  {\bibfield  {journal} {\bibinfo  {journal} {Nature Nanotech.}\ }\textbf
  {\bibinfo {volume} {12}},\ \bibinfo {pages} {744–749} (\bibinfo {year}
  {2017})}\BibitemShut {NoStop}%
\bibitem [{\citenamefont {Fülöp}\ \emph {et~al.}(2018)\citenamefont
  {Fülöp}, \citenamefont {Tajkov}, \citenamefont {Pet{\H{o}}}, \citenamefont
  {Kun}, \citenamefont {Koltai}, \citenamefont {Oroszl{\'{a}}ny}, \citenamefont
  {T{\'{o}}v{\'{a}}ri}, \citenamefont {Murakawa}, \citenamefont {Tokura},
  \citenamefont {Bord{\'{a}}cs}, \citenamefont {Tapaszt{\'{o}}},\ and\
  \citenamefont {Csonka}}]{Fulop18_Janus}%
  \BibitemOpen
  \bibfield  {author} {\bibinfo {author} {\bibfnamefont {B.}~\bibnamefont
  {Fülöp}}, \bibinfo {author} {\bibfnamefont {Z.}~\bibnamefont {Tajkov}},
  \bibinfo {author} {\bibfnamefont {J.}~\bibnamefont {Pet{\H{o}}}}, \bibinfo
  {author} {\bibfnamefont {P.}~\bibnamefont {Kun}}, \bibinfo {author}
  {\bibfnamefont {J.}~\bibnamefont {Koltai}}, \bibinfo {author} {\bibfnamefont
  {L.}~\bibnamefont {Oroszl{\'{a}}ny}}, \bibinfo {author} {\bibfnamefont
  {E.}~\bibnamefont {T{\'{o}}v{\'{a}}ri}}, \bibinfo {author} {\bibfnamefont
  {H.}~\bibnamefont {Murakawa}}, \bibinfo {author} {\bibfnamefont
  {Y.}~\bibnamefont {Tokura}}, \bibinfo {author} {\bibfnamefont
  {S.}~\bibnamefont {Bord{\'{a}}cs}}, \bibinfo {author} {\bibfnamefont
  {L.}~\bibnamefont {Tapaszt{\'{o}}}}, \ and\ \bibinfo {author} {\bibfnamefont
  {S.}~\bibnamefont {Csonka}},\ }\bibfield  {title} {\enquote {\bibinfo {title}
  {{Exfoliation of single layer {BiTeI} flakes}},}\ }\href {\doibase
  10.1088/2053-1583/aac652} {\bibfield  {journal} {\bibinfo  {journal} {2D
  Materials}\ }\textbf {\bibinfo {volume} {5}},\ \bibinfo {pages} {031013}
  (\bibinfo {year} {2018})}\BibitemShut {NoStop}%
\bibitem [{\citenamefont {Riis-Jensen}, \citenamefont {Manti},\ and\
  \citenamefont {Thygesen}(2020)}]{Riis-Jensen20}%
  \BibitemOpen
  \bibfield  {author} {\bibinfo {author} {\bibfnamefont {A.~C.}\ \bibnamefont
  {Riis-Jensen}}, \bibinfo {author} {\bibfnamefont {S.}~\bibnamefont {Manti}},
  \ and\ \bibinfo {author} {\bibfnamefont {K.~S.}\ \bibnamefont {Thygesen}},\
  }\bibfield  {title} {\enquote {\bibinfo {title} {{Engineering Atomically
  Sharp Potential Steps and Band Alignment at Solid Interfaces using 2D Janus
  Layers}},}\ }\href {\doibase 10.1021/acs.jpcc.0c01286} {\bibfield  {journal}
  {\bibinfo  {journal} {The Journal of Physical Chemistry C}\ }\textbf
  {\bibinfo {volume} {124}},\ \bibinfo {pages} {9572--9580} (\bibinfo {year}
  {2020})}\BibitemShut {NoStop}%
\bibitem [{\citenamefont {Huang}\ \emph {et~al.}(2017)\citenamefont {Huang},
  \citenamefont {Clark}, \citenamefont {Navarro-Moratalla}, \citenamefont
  {Klein}, \citenamefont {Cheng}, \citenamefont {Seyler}, \citenamefont
  {Zhong}, \citenamefont {Schmidgall}, \citenamefont {McGuire}, \citenamefont
  {Cobden} \emph {et~al.}}]{Huang2017_FM}%
  \BibitemOpen
  \bibfield  {author} {\bibinfo {author} {\bibfnamefont {B.}~\bibnamefont
  {Huang}}, \bibinfo {author} {\bibfnamefont {G.}~\bibnamefont {Clark}},
  \bibinfo {author} {\bibfnamefont {E.}~\bibnamefont {Navarro-Moratalla}},
  \bibinfo {author} {\bibfnamefont {D.~R.}\ \bibnamefont {Klein}}, \bibinfo
  {author} {\bibfnamefont {R.}~\bibnamefont {Cheng}}, \bibinfo {author}
  {\bibfnamefont {K.~L.}\ \bibnamefont {Seyler}}, \bibinfo {author}
  {\bibfnamefont {D.}~\bibnamefont {Zhong}}, \bibinfo {author} {\bibfnamefont
  {E.}~\bibnamefont {Schmidgall}}, \bibinfo {author} {\bibfnamefont {M.~A.}\
  \bibnamefont {McGuire}}, \bibinfo {author} {\bibfnamefont {D.~H.}\
  \bibnamefont {Cobden}},  \emph {et~al.},\ }\bibfield  {title} {\enquote
  {\bibinfo {title} {{Layer-dependent ferromagnetism in a van der Waals crystal
  down to the monolayer limit}},}\ }\href {\doibase 10.1038/nature22391}
  {\bibfield  {journal} {\bibinfo  {journal} {Nature}\ }\textbf {\bibinfo
  {volume} {546}},\ \bibinfo {pages} {270--273} (\bibinfo {year}
  {2017})}\BibitemShut {NoStop}%
\bibitem [{\citenamefont {{Tang}}\ \emph {et~al.}(2017)\citenamefont {{Tang}},
  \citenamefont {{Zhang}}, \citenamefont {{Wong}}, \citenamefont
  {{Pedramrazi}}, \citenamefont {{Tsai}}, \citenamefont {{Jia}}, \citenamefont
  {{Moritz}}, \citenamefont {{Claassen}}, \citenamefont {{Ryu}}, \citenamefont
  {{Kahn}}, \citenamefont {{Jiang}}, \citenamefont {{Yan}}, \citenamefont
  {{Hashimoto}}, \citenamefont {{Lu}}, \citenamefont {{Moore}}, \citenamefont
  {{Hwang}}, \citenamefont {{Hwang}}, \citenamefont {{Hussain}}, \citenamefont
  {{Chen}}, \citenamefont {{Ugeda}}, \citenamefont {{Liu}}, \citenamefont
  {{Xie}}, \citenamefont {{Devereaux}}, \citenamefont {{Crommie}},
  \citenamefont {{Mo}},\ and\ \citenamefont {{Shen}}}]{Tang17_WTe2}%
  \BibitemOpen
  \bibfield  {author} {\bibinfo {author} {\bibfnamefont {S.}~\bibnamefont
  {{Tang}}}, \bibinfo {author} {\bibfnamefont {C.}~\bibnamefont {{Zhang}}},
  \bibinfo {author} {\bibfnamefont {D.}~\bibnamefont {{Wong}}}, \bibinfo
  {author} {\bibfnamefont {Z.}~\bibnamefont {{Pedramrazi}}}, \bibinfo {author}
  {\bibfnamefont {H.-Z.}\ \bibnamefont {{Tsai}}}, \bibinfo {author}
  {\bibfnamefont {C.}~\bibnamefont {{Jia}}}, \bibinfo {author} {\bibfnamefont
  {B.}~\bibnamefont {{Moritz}}}, \bibinfo {author} {\bibfnamefont
  {M.}~\bibnamefont {{Claassen}}}, \bibinfo {author} {\bibfnamefont
  {H.}~\bibnamefont {{Ryu}}}, \bibinfo {author} {\bibfnamefont
  {S.}~\bibnamefont {{Kahn}}}, \bibinfo {author} {\bibfnamefont
  {J.}~\bibnamefont {{Jiang}}}, \bibinfo {author} {\bibfnamefont
  {H.}~\bibnamefont {{Yan}}}, \bibinfo {author} {\bibfnamefont
  {M.}~\bibnamefont {{Hashimoto}}}, \bibinfo {author} {\bibfnamefont
  {D.}~\bibnamefont {{Lu}}}, \bibinfo {author} {\bibfnamefont {R.~G.}\
  \bibnamefont {{Moore}}}, \bibinfo {author} {\bibfnamefont {C.-C.}\
  \bibnamefont {{Hwang}}}, \bibinfo {author} {\bibfnamefont {C.}~\bibnamefont
  {{Hwang}}}, \bibinfo {author} {\bibfnamefont {Z.}~\bibnamefont {{Hussain}}},
  \bibinfo {author} {\bibfnamefont {Y.}~\bibnamefont {{Chen}}}, \bibinfo
  {author} {\bibfnamefont {M.~M.}\ \bibnamefont {{Ugeda}}}, \bibinfo {author}
  {\bibfnamefont {Z.}~\bibnamefont {{Liu}}}, \bibinfo {author} {\bibfnamefont
  {X.}~\bibnamefont {{Xie}}}, \bibinfo {author} {\bibfnamefont {T.~P.}\
  \bibnamefont {{Devereaux}}}, \bibinfo {author} {\bibfnamefont {M.~F.}\
  \bibnamefont {{Crommie}}}, \bibinfo {author} {\bibfnamefont {S.-K.}\
  \bibnamefont {{Mo}}}, \ and\ \bibinfo {author} {\bibfnamefont {Z.-X.}\
  \bibnamefont {{Shen}}},\ }\bibfield  {title} {\enquote {\bibinfo {title}
  {{Quantum spin Hall state in monolayer 1T'-WTe$_{2}$}},}\ }\href {\doibase
  10.1038/nphys4174} {\bibfield  {journal} {\bibinfo  {journal} {Nature
  Physics}\ }\textbf {\bibinfo {volume} {13}},\ \bibinfo {pages} {683--687}
  (\bibinfo {year} {2017})}\BibitemShut {NoStop}%
\bibitem [{\citenamefont {Kamal}\ and\ \citenamefont
  {Ezawa}(2015)}]{Kamal15_As}%
  \BibitemOpen
  \bibfield  {author} {\bibinfo {author} {\bibfnamefont {C.}~\bibnamefont
  {Kamal}}\ and\ \bibinfo {author} {\bibfnamefont {M.}~\bibnamefont {Ezawa}},\
  }\bibfield  {title} {\enquote {\bibinfo {title} {{Arsenene: Two-dimensional
  buckled and puckered honeycomb arsenic systems}},}\ }\href {\doibase
  10.1103/PhysRevB.91.085423} {\bibfield  {journal} {\bibinfo  {journal} {Phys.
  Rev. B}\ }\textbf {\bibinfo {volume} {91}},\ \bibinfo {pages} {085423}
  (\bibinfo {year} {2015})}\BibitemShut {NoStop}%
\bibitem [{\citenamefont {{Telford}}\ \emph {et~al.}(2020)\citenamefont
  {{Telford}}, \citenamefont {{Dismukes}}, \citenamefont {{Lee}}, \citenamefont
  {{Cheng}}, \citenamefont {{Wieteska}}, \citenamefont {{Bartholomew}},
  \citenamefont {{Chen}}, \citenamefont {{Xu}}, \citenamefont {{Pasupathy}},
  \citenamefont {{Zhu}}, \citenamefont {{Dean}},\ and\ \citenamefont
  {{Roy}}}]{Telford2020_arxiv_CrSBr}%
  \BibitemOpen
  \bibfield  {author} {\bibinfo {author} {\bibfnamefont {E.~J.}\ \bibnamefont
  {{Telford}}}, \bibinfo {author} {\bibfnamefont {A.~H.}\ \bibnamefont
  {{Dismukes}}}, \bibinfo {author} {\bibfnamefont {K.}~\bibnamefont {{Lee}}},
  \bibinfo {author} {\bibfnamefont {M.}~\bibnamefont {{Cheng}}}, \bibinfo
  {author} {\bibfnamefont {A.}~\bibnamefont {{Wieteska}}}, \bibinfo {author}
  {\bibfnamefont {A.~K.}\ \bibnamefont {{Bartholomew}}}, \bibinfo {author}
  {\bibfnamefont {Y.-S.}\ \bibnamefont {{Chen}}}, \bibinfo {author}
  {\bibfnamefont {X.}~\bibnamefont {{Xu}}}, \bibinfo {author} {\bibfnamefont
  {A.~N.}\ \bibnamefont {{Pasupathy}}}, \bibinfo {author} {\bibfnamefont
  {X.}~\bibnamefont {{Zhu}}}, \bibinfo {author} {\bibfnamefont {C.~R.}\
  \bibnamefont {{Dean}}}, \ and\ \bibinfo {author} {\bibfnamefont
  {X.}~\bibnamefont {{Roy}}},\ }\bibfield  {title} {\enquote {\bibinfo {title}
  {{Layered Antiferromagnetism Induces Large Negative Magnetoresistance in the
  van der Waals Semiconductor CrSBr}},}\ }\href@noop {} {\bibfield  {journal}
  {\bibinfo  {journal} {arXiv e-prints}\ ,\ \bibinfo {eid} {arXiv:2005.06110}}
  (\bibinfo {year} {2020})},\ \Eprint {http://arxiv.org/abs/2005.06110}
  {arXiv:2005.06110} \BibitemShut {NoStop}%
\bibitem [{\citenamefont {{Lee}}\ \emph {et~al.}(2020)\citenamefont {{Lee}},
  \citenamefont {{Dismukes}}, \citenamefont {{Telford}}, \citenamefont
  {{Wiscons}}, \citenamefont {{Xu}}, \citenamefont {{Nuckolls}}, \citenamefont
  {{Dean}}, \citenamefont {{Roy}},\ and\ \citenamefont
  {{Zhu}}}]{Lee2020_arxiv_CrSBr}%
  \BibitemOpen
  \bibfield  {author} {\bibinfo {author} {\bibfnamefont {K.}~\bibnamefont
  {{Lee}}}, \bibinfo {author} {\bibfnamefont {A.~H.}\ \bibnamefont
  {{Dismukes}}}, \bibinfo {author} {\bibfnamefont {E.~J.}\ \bibnamefont
  {{Telford}}}, \bibinfo {author} {\bibfnamefont {R.~A.}\ \bibnamefont
  {{Wiscons}}}, \bibinfo {author} {\bibfnamefont {X.}~\bibnamefont {{Xu}}},
  \bibinfo {author} {\bibfnamefont {C.}~\bibnamefont {{Nuckolls}}}, \bibinfo
  {author} {\bibfnamefont {C.~R.}\ \bibnamefont {{Dean}}}, \bibinfo {author}
  {\bibfnamefont {X.}~\bibnamefont {{Roy}}}, \ and\ \bibinfo {author}
  {\bibfnamefont {X.}~\bibnamefont {{Zhu}}},\ }\bibfield  {title} {\enquote
  {\bibinfo {title} {{Magnetic Order and Symmetry in the 2D Semiconductor
  CrSBr}},}\ }\href@noop {} {\bibfield  {journal} {\bibinfo  {journal} {arXiv
  e-prints}\ ,\ \bibinfo {eid} {arXiv:2007.10715}} (\bibinfo {year} {2020})},\
  \Eprint {http://arxiv.org/abs/2007.10715} {arXiv:2007.10715} \BibitemShut
  {NoStop}%
\bibitem [{\citenamefont {Komsa}\ and\ \citenamefont
  {Krasheninnikov}(2012)}]{Komsa12}%
  \BibitemOpen
  \bibfield  {author} {\bibinfo {author} {\bibfnamefont {H.-P.}\ \bibnamefont
  {Komsa}}\ and\ \bibinfo {author} {\bibfnamefont {A.~V.}\ \bibnamefont
  {Krasheninnikov}},\ }\bibfield  {title} {\enquote {\bibinfo {title}
  {{Two-Dimensional Transition Metal Dichalcogenide Alloys: Stability and
  Electronic Properties}},}\ }\href {\doibase 10.1021/jz301673x} {\bibfield
  {journal} {\bibinfo  {journal} {The Journal of Physical Chemistry Letters}\
  }\textbf {\bibinfo {volume} {3}},\ \bibinfo {pages} {3652--3656} (\bibinfo
  {year} {2012})}\BibitemShut {NoStop}%
\bibitem [{\citenamefont {Xie}(2015)}]{Xie15_alloys}%
  \BibitemOpen
  \bibfield  {author} {\bibinfo {author} {\bibfnamefont {L.~M.}\ \bibnamefont
  {Xie}},\ }\bibfield  {title} {\enquote {\bibinfo {title} {Two-dimensional
  transition metal dichalcogenide alloys: preparation{,} characterization and
  applications},}\ }\href {\doibase 10.1039/C5NR05712D} {\bibfield  {journal}
  {\bibinfo  {journal} {Nanoscale}\ }\textbf {\bibinfo {volume} {7}},\ \bibinfo
  {pages} {18392--18401} (\bibinfo {year} {2015})}\BibitemShut {NoStop}%
\bibitem [{\citenamefont {Song}\ \emph {et~al.}(2014)\citenamefont {Song},
  \citenamefont {Liu}, \citenamefont {Yang}, \citenamefont {Han}, \citenamefont
  {Ye}, \citenamefont {Fu}, \citenamefont {Yang}, \citenamefont {Niu},
  \citenamefont {Lu},\ and\ \citenamefont {Yao}}]{Song14_BiX_SbX}%
  \BibitemOpen
  \bibfield  {author} {\bibinfo {author} {\bibfnamefont {Z.}~\bibnamefont
  {Song}}, \bibinfo {author} {\bibfnamefont {C.-C.}\ \bibnamefont {Liu}},
  \bibinfo {author} {\bibfnamefont {J.}~\bibnamefont {Yang}}, \bibinfo {author}
  {\bibfnamefont {J.}~\bibnamefont {Han}}, \bibinfo {author} {\bibfnamefont
  {M.}~\bibnamefont {Ye}}, \bibinfo {author} {\bibfnamefont {B.}~\bibnamefont
  {Fu}}, \bibinfo {author} {\bibfnamefont {Y.}~\bibnamefont {Yang}}, \bibinfo
  {author} {\bibfnamefont {Q.}~\bibnamefont {Niu}}, \bibinfo {author}
  {\bibfnamefont {J.}~\bibnamefont {Lu}}, \ and\ \bibinfo {author}
  {\bibfnamefont {Y.}~\bibnamefont {Yao}},\ }\bibfield  {title} {\enquote
  {\bibinfo {title} {{Quantum spin Hall insulators and quantum valley Hall
  insulators of BiX/SbX (X= H, F, Cl and Br) monolayers with a record bulk band
  gap}},}\ }\href {\doibase 10.1038/am.2014.113} {\bibfield  {journal}
  {\bibinfo  {journal} {NPG Asia Mater.}\ }\textbf {\bibinfo {volume} {6}},\
  \bibinfo {pages} {e147} (\bibinfo {year} {2014})}\BibitemShut {NoStop}%
\bibitem [{\citenamefont {Vannucci}, \citenamefont {Olsen},\ and\ \citenamefont
  {Thygesen}(2020)}]{Vannucci20}%
  \BibitemOpen
  \bibfield  {author} {\bibinfo {author} {\bibfnamefont {L.}~\bibnamefont
  {Vannucci}}, \bibinfo {author} {\bibfnamefont {T.}~\bibnamefont {Olsen}}, \
  and\ \bibinfo {author} {\bibfnamefont {K.~S.}\ \bibnamefont {Thygesen}},\
  }\bibfield  {title} {\enquote {\bibinfo {title} {{Conductance of quantum spin
  Hall edge states from first principles: The critical role of magnetic
  impurities and inter-edge scattering}},}\ }\href {\doibase
  10.1103/PhysRevB.101.155404} {\bibfield  {journal} {\bibinfo  {journal}
  {Phys. Rev. B}\ }\textbf {\bibinfo {volume} {101}},\ \bibinfo {pages}
  {155404} (\bibinfo {year} {2020})}\BibitemShut {NoStop}%
\bibitem [{\citenamefont {Mermin}\ and\ \citenamefont
  {Wagner}(1966)}]{Mermin1966}%
  \BibitemOpen
  \bibfield  {author} {\bibinfo {author} {\bibfnamefont {N.~D.}\ \bibnamefont
  {Mermin}}\ and\ \bibinfo {author} {\bibfnamefont {H.}~\bibnamefont
  {Wagner}},\ }\bibfield  {title} {\enquote {\bibinfo {title} {{Absence of
  Ferromagnetism or Antiferromagnetism in One- or Two-Dimensional Isotropic
  Heisenberg Models}},}\ }\href {\doibase 10.1103/PhysRevLett.17.1133}
  {\bibfield  {journal} {\bibinfo  {journal} {Phys. Rev. Lett.}\ }\textbf
  {\bibinfo {volume} {17}},\ \bibinfo {pages} {1133} (\bibinfo {year}
  {1966})}\BibitemShut {NoStop}%
\bibitem [{\citenamefont {Fei}\ \emph {et~al.}(2018)\citenamefont {Fei},
  \citenamefont {Huang}, \citenamefont {Malinowski}, \citenamefont {Wang},
  \citenamefont {Song}, \citenamefont {Sanchez}, \citenamefont {Yao},
  \citenamefont {Xiao}, \citenamefont {Zhu}, \citenamefont {May}, \citenamefont
  {Wu}, \citenamefont {Cobden}, \citenamefont {Chu},\ and\ \citenamefont
  {Xu}}]{Fei2018}%
  \BibitemOpen
  \bibfield  {author} {\bibinfo {author} {\bibfnamefont {Z.}~\bibnamefont
  {Fei}}, \bibinfo {author} {\bibfnamefont {B.}~\bibnamefont {Huang}}, \bibinfo
  {author} {\bibfnamefont {P.}~\bibnamefont {Malinowski}}, \bibinfo {author}
  {\bibfnamefont {W.}~\bibnamefont {Wang}}, \bibinfo {author} {\bibfnamefont
  {T.}~\bibnamefont {Song}}, \bibinfo {author} {\bibfnamefont {J.}~\bibnamefont
  {Sanchez}}, \bibinfo {author} {\bibfnamefont {W.}~\bibnamefont {Yao}},
  \bibinfo {author} {\bibfnamefont {D.}~\bibnamefont {Xiao}}, \bibinfo {author}
  {\bibfnamefont {X.}~\bibnamefont {Zhu}}, \bibinfo {author} {\bibfnamefont
  {A.~F.}\ \bibnamefont {May}}, \bibinfo {author} {\bibfnamefont
  {W.}~\bibnamefont {Wu}}, \bibinfo {author} {\bibfnamefont {D.~H.}\
  \bibnamefont {Cobden}}, \bibinfo {author} {\bibfnamefont {J.-H.}\
  \bibnamefont {Chu}}, \ and\ \bibinfo {author} {\bibfnamefont
  {X.}~\bibnamefont {Xu}},\ }\bibfield  {title} {\enquote {\bibinfo {title}
  {{Two-dimensional itinerant ferromagnetism in atomically thin
  Fe$_3$GeTe$_2$}},}\ }\href {\doibase 10.1038/s41563-018-0149-7} {\bibfield
  {journal} {\bibinfo  {journal} {Nature Mater.}\ }\textbf {\bibinfo {volume}
  {17}},\ \bibinfo {pages} {778–782} (\bibinfo {year} {2018})}\BibitemShut
  {NoStop}%
\bibitem [{\citenamefont {Zhang}\ \emph
  {et~al.}(2015{\natexlab{a}})\citenamefont {Zhang}, \citenamefont {Qu},
  \citenamefont {Zhu},\ and\ \citenamefont {Lam}}]{Zhang2015}%
  \BibitemOpen
  \bibfield  {author} {\bibinfo {author} {\bibfnamefont {W.-B.}\ \bibnamefont
  {Zhang}}, \bibinfo {author} {\bibfnamefont {Q.}~\bibnamefont {Qu}}, \bibinfo
  {author} {\bibfnamefont {P.}~\bibnamefont {Zhu}}, \ and\ \bibinfo {author}
  {\bibfnamefont {C.-H.}\ \bibnamefont {Lam}},\ }\bibfield  {title} {\enquote
  {\bibinfo {title} {{Robust intrinsic ferromagnetism and half semiconductivity
  in stable two-dimensional single-layer chromium trihalides}},}\ }\href
  {\doibase 10.1039/C5TC02840J} {\bibfield  {journal} {\bibinfo  {journal} {J.
  Mater. Chem. C}\ }\textbf {\bibinfo {volume} {3}},\ \bibinfo {pages}
  {12457--12468} (\bibinfo {year} {2015}{\natexlab{a}})}\BibitemShut {NoStop}%
\bibitem [{\citenamefont {Lado}\ and\ \citenamefont
  {Fernández-Rossier}(2017)}]{Rossier2017}%
  \BibitemOpen
  \bibfield  {author} {\bibinfo {author} {\bibfnamefont {J.~L.}\ \bibnamefont
  {Lado}}\ and\ \bibinfo {author} {\bibfnamefont {J.}~\bibnamefont
  {Fernández-Rossier}},\ }\bibfield  {title} {\enquote {\bibinfo {title} {{On
  the origin of magnetic anisotropy in two dimensional CrI$_3$}},}\ }\href
  {\doibase 10.1088/2053-1583/aa75ed} {\bibfield  {journal} {\bibinfo
  {journal} {2D Mater.}\ }\textbf {\bibinfo {volume} {4}},\ \bibinfo {pages}
  {035002} (\bibinfo {year} {2017})}\BibitemShut {NoStop}%
\bibitem [{\citenamefont {Torelli}\ and\ \citenamefont
  {Olsen}(2018)}]{Torelli2018}%
  \BibitemOpen
  \bibfield  {author} {\bibinfo {author} {\bibfnamefont {D.}~\bibnamefont
  {Torelli}}\ and\ \bibinfo {author} {\bibfnamefont {T.}~\bibnamefont
  {Olsen}},\ }\bibfield  {title} {\enquote {\bibinfo {title} {{Calculating
  critical temperatures for ferromagnetic order in two-dimensional
  materials}},}\ }\href {\doibase 10.1088/2053-1583/aaf06d} {\bibfield
  {journal} {\bibinfo  {journal} {2D Mater.}\ }\textbf {\bibinfo {volume}
  {6}},\ \bibinfo {pages} {015028} (\bibinfo {year} {2018})}\BibitemShut
  {NoStop}%
\bibitem [{\citenamefont {Torelli}, \citenamefont {Thygesen},\ and\
  \citenamefont {Olsen}(2019)}]{Torelli2019}%
  \BibitemOpen
  \bibfield  {author} {\bibinfo {author} {\bibfnamefont {D.}~\bibnamefont
  {Torelli}}, \bibinfo {author} {\bibfnamefont {K.~S.}\ \bibnamefont
  {Thygesen}}, \ and\ \bibinfo {author} {\bibfnamefont {T.}~\bibnamefont
  {Olsen}},\ }\bibfield  {title} {\enquote {\bibinfo {title} {{High throughput
  computational screening for 2D ferromagnetic materials: the critical role of
  anisotropy and local correlations}},}\ }\href {\doibase
  10.1088/2053-1583/ab2c43} {\bibfield  {journal} {\bibinfo  {journal} {2D
  Mater.}\ }\textbf {\bibinfo {volume} {6}},\ \bibinfo {pages} {045018}
  (\bibinfo {year} {2019})}\BibitemShut {NoStop}%
\bibitem [{\citenamefont {Gong}\ and\ \citenamefont {Zhang}(2019)}]{Gong2019}%
  \BibitemOpen
  \bibfield  {author} {\bibinfo {author} {\bibfnamefont {C.}~\bibnamefont
  {Gong}}\ and\ \bibinfo {author} {\bibfnamefont {X.}~\bibnamefont {Zhang}},\
  }\bibfield  {title} {\enquote {\bibinfo {title} {Two-dimensional magnetic
  crystals and emergent heterostructure devices},}\ }\href {\doibase
  10.1126/science.aav4450} {\bibfield  {journal} {\bibinfo  {journal}
  {Science}\ }\textbf {\bibinfo {volume} {363}},\ \bibinfo {pages} {eaav4450}
  (\bibinfo {year} {2019})}\BibitemShut {NoStop}%
\bibitem [{\citenamefont {Cardoso}\ \emph {et~al.}(2018)\citenamefont
  {Cardoso}, \citenamefont {Soriano}, \citenamefont {García-Martínez},\ and\
  \citenamefont {Fernández-Rossier}}]{Cardoso2018}%
  \BibitemOpen
  \bibfield  {author} {\bibinfo {author} {\bibfnamefont {C.}~\bibnamefont
  {Cardoso}}, \bibinfo {author} {\bibfnamefont {D.}~\bibnamefont {Soriano}},
  \bibinfo {author} {\bibfnamefont {N.~A.}\ \bibnamefont {García-Martínez}},
  \ and\ \bibinfo {author} {\bibfnamefont {J.}~\bibnamefont
  {Fernández-Rossier}},\ }\bibfield  {title} {\enquote {\bibinfo {title} {{Van
  der Waals Spin Valves}},}\ }\href {\doibase 10.1103/PhysRevLett.121.067701}
  {\bibfield  {journal} {\bibinfo  {journal} {Phys. Rev. Lett.}\ }\textbf
  {\bibinfo {volume} {121}},\ \bibinfo {pages} {067701} (\bibinfo {year}
  {2018})}\BibitemShut {NoStop}%
\bibitem [{\citenamefont {Zhong}\ \emph {et~al.}(2017)\citenamefont {Zhong},
  \citenamefont {Seyler}, \citenamefont {Linpeng}, \citenamefont {Cheng},
  \citenamefont {Sivadas}, \citenamefont {Huang}, \citenamefont {Schmidgall},
  \citenamefont {Taniguchi}, \citenamefont {Watanabe}, \citenamefont {McGuire},
  \citenamefont {Yao}, \citenamefont {Xiao}, \citenamefont {Fu},\ and\
  \citenamefont {Xu}}]{Zhong2017}%
  \BibitemOpen
  \bibfield  {author} {\bibinfo {author} {\bibfnamefont {D.}~\bibnamefont
  {Zhong}}, \bibinfo {author} {\bibfnamefont {K.~L.}\ \bibnamefont {Seyler}},
  \bibinfo {author} {\bibfnamefont {X.}~\bibnamefont {Linpeng}}, \bibinfo
  {author} {\bibfnamefont {R.}~\bibnamefont {Cheng}}, \bibinfo {author}
  {\bibfnamefont {N.}~\bibnamefont {Sivadas}}, \bibinfo {author} {\bibfnamefont
  {B.}~\bibnamefont {Huang}}, \bibinfo {author} {\bibfnamefont
  {E.}~\bibnamefont {Schmidgall}}, \bibinfo {author} {\bibfnamefont
  {T.}~\bibnamefont {Taniguchi}}, \bibinfo {author} {\bibfnamefont
  {K.}~\bibnamefont {Watanabe}}, \bibinfo {author} {\bibfnamefont {M.~A.}\
  \bibnamefont {McGuire}}, \bibinfo {author} {\bibfnamefont {W.}~\bibnamefont
  {Yao}}, \bibinfo {author} {\bibfnamefont {D.}~\bibnamefont {Xiao}}, \bibinfo
  {author} {\bibfnamefont {K.-M.~C.}\ \bibnamefont {Fu}}, \ and\ \bibinfo
  {author} {\bibfnamefont {X.}~\bibnamefont {Xu}},\ }\bibfield  {title}
  {\enquote {\bibinfo {title} {{Van der Waals engineering of ferromagnetic
  semiconductor heterostructures for spin and valleytronics}},}\ }\href
  {\doibase 10.1126/sciadv.1603113} {\bibfield  {journal} {\bibinfo  {journal}
  {Sci. Adv.}\ }\textbf {\bibinfo {volume} {3}},\ \bibinfo {pages} {e1603113}
  (\bibinfo {year} {2017})}\BibitemShut {NoStop}%
\bibitem [{\citenamefont {Burch}, \citenamefont {Mandrus},\ and\ \citenamefont
  {Park}(2018)}]{Burch2018}%
  \BibitemOpen
  \bibfield  {author} {\bibinfo {author} {\bibfnamefont {K.}~\bibnamefont
  {Burch}}, \bibinfo {author} {\bibfnamefont {D.}~\bibnamefont {Mandrus}}, \
  and\ \bibinfo {author} {\bibfnamefont {J.}~\bibnamefont {Park}},\ }\bibfield
  {title} {\enquote {\bibinfo {title} {{Magnetism in two-dimensional van der
  Waals materials}},}\ }\href {\doibase 10.1038/s41586-018-0631-z} {\bibfield
  {journal} {\bibinfo  {journal} {Nature}\ }\textbf {\bibinfo {volume} {563}},\
  \bibinfo {pages} {47–52} (\bibinfo {year} {2018})}\BibitemShut {NoStop}%
\bibitem [{\citenamefont {Wang}\ \emph {et~al.}(2018)\citenamefont {Wang},
  \citenamefont {Gutiérrez-Lezama}, \citenamefont {Ubrig}, \citenamefont
  {Kroner}, \citenamefont {Gibertini}, \citenamefont {Taniguchi}, \citenamefont
  {Watanabe}, \citenamefont {Imamoğlu}, \citenamefont {Giannini},\ and\
  \citenamefont {Morpurgo}}]{WangMorpurgo2018}%
  \BibitemOpen
  \bibfield  {author} {\bibinfo {author} {\bibfnamefont {Z.}~\bibnamefont
  {Wang}}, \bibinfo {author} {\bibfnamefont {I.}~\bibnamefont
  {Gutiérrez-Lezama}}, \bibinfo {author} {\bibfnamefont {N.}~\bibnamefont
  {Ubrig}}, \bibinfo {author} {\bibfnamefont {M.}~\bibnamefont {Kroner}},
  \bibinfo {author} {\bibfnamefont {M.}~\bibnamefont {Gibertini}}, \bibinfo
  {author} {\bibfnamefont {T.}~\bibnamefont {Taniguchi}}, \bibinfo {author}
  {\bibfnamefont {K.}~\bibnamefont {Watanabe}}, \bibinfo {author}
  {\bibfnamefont {A.}~\bibnamefont {Imamoğlu}}, \bibinfo {author}
  {\bibfnamefont {E.}~\bibnamefont {Giannini}}, \ and\ \bibinfo {author}
  {\bibfnamefont {A.~F.}\ \bibnamefont {Morpurgo}},\ }\bibfield  {title}
  {\enquote {\bibinfo {title} {{Very large tunneling magnetoresistance in
  layered magnetic semiconductor CrI$_3$}},}\ }\href {\doibase
  10.1038/s41467-018-04953-8} {\bibfield  {journal} {\bibinfo  {journal} {Nat
  Commun}\ }\textbf {\bibinfo {volume} {9}},\ \bibinfo {pages} {2516} (\bibinfo
  {year} {2018})}\BibitemShut {NoStop}%
\bibitem [{\citenamefont {Olsen}(2016)}]{Olsen16_SOC}%
  \BibitemOpen
  \bibfield  {author} {\bibinfo {author} {\bibfnamefont {T.}~\bibnamefont
  {Olsen}},\ }\bibfield  {title} {\enquote {\bibinfo {title} {{Designing
  in-plane heterostructures of quantum spin Hall insulators from first
  principles: $1{\mathrm{T}}^{\ensuremath{'}}\ensuremath{-}{\mathrm{MoS}}_{2}$
  with adsorbates}},}\ }\href {\doibase 10.1103/PhysRevB.94.235106} {\bibfield
  {journal} {\bibinfo  {journal} {Phys. Rev. B}\ }\textbf {\bibinfo {volume}
  {94}},\ \bibinfo {pages} {235106} (\bibinfo {year} {2016})}\BibitemShut
  {NoStop}%
\bibitem [{\citenamefont {Olsen}\ \emph {et~al.}(2019)\citenamefont {Olsen},
  \citenamefont {Andersen}, \citenamefont {Okugawa}, \citenamefont {Torelli},
  \citenamefont {Deilmann},\ and\ \citenamefont
  {Thygesen}}]{Olsen19_topological}%
  \BibitemOpen
  \bibfield  {author} {\bibinfo {author} {\bibfnamefont {T.}~\bibnamefont
  {Olsen}}, \bibinfo {author} {\bibfnamefont {E.}~\bibnamefont {Andersen}},
  \bibinfo {author} {\bibfnamefont {T.}~\bibnamefont {Okugawa}}, \bibinfo
  {author} {\bibfnamefont {D.}~\bibnamefont {Torelli}}, \bibinfo {author}
  {\bibfnamefont {T.}~\bibnamefont {Deilmann}}, \ and\ \bibinfo {author}
  {\bibfnamefont {K.~S.}\ \bibnamefont {Thygesen}},\ }\bibfield  {title}
  {\enquote {\bibinfo {title} {Discovering two-dimensional topological
  insulators from high-throughput computations},}\ }\href {\doibase
  10.1103/PhysRevMaterials.3.024005} {\bibfield  {journal} {\bibinfo  {journal}
  {Phys. Rev. Materials}\ }\textbf {\bibinfo {volume} {3}},\ \bibinfo {pages}
  {024005} (\bibinfo {year} {2019})}\BibitemShut {NoStop}%
\bibitem [{\citenamefont {Kappera}\ \emph {et~al.}(2014)\citenamefont
  {Kappera}, \citenamefont {Voiry}, \citenamefont {Yalcin}, \citenamefont
  {Branch}, \citenamefont {Gupta}, \citenamefont {Mohite},\ and\ \citenamefont
  {Chhowalla}}]{Kappera2014}%
  \BibitemOpen
  \bibfield  {author} {\bibinfo {author} {\bibfnamefont {R.}~\bibnamefont
  {Kappera}}, \bibinfo {author} {\bibfnamefont {D.}~\bibnamefont {Voiry}},
  \bibinfo {author} {\bibfnamefont {S.~E.}\ \bibnamefont {Yalcin}}, \bibinfo
  {author} {\bibfnamefont {B.}~\bibnamefont {Branch}}, \bibinfo {author}
  {\bibfnamefont {G.}~\bibnamefont {Gupta}}, \bibinfo {author} {\bibfnamefont
  {A.~D.}\ \bibnamefont {Mohite}}, \ and\ \bibinfo {author} {\bibfnamefont
  {M.}~\bibnamefont {Chhowalla}},\ }\bibfield  {title} {\enquote {\bibinfo
  {title} {{Phase-engineered low-resistance contacts for ultrathin MoS$_2$
  transistors}},}\ }\href {\doibase 10.1038/nmat4080} {\bibfield  {journal}
  {\bibinfo  {journal} {Nature Mater.}\ }\textbf {\bibinfo {volume} {13}},\
  \bibinfo {pages} {1128–1134} (\bibinfo {year} {2014})}\BibitemShut
  {NoStop}%
\bibitem [{\citenamefont {Akinwande}\ \emph {et~al.}(2017)\citenamefont
  {Akinwande}, \citenamefont {Brennan}, \citenamefont {Bunch}, \citenamefont
  {Egberts}, \citenamefont {Felts}, \citenamefont {Gao}, \citenamefont {Huang},
  \citenamefont {Kim}, \citenamefont {Li}, \citenamefont {Li}, \citenamefont
  {Liechti}, \citenamefont {Lu}, \citenamefont {Park}, \citenamefont {Reed},
  \citenamefont {Wang}, \citenamefont {Yakobson}, \citenamefont {Zhang},
  \citenamefont {Zhang}, \citenamefont {Zhou},\ and\ \citenamefont
  {Zhu}}]{Akinwande17}%
  \BibitemOpen
  \bibfield  {author} {\bibinfo {author} {\bibfnamefont {D.}~\bibnamefont
  {Akinwande}}, \bibinfo {author} {\bibfnamefont {C.~J.}\ \bibnamefont
  {Brennan}}, \bibinfo {author} {\bibfnamefont {J.~S.}\ \bibnamefont {Bunch}},
  \bibinfo {author} {\bibfnamefont {P.}~\bibnamefont {Egberts}}, \bibinfo
  {author} {\bibfnamefont {J.~R.}\ \bibnamefont {Felts}}, \bibinfo {author}
  {\bibfnamefont {H.}~\bibnamefont {Gao}}, \bibinfo {author} {\bibfnamefont
  {R.}~\bibnamefont {Huang}}, \bibinfo {author} {\bibfnamefont {J.-S.}\
  \bibnamefont {Kim}}, \bibinfo {author} {\bibfnamefont {T.}~\bibnamefont
  {Li}}, \bibinfo {author} {\bibfnamefont {Y.}~\bibnamefont {Li}}, \bibinfo
  {author} {\bibfnamefont {K.~M.}\ \bibnamefont {Liechti}}, \bibinfo {author}
  {\bibfnamefont {N.}~\bibnamefont {Lu}}, \bibinfo {author} {\bibfnamefont
  {H.~S.}\ \bibnamefont {Park}}, \bibinfo {author} {\bibfnamefont {E.~J.}\
  \bibnamefont {Reed}}, \bibinfo {author} {\bibfnamefont {P.}~\bibnamefont
  {Wang}}, \bibinfo {author} {\bibfnamefont {B.~I.}\ \bibnamefont {Yakobson}},
  \bibinfo {author} {\bibfnamefont {T.}~\bibnamefont {Zhang}}, \bibinfo
  {author} {\bibfnamefont {Y.-W.}\ \bibnamefont {Zhang}}, \bibinfo {author}
  {\bibfnamefont {Y.}~\bibnamefont {Zhou}}, \ and\ \bibinfo {author}
  {\bibfnamefont {Y.}~\bibnamefont {Zhu}},\ }\bibfield  {title} {\enquote
  {\bibinfo {title} {{A review on mechanics and mechanical properties of 2D
  materials---Graphene and beyond}},}\ }\href {\doibase
  https://doi.org/10.1016/j.eml.2017.01.008} {\bibfield  {journal} {\bibinfo
  {journal} {Extreme Mechanics Letters}\ }\textbf {\bibinfo {volume} {13}},\
  \bibinfo {pages} {42 -- 77} (\bibinfo {year} {2017})}\BibitemShut {NoStop}%
\bibitem [{\citenamefont {Androulidakis}\ \emph {et~al.}(2018)\citenamefont
  {Androulidakis}, \citenamefont {Zhang}, \citenamefont {Robertson},\ and\
  \citenamefont {Tawfick}}]{Androulidakis18}%
  \BibitemOpen
  \bibfield  {author} {\bibinfo {author} {\bibfnamefont {C.}~\bibnamefont
  {Androulidakis}}, \bibinfo {author} {\bibfnamefont {K.}~\bibnamefont
  {Zhang}}, \bibinfo {author} {\bibfnamefont {M.}~\bibnamefont {Robertson}}, \
  and\ \bibinfo {author} {\bibfnamefont {S.}~\bibnamefont {Tawfick}},\
  }\bibfield  {title} {\enquote {\bibinfo {title} {{Tailoring the mechanical
  properties of 2D materials and heterostructures}},}\ }\href {\doibase
  10.1088/2053-1583/aac764} {\bibfield  {journal} {\bibinfo  {journal} {2D
  Materials}\ }\textbf {\bibinfo {volume} {5}},\ \bibinfo {pages} {032005}
  (\bibinfo {year} {2018})}\BibitemShut {NoStop}%
\bibitem [{\citenamefont {Landau}\ and\ \citenamefont
  {Lifshitz}(1970)}]{Landau_elasticity}%
  \BibitemOpen
  \bibfield  {author} {\bibinfo {author} {\bibfnamefont {L.~D.}\ \bibnamefont
  {Landau}}\ and\ \bibinfo {author} {\bibfnamefont {E.~M.}\ \bibnamefont
  {Lifshitz}},\ }\href@noop {} {\emph {\bibinfo {title} {Theory of
  elasticity}}},\ \bibinfo {edition} {2nd}\ ed.,\ \bibinfo {series} {Course of
  theoretical Physics}, Vol.~\bibinfo {volume} {7}\ (\bibinfo  {publisher}
  {Pergamon Press},\ \bibinfo {address} {Oxford},\ \bibinfo {year}
  {1970})\BibitemShut {NoStop}%
\bibitem [{\citenamefont {{Jiang}}, \citenamefont {{Kim}},\ and\ \citenamefont
  {{Park}}(2016)}]{Jiang16}%
  \BibitemOpen
  \bibfield  {author} {\bibinfo {author} {\bibfnamefont {J.-W.}\ \bibnamefont
  {{Jiang}}}, \bibinfo {author} {\bibfnamefont {S.~Y.}\ \bibnamefont {{Kim}}},
  \ and\ \bibinfo {author} {\bibfnamefont {H.~S.}\ \bibnamefont {{Park}}},\
  }\bibfield  {title} {\enquote {\bibinfo {title} {{Auxetic nanomaterials:
  Recent progress and future development}},}\ }\href {\doibase
  10.1063/1.4964479} {\bibfield  {journal} {\bibinfo  {journal} {Applied
  Physics Reviews}\ }\textbf {\bibinfo {volume} {3}},\ \bibinfo {pages}
  {041101} (\bibinfo {year} {2016})}\BibitemShut {NoStop}%
\bibitem [{\citenamefont {Ting}\ and\ \citenamefont {Chen}(2005)}]{Ting05}%
  \BibitemOpen
  \bibfield  {author} {\bibinfo {author} {\bibfnamefont {T.~C.~T.}\
  \bibnamefont {Ting}}\ and\ \bibinfo {author} {\bibfnamefont {T.}~\bibnamefont
  {Chen}},\ }\bibfield  {title} {\enquote {\bibinfo {title} {{Poisson's ratio
  for anisotropic elastic materials can have no bounds}},}\ }\href {\doibase
  10.1093/qjmamj/hbh021} {\bibfield  {journal} {\bibinfo  {journal} {{The
  Quarterly Journal of Mechanics and Applied Mathematics}}\ }\textbf {\bibinfo
  {volume} {58}},\ \bibinfo {pages} {73--82} (\bibinfo {year}
  {2005})}\BibitemShut {NoStop}%
\bibitem [{\citenamefont {Liu}\ \emph {et~al.}(2019)\citenamefont {Liu},
  \citenamefont {Niu}, \citenamefont {Fu}, \citenamefont {Xi}, \citenamefont
  {Lei},\ and\ \citenamefont {Quhe}}]{Liu19}%
  \BibitemOpen
  \bibfield  {author} {\bibinfo {author} {\bibfnamefont {B.}~\bibnamefont
  {Liu}}, \bibinfo {author} {\bibfnamefont {M.}~\bibnamefont {Niu}}, \bibinfo
  {author} {\bibfnamefont {J.}~\bibnamefont {Fu}}, \bibinfo {author}
  {\bibfnamefont {Z.}~\bibnamefont {Xi}}, \bibinfo {author} {\bibfnamefont
  {M.}~\bibnamefont {Lei}}, \ and\ \bibinfo {author} {\bibfnamefont
  {R.}~\bibnamefont {Quhe}},\ }\bibfield  {title} {\enquote {\bibinfo {title}
  {{Negative Poisson's ratio in puckered two-dimensional materials}},}\ }\href
  {\doibase 10.1103/PhysRevMaterials.3.054002} {\bibfield  {journal} {\bibinfo
  {journal} {Phys. Rev. Materials}\ }\textbf {\bibinfo {volume} {3}},\ \bibinfo
  {pages} {054002} (\bibinfo {year} {2019})}\BibitemShut {NoStop}%
\bibitem [{\citenamefont {Kong}\ \emph {et~al.}(2018)\citenamefont {Kong},
  \citenamefont {Deng}, \citenamefont {Li}, \citenamefont {Liu}, \citenamefont
  {Ding}, \citenamefont {Sun},\ and\ \citenamefont {Liu}}]{Kong18}%
  \BibitemOpen
  \bibfield  {author} {\bibinfo {author} {\bibfnamefont {X.}~\bibnamefont
  {Kong}}, \bibinfo {author} {\bibfnamefont {J.}~\bibnamefont {Deng}}, \bibinfo
  {author} {\bibfnamefont {L.}~\bibnamefont {Li}}, \bibinfo {author}
  {\bibfnamefont {Y.}~\bibnamefont {Liu}}, \bibinfo {author} {\bibfnamefont
  {X.}~\bibnamefont {Ding}}, \bibinfo {author} {\bibfnamefont {J.}~\bibnamefont
  {Sun}}, \ and\ \bibinfo {author} {\bibfnamefont {J.~Z.}\ \bibnamefont
  {Liu}},\ }\bibfield  {title} {\enquote {\bibinfo {title} {{Tunable auxetic
  properties in group-IV monochalcogenide monolayers}},}\ }\href {\doibase
  10.1103/PhysRevB.98.184104} {\bibfield  {journal} {\bibinfo  {journal} {Phys.
  Rev. B}\ }\textbf {\bibinfo {volume} {98}},\ \bibinfo {pages} {184104}
  (\bibinfo {year} {2018})}\BibitemShut {NoStop}%
\bibitem [{\citenamefont {Gomes}, \citenamefont {Carvalho},\ and\ \citenamefont
  {Castro~Neto}(2015)}]{Gomes15}%
  \BibitemOpen
  \bibfield  {author} {\bibinfo {author} {\bibfnamefont {L.~C.}\ \bibnamefont
  {Gomes}}, \bibinfo {author} {\bibfnamefont {A.}~\bibnamefont {Carvalho}}, \
  and\ \bibinfo {author} {\bibfnamefont {A.~H.}\ \bibnamefont {Castro~Neto}},\
  }\bibfield  {title} {\enquote {\bibinfo {title} {{Enhanced piezoelectricity
  and modified dielectric screening of two-dimensional group-IV
  monochalcogenides}},}\ }\href {\doibase 10.1103/PhysRevB.92.214103}
  {\bibfield  {journal} {\bibinfo  {journal} {Phys. Rev. B}\ }\textbf {\bibinfo
  {volume} {92}},\ \bibinfo {pages} {214103} (\bibinfo {year}
  {2015})}\BibitemShut {NoStop}%
\bibitem [{\citenamefont {{Jiang}}\ and\ \citenamefont
  {{Park}}(2014)}]{Jiang14}%
  \BibitemOpen
  \bibfield  {author} {\bibinfo {author} {\bibfnamefont {J.-W.}\ \bibnamefont
  {{Jiang}}}\ and\ \bibinfo {author} {\bibfnamefont {H.~S.}\ \bibnamefont
  {{Park}}},\ }\bibfield  {title} {\enquote {\bibinfo {title} {{Negative
  poisson{\textquoteright}s ratio in single-layer black phosphorus}},}\ }\href
  {\doibase 10.1038/ncomms5727} {\bibfield  {journal} {\bibinfo  {journal}
  {Nature Communications}\ }\textbf {\bibinfo {volume} {5}},\ \bibinfo {pages}
  {4727} (\bibinfo {year} {2014})}\BibitemShut {NoStop}%
\bibitem [{\citenamefont {{Du}}\ \emph {et~al.}(2016)\citenamefont {{Du}},
  \citenamefont {{Maassen}}, \citenamefont {{Wu}}, \citenamefont {{Luo}},
  \citenamefont {{Xu}},\ and\ \citenamefont {{Ye}}}]{Du16}%
  \BibitemOpen
  \bibfield  {author} {\bibinfo {author} {\bibfnamefont {Y.}~\bibnamefont
  {{Du}}}, \bibinfo {author} {\bibfnamefont {J.}~\bibnamefont {{Maassen}}},
  \bibinfo {author} {\bibfnamefont {W.}~\bibnamefont {{Wu}}}, \bibinfo {author}
  {\bibfnamefont {Z.}~\bibnamefont {{Luo}}}, \bibinfo {author} {\bibfnamefont
  {X.}~\bibnamefont {{Xu}}}, \ and\ \bibinfo {author} {\bibfnamefont {P.~D.}\
  \bibnamefont {{Ye}}},\ }\bibfield  {title} {\enquote {\bibinfo {title}
  {{Auxetic Black Phosphorus: A 2D Material with Negative Poisson's Ratio}},}\
  }\href {\doibase 10.1021/acs.nanolett.6b03607} {\bibfield  {journal}
  {\bibinfo  {journal} {Nano Letters}\ }\textbf {\bibinfo {volume} {16}},\
  \bibinfo {pages} {6701--6708} (\bibinfo {year} {2016})}\BibitemShut {NoStop}%
\bibitem [{\citenamefont {{Yu}}, \citenamefont {{Yan}},\ and\ \citenamefont
  {{Ruzsinszky}}(2017)}]{Yu17}%
  \BibitemOpen
  \bibfield  {author} {\bibinfo {author} {\bibfnamefont {L.}~\bibnamefont
  {{Yu}}}, \bibinfo {author} {\bibfnamefont {Q.}~\bibnamefont {{Yan}}}, \ and\
  \bibinfo {author} {\bibfnamefont {A.}~\bibnamefont {{Ruzsinszky}}},\
  }\bibfield  {title} {\enquote {\bibinfo {title} {{Negative Poisson's ratio in
  1T-type crystalline two-dimensional transition metal dichalcogenides}},}\
  }\href {\doibase 10.1038/ncomms15224} {\bibfield  {journal} {\bibinfo
  {journal} {Nature Communications}\ }\textbf {\bibinfo {volume} {8}},\
  \bibinfo {pages} {15224} (\bibinfo {year} {2017})}\BibitemShut {NoStop}%
\bibitem [{\citenamefont {Qin}\ and\ \citenamefont {Qin}(2020)}]{Qin20}%
  \BibitemOpen
  \bibfield  {author} {\bibinfo {author} {\bibfnamefont {G.}~\bibnamefont
  {Qin}}\ and\ \bibinfo {author} {\bibfnamefont {Z.}~\bibnamefont {Qin}},\
  }\bibfield  {title} {\enquote {\bibinfo {title} {{Negative Poisson's ratio in
  two-dimensional honeycomb structures}},}\ }\href {\doibase
  10.1038/s41524-020-0313-x} {\bibfield  {journal} {\bibinfo  {journal} {npj
  Comput. Mater.}\ }\textbf {\bibinfo {volume} {6}},\ \bibinfo {pages} {51}
  (\bibinfo {year} {2020})}\BibitemShut {NoStop}%
\bibitem [{\citenamefont {{Cao}}\ \emph {et~al.}(2012)\citenamefont {{Cao}},
  \citenamefont {{Wang}}, \citenamefont {{Han}}, \citenamefont {{Ye}},
  \citenamefont {{Zhu}}, \citenamefont {{Shi}}, \citenamefont {{Niu}},
  \citenamefont {{Tan}}, \citenamefont {{Wang}}, \citenamefont {{Liu}},\ and\
  \citenamefont {{Feng}}}]{Cao2012_ValleyHall}%
  \BibitemOpen
  \bibfield  {author} {\bibinfo {author} {\bibfnamefont {T.}~\bibnamefont
  {{Cao}}}, \bibinfo {author} {\bibfnamefont {G.}~\bibnamefont {{Wang}}},
  \bibinfo {author} {\bibfnamefont {W.}~\bibnamefont {{Han}}}, \bibinfo
  {author} {\bibfnamefont {H.}~\bibnamefont {{Ye}}}, \bibinfo {author}
  {\bibfnamefont {C.}~\bibnamefont {{Zhu}}}, \bibinfo {author} {\bibfnamefont
  {J.}~\bibnamefont {{Shi}}}, \bibinfo {author} {\bibfnamefont
  {Q.}~\bibnamefont {{Niu}}}, \bibinfo {author} {\bibfnamefont
  {P.}~\bibnamefont {{Tan}}}, \bibinfo {author} {\bibfnamefont
  {E.}~\bibnamefont {{Wang}}}, \bibinfo {author} {\bibfnamefont
  {B.}~\bibnamefont {{Liu}}}, \ and\ \bibinfo {author} {\bibfnamefont
  {J.}~\bibnamefont {{Feng}}},\ }\bibfield  {title} {\enquote {\bibinfo {title}
  {{Valley-selective circular dichroism of monolayer molybdenum disulphide}},}\
  }\href {\doibase 10.1038/ncomms1882} {\bibfield  {journal} {\bibinfo
  {journal} {Nature Communications}\ }\textbf {\bibinfo {volume} {3}},\
  \bibinfo {eid} {887} (\bibinfo {year} {2012})}\BibitemShut {NoStop}%
\bibitem [{\citenamefont {{Mak}}\ \emph {et~al.}(2014)\citenamefont {{Mak}},
  \citenamefont {{McGill}}, \citenamefont {{Park}},\ and\ \citenamefont
  {{McEuen}}}]{Mak14_ValleyHall}%
  \BibitemOpen
  \bibfield  {author} {\bibinfo {author} {\bibfnamefont {K.~F.}\ \bibnamefont
  {{Mak}}}, \bibinfo {author} {\bibfnamefont {K.~L.}\ \bibnamefont {{McGill}}},
  \bibinfo {author} {\bibfnamefont {J.}~\bibnamefont {{Park}}}, \ and\ \bibinfo
  {author} {\bibfnamefont {P.~L.}\ \bibnamefont {{McEuen}}},\ }\bibfield
  {title} {\enquote {\bibinfo {title} {{The valley Hall effect in MoS$_{2}$
  transistors}},}\ }\href {\doibase 10.1126/science.1250140} {\bibfield
  {journal} {\bibinfo  {journal} {Science}\ }\textbf {\bibinfo {volume}
  {344}},\ \bibinfo {pages} {1489--1492} (\bibinfo {year} {2014})}\BibitemShut
  {NoStop}%
\bibitem [{\citenamefont {{Lee}}, \citenamefont {{Mak}},\ and\ \citenamefont
  {{Shan}}(2016)}]{Lee2016_ValleyHall}%
  \BibitemOpen
  \bibfield  {author} {\bibinfo {author} {\bibfnamefont {J.}~\bibnamefont
  {{Lee}}}, \bibinfo {author} {\bibfnamefont {K.~F.}\ \bibnamefont {{Mak}}}, \
  and\ \bibinfo {author} {\bibfnamefont {J.}~\bibnamefont {{Shan}}},\
  }\bibfield  {title} {\enquote {\bibinfo {title} {{Electrical control of the
  valley Hall effect in bilayer MoS$_{2}$ transistors}},}\ }\href {\doibase
  10.1038/nnano.2015.337} {\bibfield  {journal} {\bibinfo  {journal} {Nature
  Nanotechnology}\ }\textbf {\bibinfo {volume} {11}},\ \bibinfo {pages}
  {421--425} (\bibinfo {year} {2016})}\BibitemShut {NoStop}%
\bibitem [{\citenamefont {{Chen}}\ \emph {et~al.}(2018)\citenamefont {{Chen}},
  \citenamefont {{Pai}}, \citenamefont {{Chan}}, \citenamefont {{Sun}},
  \citenamefont {{Xu}}, \citenamefont {{Lin}}, \citenamefont {{Chou}},
  \citenamefont {{Fedorov}},\ and\ \citenamefont {{Chiang}}}]{Chen18_W2Se4}%
  \BibitemOpen
  \bibfield  {author} {\bibinfo {author} {\bibfnamefont {P.}~\bibnamefont
  {{Chen}}}, \bibinfo {author} {\bibfnamefont {W.~W.}\ \bibnamefont {{Pai}}},
  \bibinfo {author} {\bibfnamefont {Y.~H.}\ \bibnamefont {{Chan}}}, \bibinfo
  {author} {\bibfnamefont {W.~L.}\ \bibnamefont {{Sun}}}, \bibinfo {author}
  {\bibfnamefont {C.~Z.}\ \bibnamefont {{Xu}}}, \bibinfo {author}
  {\bibfnamefont {D.~S.}\ \bibnamefont {{Lin}}}, \bibinfo {author}
  {\bibfnamefont {M.~Y.}\ \bibnamefont {{Chou}}}, \bibinfo {author}
  {\bibfnamefont {A.~V.}\ \bibnamefont {{Fedorov}}}, \ and\ \bibinfo {author}
  {\bibfnamefont {T.~C.}\ \bibnamefont {{Chiang}}},\ }\bibfield  {title}
  {\enquote {\bibinfo {title} {{Large quantum-spin-Hall gap in single-layer 1T'
  WSe$_{2}$}},}\ }\href {\doibase 10.1038/s41467-018-04395-2} {\bibfield
  {journal} {\bibinfo  {journal} {Nature Communications}\ }\textbf {\bibinfo
  {volume} {9}},\ \bibinfo {pages} {2003} (\bibinfo {year} {2018})}\BibitemShut
  {NoStop}%
\bibitem [{\citenamefont {Aretouli}\ \emph {et~al.}(2015)\citenamefont
  {Aretouli}, \citenamefont {Tsipas}, \citenamefont {Tsoutsou}, \citenamefont
  {Marquez-Velasco}, \citenamefont {Xenogiannopoulou}, \citenamefont {Giamini},
  \citenamefont {Vassalou}, \citenamefont {Kelaidis},\ and\ \citenamefont
  {Dimoulas}}]{Aretouli15_HfSe2}%
  \BibitemOpen
  \bibfield  {author} {\bibinfo {author} {\bibfnamefont {K.~E.}\ \bibnamefont
  {Aretouli}}, \bibinfo {author} {\bibfnamefont {P.}~\bibnamefont {Tsipas}},
  \bibinfo {author} {\bibfnamefont {D.}~\bibnamefont {Tsoutsou}}, \bibinfo
  {author} {\bibfnamefont {J.}~\bibnamefont {Marquez-Velasco}}, \bibinfo
  {author} {\bibfnamefont {E.}~\bibnamefont {Xenogiannopoulou}}, \bibinfo
  {author} {\bibfnamefont {S.~A.}\ \bibnamefont {Giamini}}, \bibinfo {author}
  {\bibfnamefont {E.}~\bibnamefont {Vassalou}}, \bibinfo {author}
  {\bibfnamefont {N.}~\bibnamefont {Kelaidis}}, \ and\ \bibinfo {author}
  {\bibfnamefont {A.}~\bibnamefont {Dimoulas}},\ }\bibfield  {title} {\enquote
  {\bibinfo {title} {{Two-dimensional semiconductor HfSe2 and MoSe2/HfSe2 van
  der Waals heterostructures by molecular beam epitaxy}},}\ }\href {\doibase
  10.1063/1.4917422} {\bibfield  {journal} {\bibinfo  {journal} {Applied
  Physics Letters}\ }\textbf {\bibinfo {volume} {106}},\ \bibinfo {pages}
  {143105} (\bibinfo {year} {2015})}\BibitemShut {NoStop}%
\bibitem [{\citenamefont {Melchior}\ \emph {et~al.}(2018)\citenamefont
  {Melchior}, \citenamefont {Raju}, \citenamefont {Ike}, \citenamefont
  {Erasmus}, \citenamefont {Kabongo}, \citenamefont {Sigalas}, \citenamefont
  {Iyuke},\ and\ \citenamefont {Ozoemena}}]{Melchior18_TiCO2}%
  \BibitemOpen
  \bibfield  {author} {\bibinfo {author} {\bibfnamefont {S.~A.}\ \bibnamefont
  {Melchior}}, \bibinfo {author} {\bibfnamefont {K.}~\bibnamefont {Raju}},
  \bibinfo {author} {\bibfnamefont {I.~S.}\ \bibnamefont {Ike}}, \bibinfo
  {author} {\bibfnamefont {R.~M.}\ \bibnamefont {Erasmus}}, \bibinfo {author}
  {\bibfnamefont {G.}~\bibnamefont {Kabongo}}, \bibinfo {author} {\bibfnamefont
  {I.}~\bibnamefont {Sigalas}}, \bibinfo {author} {\bibfnamefont {S.~E.}\
  \bibnamefont {Iyuke}}, \ and\ \bibinfo {author} {\bibfnamefont {K.~I.}\
  \bibnamefont {Ozoemena}},\ }\bibfield  {title} {\enquote {\bibinfo {title}
  {{High-Voltage Symmetric Supercapacitor Based on 2D Titanium Carbide
  ({MXene}, Ti2CTx)/Carbon Nanosphere Composites in a Neutral Aqueous
  Electrolyte}},}\ }\href {\doibase 10.1149/2.0401803jes} {\bibfield  {journal}
  {\bibinfo  {journal} {Journal of The Electrochemical Society}\ }\textbf
  {\bibinfo {volume} {165}},\ \bibinfo {pages} {A501--A511} (\bibinfo {year}
  {2018})}\BibitemShut {NoStop}%
\bibitem [{\citenamefont {Mañas-Valero}\ \emph {et~al.}(2016)\citenamefont
  {Mañas-Valero}, \citenamefont {García-López}, \citenamefont {Cantarero},\
  and\ \citenamefont {Galbiati}}]{Manas16_ZrSe2}%
  \BibitemOpen
  \bibfield  {author} {\bibinfo {author} {\bibfnamefont {S.}~\bibnamefont
  {Mañas-Valero}}, \bibinfo {author} {\bibfnamefont {V.}~\bibnamefont
  {García-López}}, \bibinfo {author} {\bibfnamefont {A.}~\bibnamefont
  {Cantarero}}, \ and\ \bibinfo {author} {\bibfnamefont {M.}~\bibnamefont
  {Galbiati}},\ }\bibfield  {title} {\enquote {\bibinfo {title} {{Raman Spectra
  of ZrS2 and ZrSe2 from Bulk to Atomically Thin Layers}},}\ }\href {\doibase
  10.3390/app6090264} {\bibfield  {journal} {\bibinfo  {journal} {Applied
  Sciences}\ }\textbf {\bibinfo {volume} {6}} (\bibinfo {year} {2016}),\
  10.3390/app6090264}\BibitemShut {NoStop}%
\bibitem [{\citenamefont {Sun}\ \emph {et~al.}(2012)\citenamefont {Sun},
  \citenamefont {Cheng}, \citenamefont {Gao}, \citenamefont {Sun},
  \citenamefont {Liu}, \citenamefont {Liu}, \citenamefont {Lei}, \citenamefont
  {Yao}, \citenamefont {He}, \citenamefont {Wei},\ and\ \citenamefont
  {Xie}}]{Sun12_SnS2}%
  \BibitemOpen
  \bibfield  {author} {\bibinfo {author} {\bibfnamefont {Y.}~\bibnamefont
  {Sun}}, \bibinfo {author} {\bibfnamefont {H.}~\bibnamefont {Cheng}}, \bibinfo
  {author} {\bibfnamefont {S.}~\bibnamefont {Gao}}, \bibinfo {author}
  {\bibfnamefont {Z.}~\bibnamefont {Sun}}, \bibinfo {author} {\bibfnamefont
  {Q.}~\bibnamefont {Liu}}, \bibinfo {author} {\bibfnamefont {Q.}~\bibnamefont
  {Liu}}, \bibinfo {author} {\bibfnamefont {F.}~\bibnamefont {Lei}}, \bibinfo
  {author} {\bibfnamefont {T.}~\bibnamefont {Yao}}, \bibinfo {author}
  {\bibfnamefont {J.}~\bibnamefont {He}}, \bibinfo {author} {\bibfnamefont
  {S.}~\bibnamefont {Wei}}, \ and\ \bibinfo {author} {\bibfnamefont
  {Y.}~\bibnamefont {Xie}},\ }\bibfield  {title} {\enquote {\bibinfo {title}
  {{Freestanding Tin Disulfide Single-Layers Realizing Efficient Visible-Light
  Water Splitting}},}\ }\href {\doibase 10.1002/anie.201204675} {\bibfield
  {journal} {\bibinfo  {journal} {Angewandte Chemie International Edition}\
  }\textbf {\bibinfo {volume} {51}},\ \bibinfo {pages} {8727--8731} (\bibinfo
  {year} {2012})}\BibitemShut {NoStop}%
\bibitem [{\citenamefont {Zhang}\ \emph
  {et~al.}(2015{\natexlab{b}})\citenamefont {Zhang}, \citenamefont {Zhu},
  \citenamefont {Wang}, \citenamefont {Feng}, \citenamefont {Qiao},
  \citenamefont {Wen}, \citenamefont {Chen}, \citenamefont {Cui}, \citenamefont
  {Zhang}, \citenamefont {Cai},\ and\ \citenamefont {Xie}}]{Zhang15_ZrS2}%
  \BibitemOpen
  \bibfield  {author} {\bibinfo {author} {\bibfnamefont {M.}~\bibnamefont
  {Zhang}}, \bibinfo {author} {\bibfnamefont {Y.}~\bibnamefont {Zhu}}, \bibinfo
  {author} {\bibfnamefont {X.}~\bibnamefont {Wang}}, \bibinfo {author}
  {\bibfnamefont {Q.}~\bibnamefont {Feng}}, \bibinfo {author} {\bibfnamefont
  {S.}~\bibnamefont {Qiao}}, \bibinfo {author} {\bibfnamefont {W.}~\bibnamefont
  {Wen}}, \bibinfo {author} {\bibfnamefont {Y.}~\bibnamefont {Chen}}, \bibinfo
  {author} {\bibfnamefont {M.}~\bibnamefont {Cui}}, \bibinfo {author}
  {\bibfnamefont {J.}~\bibnamefont {Zhang}}, \bibinfo {author} {\bibfnamefont
  {C.}~\bibnamefont {Cai}}, \ and\ \bibinfo {author} {\bibfnamefont
  {L.}~\bibnamefont {Xie}},\ }\bibfield  {title} {\enquote {\bibinfo {title}
  {{Controlled Synthesis of ZrS2 Monolayer and Few Layers on Hexagonal Boron
  Nitride}},}\ }\href {\doibase 10.1021/jacs.5b03807} {\bibfield  {journal}
  {\bibinfo  {journal} {Journal of the American Chemical Society}\ }\textbf
  {\bibinfo {volume} {137}},\ \bibinfo {pages} {7051--7054} (\bibinfo {year}
  {2015}{\natexlab{b}})}\BibitemShut {NoStop}%
\bibitem [{\citenamefont {Zheng}\ \emph {et~al.}(2019)\citenamefont {Zheng},
  \citenamefont {Zheng}, \citenamefont {Yan}, \citenamefont {Liu},
  \citenamefont {Sun}, \citenamefont {Qi}, \citenamefont {Yang}, \citenamefont
  {Jiang}, \citenamefont {Huang}, \citenamefont {Fan}, \citenamefont {Jiang},
  \citenamefont {Ji}, \citenamefont {Wang},\ and\ \citenamefont
  {Pan}}]{Zheng19_PbI2}%
  \BibitemOpen
  \bibfield  {author} {\bibinfo {author} {\bibfnamefont {W.}~\bibnamefont
  {Zheng}}, \bibinfo {author} {\bibfnamefont {B.}~\bibnamefont {Zheng}},
  \bibinfo {author} {\bibfnamefont {C.}~\bibnamefont {Yan}}, \bibinfo {author}
  {\bibfnamefont {Y.}~\bibnamefont {Liu}}, \bibinfo {author} {\bibfnamefont
  {X.}~\bibnamefont {Sun}}, \bibinfo {author} {\bibfnamefont {Z.}~\bibnamefont
  {Qi}}, \bibinfo {author} {\bibfnamefont {T.}~\bibnamefont {Yang}}, \bibinfo
  {author} {\bibfnamefont {Y.}~\bibnamefont {Jiang}}, \bibinfo {author}
  {\bibfnamefont {W.}~\bibnamefont {Huang}}, \bibinfo {author} {\bibfnamefont
  {P.}~\bibnamefont {Fan}}, \bibinfo {author} {\bibfnamefont {F.}~\bibnamefont
  {Jiang}}, \bibinfo {author} {\bibfnamefont {W.}~\bibnamefont {Ji}}, \bibinfo
  {author} {\bibfnamefont {X.}~\bibnamefont {Wang}}, \ and\ \bibinfo {author}
  {\bibfnamefont {A.}~\bibnamefont {Pan}},\ }\bibfield  {title} {\enquote
  {\bibinfo {title} {{Direct Vapor Growth of 2D Vertical Heterostructures with
  Tunable Band Alignments and Interfacial Charge Transfer Behaviors}},}\ }\href
  {\doibase 10.1002/advs.201802204} {\bibfield  {journal} {\bibinfo  {journal}
  {Advanced Science}\ }\textbf {\bibinfo {volume} {6}},\ \bibinfo {pages}
  {1802204} (\bibinfo {year} {2019})}\BibitemShut {NoStop}%
\bibitem [{\citenamefont {Park}\ \emph {et~al.}(2016)\citenamefont {Park},
  \citenamefont {Jerng}, \citenamefont {Jeon}, \citenamefont {Roy},
  \citenamefont {Akbar}, \citenamefont {Kim}, \citenamefont {Sim},
  \citenamefont {Seong}, \citenamefont {Kim}, \citenamefont {Lee},
  \citenamefont {Kim}, \citenamefont {Yi}, \citenamefont {Kim}, \citenamefont
  {Noh},\ and\ \citenamefont {Chun}}]{Park16_SnSe2}%
  \BibitemOpen
  \bibfield  {author} {\bibinfo {author} {\bibfnamefont {Y.~W.}\ \bibnamefont
  {Park}}, \bibinfo {author} {\bibfnamefont {S.-K.}\ \bibnamefont {Jerng}},
  \bibinfo {author} {\bibfnamefont {J.~H.}\ \bibnamefont {Jeon}}, \bibinfo
  {author} {\bibfnamefont {S.~B.}\ \bibnamefont {Roy}}, \bibinfo {author}
  {\bibfnamefont {K.}~\bibnamefont {Akbar}}, \bibinfo {author} {\bibfnamefont
  {J.}~\bibnamefont {Kim}}, \bibinfo {author} {\bibfnamefont {Y.}~\bibnamefont
  {Sim}}, \bibinfo {author} {\bibfnamefont {M.-J.}\ \bibnamefont {Seong}},
  \bibinfo {author} {\bibfnamefont {J.~H.}\ \bibnamefont {Kim}}, \bibinfo
  {author} {\bibfnamefont {Z.}~\bibnamefont {Lee}}, \bibinfo {author}
  {\bibfnamefont {M.}~\bibnamefont {Kim}}, \bibinfo {author} {\bibfnamefont
  {Y.}~\bibnamefont {Yi}}, \bibinfo {author} {\bibfnamefont {J.}~\bibnamefont
  {Kim}}, \bibinfo {author} {\bibfnamefont {D.~Y.}\ \bibnamefont {Noh}}, \ and\
  \bibinfo {author} {\bibfnamefont {S.-H.}\ \bibnamefont {Chun}},\ }\bibfield
  {title} {\enquote {\bibinfo {title} {{Molecular beam epitaxy of large-area
  {SnSe}$_2$ with monolayer thickness fluctuation}},}\ }\href {\doibase
  10.1088/2053-1583/aa51a2} {\bibfield  {journal} {\bibinfo  {journal} {2D
  Materials}\ }\textbf {\bibinfo {volume} {4}},\ \bibinfo {pages} {014006}
  (\bibinfo {year} {2016})}\BibitemShut {NoStop}%
\bibitem [{\citenamefont {Pozo-Zamudio}\ \emph {et~al.}(2015)\citenamefont
  {Pozo-Zamudio}, \citenamefont {Schwarz}, \citenamefont {Sich}, \citenamefont
  {Akimov}, \citenamefont {Bayer}, \citenamefont {Schofield}, \citenamefont
  {Chekhovich}, \citenamefont {Robinson}, \citenamefont {Kay}, \citenamefont
  {Kolosov}, \citenamefont {Dmitriev}, \citenamefont {Lashkarev}, \citenamefont
  {Borisenko}, \citenamefont {Kolesnikov},\ and\ \citenamefont
  {Tartakovskii}}]{PozoZamudio15}%
  \BibitemOpen
  \bibfield  {author} {\bibinfo {author} {\bibfnamefont {O.~D.}\ \bibnamefont
  {Pozo-Zamudio}}, \bibinfo {author} {\bibfnamefont {S.}~\bibnamefont
  {Schwarz}}, \bibinfo {author} {\bibfnamefont {M.}~\bibnamefont {Sich}},
  \bibinfo {author} {\bibfnamefont {I.~A.}\ \bibnamefont {Akimov}}, \bibinfo
  {author} {\bibfnamefont {M.}~\bibnamefont {Bayer}}, \bibinfo {author}
  {\bibfnamefont {R.~C.}\ \bibnamefont {Schofield}}, \bibinfo {author}
  {\bibfnamefont {E.~A.}\ \bibnamefont {Chekhovich}}, \bibinfo {author}
  {\bibfnamefont {B.~J.}\ \bibnamefont {Robinson}}, \bibinfo {author}
  {\bibfnamefont {N.~D.}\ \bibnamefont {Kay}}, \bibinfo {author} {\bibfnamefont
  {O.~V.}\ \bibnamefont {Kolosov}}, \bibinfo {author} {\bibfnamefont {A.~I.}\
  \bibnamefont {Dmitriev}}, \bibinfo {author} {\bibfnamefont {G.~V.}\
  \bibnamefont {Lashkarev}}, \bibinfo {author} {\bibfnamefont {D.~N.}\
  \bibnamefont {Borisenko}}, \bibinfo {author} {\bibfnamefont {N.~N.}\
  \bibnamefont {Kolesnikov}}, \ and\ \bibinfo {author} {\bibfnamefont {A.~I.}\
  \bibnamefont {Tartakovskii}},\ }\bibfield  {title} {\enquote {\bibinfo
  {title} {{Photoluminescence of two-dimensional {GaTe} and {GaSe} films}},}\
  }\href {\doibase 10.1088/2053-1583/2/3/035010} {\bibfield  {journal}
  {\bibinfo  {journal} {2D Materials}\ }\textbf {\bibinfo {volume} {2}},\
  \bibinfo {pages} {035010} (\bibinfo {year} {2015})}\BibitemShut {NoStop}%
\bibitem [{\citenamefont {Le~Ru}\ and\ \citenamefont
  {Etchegoin}(2008)}]{LeRu2008}%
  \BibitemOpen
  \bibfield  {author} {\bibinfo {author} {\bibfnamefont {E.~C.}\ \bibnamefont
  {Le~Ru}}\ and\ \bibinfo {author} {\bibfnamefont {P.~G.}\ \bibnamefont
  {Etchegoin}},\ }\href@noop {} {\emph {\bibinfo {title} {{Principles of
  surface-enhanced Raman spectroscopy and related plasmonic effects}}}},\
  \bibinfo {edition} {1st}\ ed.,\ Vol.~\bibinfo {volume} {7}\ (\bibinfo
  {publisher} {Elsevier},\ \bibinfo {year} {2008})\BibitemShut {NoStop}%
\bibitem [{\citenamefont {Hüser}, \citenamefont {Olsen},\ and\ \citenamefont
  {Thygesen}(2013)}]{Huser2013}%
  \BibitemOpen
  \bibfield  {author} {\bibinfo {author} {\bibfnamefont {F.}~\bibnamefont
  {Hüser}}, \bibinfo {author} {\bibfnamefont {T.}~\bibnamefont {Olsen}}, \
  and\ \bibinfo {author} {\bibfnamefont {K.~S.}\ \bibnamefont {Thygesen}},\
  }\bibfield  {title} {\enquote {\bibinfo {title} {{How dielectric screening in
  two-dimensional crystals affects the convergence of excited-state
  calculations: Monolayer MoS2}},}\ }\href {\doibase
  10.1103/PhysRevB.88.245309} {\bibfield  {journal} {\bibinfo  {journal} {Phys.
  Rev. B}\ }\textbf {\bibinfo {volume} {88}},\ \bibinfo {pages} {245309}
  (\bibinfo {year} {2013})}\BibitemShut {NoStop}%
\bibitem [{\citenamefont {Tian}\ \emph {et~al.}(2020)\citenamefont {Tian},
  \citenamefont {Scullion}, \citenamefont {Hughes}, \citenamefont {Li},
  \citenamefont {Shih}, \citenamefont {Coleman}, \citenamefont {Chhowalla},\
  and\ \citenamefont {Santos}}]{TianSantos2019}%
  \BibitemOpen
  \bibfield  {author} {\bibinfo {author} {\bibfnamefont {T.}~\bibnamefont
  {Tian}}, \bibinfo {author} {\bibfnamefont {D.}~\bibnamefont {Scullion}},
  \bibinfo {author} {\bibfnamefont {D.}~\bibnamefont {Hughes}}, \bibinfo
  {author} {\bibfnamefont {L.~H.}\ \bibnamefont {Li}}, \bibinfo {author}
  {\bibfnamefont {C.-J.}\ \bibnamefont {Shih}}, \bibinfo {author}
  {\bibfnamefont {J.}~\bibnamefont {Coleman}}, \bibinfo {author} {\bibfnamefont
  {M.}~\bibnamefont {Chhowalla}}, \ and\ \bibinfo {author} {\bibfnamefont
  {E.~J.~G.}\ \bibnamefont {Santos}},\ }\bibfield  {title} {\enquote {\bibinfo
  {title} {Electronic polarizability as the fundamental variable in the
  dielectric properties of two-dimensional materials},}\ }\href {\doibase
  10.1021/acs.nanolett.9b02982} {\bibfield  {journal} {\bibinfo  {journal}
  {Nano Letters}\ }\textbf {\bibinfo {volume} {20}},\ \bibinfo {pages}
  {841--851} (\bibinfo {year} {2020})}\BibitemShut {NoStop}%
\bibitem [{\citenamefont {Castro~Neto}\ \emph {et~al.}(2009)\citenamefont
  {Castro~Neto}, \citenamefont {Guinea}, \citenamefont {Peres}, \citenamefont
  {Novoselov},\ and\ \citenamefont {Geim}}]{CastroNeto2009}%
  \BibitemOpen
  \bibfield  {author} {\bibinfo {author} {\bibfnamefont {A.~H.}\ \bibnamefont
  {Castro~Neto}}, \bibinfo {author} {\bibfnamefont {F.}~\bibnamefont {Guinea}},
  \bibinfo {author} {\bibfnamefont {N.~M.~R.}\ \bibnamefont {Peres}}, \bibinfo
  {author} {\bibfnamefont {K.~S.}\ \bibnamefont {Novoselov}}, \ and\ \bibinfo
  {author} {\bibfnamefont {A.~K.}\ \bibnamefont {Geim}},\ }\bibfield  {title}
  {\enquote {\bibinfo {title} {The electronic properties of graphene},}\ }\href
  {\doibase 10.1103/RevModPhys.81.109} {\bibfield  {journal} {\bibinfo
  {journal} {Rev. Mod. Phys.}\ }\textbf {\bibinfo {volume} {81}},\ \bibinfo
  {pages} {109} (\bibinfo {year} {2009})}\BibitemShut {NoStop}%
\bibitem [{\citenamefont {Thygesen}(2017)}]{thygesen2017calculating}%
  \BibitemOpen
  \bibfield  {author} {\bibinfo {author} {\bibfnamefont {K.~S.}\ \bibnamefont
  {Thygesen}},\ }\bibfield  {title} {\enquote {\bibinfo {title} {{Calculating
  excitons, plasmons, and quasiparticles in 2D materials and van der Waals
  heterostructures}},}\ }\href {\doibase 10.1088/2053-1583/aa6432} {\bibfield
  {journal} {\bibinfo  {journal} {2D Materials}\ }\textbf {\bibinfo {volume}
  {4}},\ \bibinfo {pages} {022004} (\bibinfo {year} {2017})}\BibitemShut
  {NoStop}%
\bibitem [{\citenamefont {Liu}\ \emph {et~al.}(2014)\citenamefont {Liu},
  \citenamefont {Neal}, \citenamefont {Zhu}, \citenamefont {Luo}, \citenamefont
  {Xu}, \citenamefont {Tománek},\ and\ \citenamefont {D.}}]{LiuYe2014}%
  \BibitemOpen
  \bibfield  {author} {\bibinfo {author} {\bibfnamefont {H.}~\bibnamefont
  {Liu}}, \bibinfo {author} {\bibfnamefont {A.~T.}\ \bibnamefont {Neal}},
  \bibinfo {author} {\bibfnamefont {Z.}~\bibnamefont {Zhu}}, \bibinfo {author}
  {\bibfnamefont {Z.}~\bibnamefont {Luo}}, \bibinfo {author} {\bibfnamefont
  {X.}~\bibnamefont {Xu}}, \bibinfo {author} {\bibfnamefont {D.}~\bibnamefont
  {Tománek}}, \ and\ \bibinfo {author} {\bibfnamefont {Y.~P.}\ \bibnamefont
  {D.}},\ }\bibfield  {title} {\enquote {\bibinfo {title} {{Phosphorene: An
  Unexplored 2D Semiconductor with a High Hole Mobility}},}\ }\href {\doibase
  10.1021/nn501226z} {\bibfield  {journal} {\bibinfo  {journal} {ACS Nano}\
  }\textbf {\bibinfo {volume} {8}},\ \bibinfo {pages} {4033–4041} (\bibinfo
  {year} {2014})}\BibitemShut {NoStop}%
\bibitem [{\citenamefont {Wy}\ \emph {et~al.}(2014)\citenamefont {Wy},
  \citenamefont {Soklaski}, \citenamefont {Liang},\ and\ \citenamefont
  {Yang}}]{TranYang2014}%
  \BibitemOpen
  \bibfield  {author} {\bibinfo {author} {\bibfnamefont {T.}~\bibnamefont
  {Wy}}, \bibinfo {author} {\bibfnamefont {R.}~\bibnamefont {Soklaski}},
  \bibinfo {author} {\bibfnamefont {Y.}~\bibnamefont {Liang}}, \ and\ \bibinfo
  {author} {\bibfnamefont {L.}~\bibnamefont {Yang}},\ }\bibfield  {title}
  {\enquote {\bibinfo {title} {{Layer-controlled band gap and anisotropic
  excitons in few-layer black phosphorus}},}\ }\href {\doibase
  10.1103/PhysRevB.89.235319} {\bibfield  {journal} {\bibinfo  {journal} {Phys.
  Rev. B}\ }\textbf {\bibinfo {volume} {89}},\ \bibinfo {pages} {235319}
  (\bibinfo {year} {2014})}\BibitemShut {NoStop}%
\bibitem [{\citenamefont {Knill}, \citenamefont {Laflamme},\ and\ \citenamefont
  {Milburn}(2001)}]{Knill2001}%
  \BibitemOpen
  \bibfield  {author} {\bibinfo {author} {\bibfnamefont {E.}~\bibnamefont
  {Knill}}, \bibinfo {author} {\bibfnamefont {R.}~\bibnamefont {Laflamme}}, \
  and\ \bibinfo {author} {\bibfnamefont {G.~A.}\ \bibnamefont {Milburn}},\
  }\bibfield  {title} {\enquote {\bibinfo {title} {{A scheme for efficient
  quantum computation with linear optics}},}\ }\href {\doibase
  10.1038/35051009} {\bibfield  {journal} {\bibinfo  {journal} {Nature}\
  }\textbf {\bibinfo {volume} {409}},\ \bibinfo {pages} {46–52} (\bibinfo
  {year} {2001})}\BibitemShut {NoStop}%
\bibitem [{\citenamefont {Zeng}\ \emph {et~al.}(2009)\citenamefont {Zeng},
  \citenamefont {Jiang}, \citenamefont {Gao}, \citenamefont {He},\ and\
  \citenamefont {Ma}}]{Zeng2009}%
  \BibitemOpen
  \bibfield  {author} {\bibinfo {author} {\bibfnamefont {N.}~\bibnamefont
  {Zeng}}, \bibinfo {author} {\bibfnamefont {X.}~\bibnamefont {Jiang}},
  \bibinfo {author} {\bibfnamefont {Q.}~\bibnamefont {Gao}}, \bibinfo {author}
  {\bibfnamefont {Y.}~\bibnamefont {He}}, \ and\ \bibinfo {author}
  {\bibfnamefont {H.}~\bibnamefont {Ma}},\ }\bibfield  {title} {\enquote
  {\bibinfo {title} {{Linear polarization difference imaging and its potential
  applications}},}\ }\href {\doibase 10.1364/AO.48.006734} {\bibfield
  {journal} {\bibinfo  {journal} {Applied Optics}\ }\textbf {\bibinfo {volume}
  {48}},\ \bibinfo {pages} {6734--6739} (\bibinfo {year} {2009})}\BibitemShut
  {NoStop}%
\bibitem [{\citenamefont {Ling}\ \emph {et~al.}(2016)\citenamefont {Ling},
  \citenamefont {Huang}, \citenamefont {Hasdeo}, \citenamefont {Liang},
  \citenamefont {Parkin}, \citenamefont {Tatsumi}, \citenamefont {Nugraha},
  \citenamefont {Puretzky}, \citenamefont {Das}, \citenamefont {Sumpter},
  \citenamefont {Geohegan}, \citenamefont {Kong}, \citenamefont {Saito},
  \citenamefont {Drndic}, \citenamefont {Meunier},\ and\ \citenamefont
  {Dresselhaus}}]{Dresselhaus2016}%
  \BibitemOpen
  \bibfield  {author} {\bibinfo {author} {\bibfnamefont {X.}~\bibnamefont
  {Ling}}, \bibinfo {author} {\bibfnamefont {S.}~\bibnamefont {Huang}},
  \bibinfo {author} {\bibfnamefont {E.~H.}\ \bibnamefont {Hasdeo}}, \bibinfo
  {author} {\bibfnamefont {L.}~\bibnamefont {Liang}}, \bibinfo {author}
  {\bibfnamefont {W.~M.}\ \bibnamefont {Parkin}}, \bibinfo {author}
  {\bibfnamefont {Y.}~\bibnamefont {Tatsumi}}, \bibinfo {author} {\bibfnamefont
  {A.~R.~T.}\ \bibnamefont {Nugraha}}, \bibinfo {author} {\bibfnamefont
  {A.~A.}\ \bibnamefont {Puretzky}}, \bibinfo {author} {\bibfnamefont {P.~M.}\
  \bibnamefont {Das}}, \bibinfo {author} {\bibfnamefont {B.~G.}\ \bibnamefont
  {Sumpter}}, \bibinfo {author} {\bibfnamefont {D.~B.}\ \bibnamefont
  {Geohegan}}, \bibinfo {author} {\bibfnamefont {J.}~\bibnamefont {Kong}},
  \bibinfo {author} {\bibfnamefont {R.}~\bibnamefont {Saito}}, \bibinfo
  {author} {\bibfnamefont {M.}~\bibnamefont {Drndic}}, \bibinfo {author}
  {\bibfnamefont {V.}~\bibnamefont {Meunier}}, \ and\ \bibinfo {author}
  {\bibfnamefont {M.~S.}\ \bibnamefont {Dresselhaus}},\ }\bibfield  {title}
  {\enquote {\bibinfo {title} {{Anisotropic Electron-Photon and Electron-Phonon
  Interactions in Black Phosphorus}},}\ }\href {\doibase
  10.1021/acs.nanolett.5b04540} {\bibfield  {journal} {\bibinfo  {journal}
  {Nano Letters}\ }\textbf {\bibinfo {volume} {16}},\ \bibinfo {pages}
  {2260--2267} (\bibinfo {year} {2016})}\BibitemShut {NoStop}%
\bibitem [{\citenamefont {Li}\ \emph {et~al.}(2015)\citenamefont {Li},
  \citenamefont {Santos}, \citenamefont {Xing}, \citenamefont {Cappelluti},
  \citenamefont {Roldán}, \citenamefont {Chen}, \citenamefont {Watanabe},\
  and\ \citenamefont {Taniguchi}}]{LiTaniguchi2016}%
  \BibitemOpen
  \bibfield  {author} {\bibinfo {author} {\bibfnamefont {L.~H.}\ \bibnamefont
  {Li}}, \bibinfo {author} {\bibfnamefont {E.~J.~G.}\ \bibnamefont {Santos}},
  \bibinfo {author} {\bibfnamefont {T.}~\bibnamefont {Xing}}, \bibinfo {author}
  {\bibfnamefont {E.}~\bibnamefont {Cappelluti}}, \bibinfo {author}
  {\bibfnamefont {R.}~\bibnamefont {Roldán}}, \bibinfo {author} {\bibfnamefont
  {Y.}~\bibnamefont {Chen}}, \bibinfo {author} {\bibfnamefont {K.}~\bibnamefont
  {Watanabe}}, \ and\ \bibinfo {author} {\bibfnamefont {T.}~\bibnamefont
  {Taniguchi}},\ }\bibfield  {title} {\enquote {\bibinfo {title} {{Dielectric
  Screening in Atomically Thin Boron Nitride Nanosheets}},}\ }\href {\doibase
  10.1021/nl503411a} {\bibfield  {journal} {\bibinfo  {journal} {Nano Lett.}\
  }\textbf {\bibinfo {volume} {15}},\ \bibinfo {pages} {218–223} (\bibinfo
  {year} {2015})}\BibitemShut {NoStop}%
\bibitem [{\citenamefont {Xu}\ \emph {et~al.}(2016)\citenamefont {Xu},
  \citenamefont {Zhang}, \citenamefont {Wang}, \citenamefont {Yang},
  \citenamefont {Wang}, \citenamefont {Pei}, \citenamefont {Myint},
  \citenamefont {Xing}, \citenamefont {Yu}, \citenamefont {Fu} \emph
  {et~al.}}]{xu2016extraordinarily}%
  \BibitemOpen
  \bibfield  {author} {\bibinfo {author} {\bibfnamefont {R.}~\bibnamefont
  {Xu}}, \bibinfo {author} {\bibfnamefont {S.}~\bibnamefont {Zhang}}, \bibinfo
  {author} {\bibfnamefont {F.}~\bibnamefont {Wang}}, \bibinfo {author}
  {\bibfnamefont {J.}~\bibnamefont {Yang}}, \bibinfo {author} {\bibfnamefont
  {Z.}~\bibnamefont {Wang}}, \bibinfo {author} {\bibfnamefont {J.}~\bibnamefont
  {Pei}}, \bibinfo {author} {\bibfnamefont {Y.~W.}\ \bibnamefont {Myint}},
  \bibinfo {author} {\bibfnamefont {B.}~\bibnamefont {Xing}}, \bibinfo {author}
  {\bibfnamefont {Z.}~\bibnamefont {Yu}}, \bibinfo {author} {\bibfnamefont
  {L.}~\bibnamefont {Fu}},  \emph {et~al.},\ }\bibfield  {title} {\enquote
  {\bibinfo {title} {Extraordinarily bound quasi-one-dimensional trions in
  two-dimensional phosphorene atomic semiconductors},}\ }\href {\doibase
  10.1021/acsnano.5b06193} {\bibfield  {journal} {\bibinfo  {journal} {ACS
  nano}\ }\textbf {\bibinfo {volume} {10}},\ \bibinfo {pages} {2046--2053}
  (\bibinfo {year} {2016})}\BibitemShut {NoStop}%
\bibitem [{\citenamefont {Yang}\ \emph {et~al.}(2015)\citenamefont {Yang},
  \citenamefont {Xu}, \citenamefont {Pei}, \citenamefont {Myint}, \citenamefont
  {Wang}, \citenamefont {Wang}, \citenamefont {Zhang}, \citenamefont {Yu},\
  and\ \citenamefont {Lu}}]{yang2015optical}%
  \BibitemOpen
  \bibfield  {author} {\bibinfo {author} {\bibfnamefont {J.}~\bibnamefont
  {Yang}}, \bibinfo {author} {\bibfnamefont {R.}~\bibnamefont {Xu}}, \bibinfo
  {author} {\bibfnamefont {J.}~\bibnamefont {Pei}}, \bibinfo {author}
  {\bibfnamefont {Y.~W.}\ \bibnamefont {Myint}}, \bibinfo {author}
  {\bibfnamefont {F.}~\bibnamefont {Wang}}, \bibinfo {author} {\bibfnamefont
  {Z.}~\bibnamefont {Wang}}, \bibinfo {author} {\bibfnamefont {S.}~\bibnamefont
  {Zhang}}, \bibinfo {author} {\bibfnamefont {Z.}~\bibnamefont {Yu}}, \ and\
  \bibinfo {author} {\bibfnamefont {Y.}~\bibnamefont {Lu}},\ }\bibfield
  {title} {\enquote {\bibinfo {title} {Optical tuning of exciton and trion
  emissions in monolayer phosphorene},}\ }\href {\doibase 10.1038/lsa.2015.85}
  {\bibfield  {journal} {\bibinfo  {journal} {Light: Science \& Applications}\
  }\textbf {\bibinfo {volume} {4}},\ \bibinfo {pages} {e312--e312} (\bibinfo
  {year} {2015})}\BibitemShut {NoStop}%
\bibitem [{\citenamefont {Deilmann}\ and\ \citenamefont
  {Thygesen}(2018)}]{deilmann2018unraveling}%
  \BibitemOpen
  \bibfield  {author} {\bibinfo {author} {\bibfnamefont {T.}~\bibnamefont
  {Deilmann}}\ and\ \bibinfo {author} {\bibfnamefont {K.~S.}\ \bibnamefont
  {Thygesen}},\ }\bibfield  {title} {\enquote {\bibinfo {title} {Unraveling the
  not-so-large trion binding energy in monolayer black phosphorus},}\ }\href
  {\doibase 10.1088/2053-1583/aadc28} {\bibfield  {journal} {\bibinfo
  {journal} {2D Materials}\ }\textbf {\bibinfo {volume} {5}},\ \bibinfo {pages}
  {041007} (\bibinfo {year} {2018})}\BibitemShut {NoStop}%
\bibitem [{\citenamefont {Gjerding}\ \emph {et~al.}(2020)\citenamefont
  {Gjerding}, \citenamefont {Cavalcante}, \citenamefont {Chaves},\ and\
  \citenamefont {Thygesen}}]{gjerding2020efficient}%
  \BibitemOpen
  \bibfield  {author} {\bibinfo {author} {\bibfnamefont {M.~N.}\ \bibnamefont
  {Gjerding}}, \bibinfo {author} {\bibfnamefont {L.~S.}\ \bibnamefont
  {Cavalcante}}, \bibinfo {author} {\bibfnamefont {A.}~\bibnamefont {Chaves}},
  \ and\ \bibinfo {author} {\bibfnamefont {K.~S.}\ \bibnamefont {Thygesen}},\
  }\bibfield  {title} {\enquote {\bibinfo {title} {{Efficient Ab-Initio Based
  Modeling of Dielectric Screening in 2D Van Der Waals Materials: Including
  Phonons, Substrates, and Doping}},}\ }\href {\doibase
  10.1021/acs.jpcc.0c01635} {\bibfield  {journal} {\bibinfo  {journal} {The
  Journal of Physical Chemistry C}\ }\textbf {\bibinfo {volume} {124}},\
  \bibinfo {pages} {11609--11616} (\bibinfo {year} {2020})}\BibitemShut
  {NoStop}%
\bibitem [{\citenamefont {Langreth}\ and\ \citenamefont {Perdew}(1975)}]{RPA}%
  \BibitemOpen
  \bibfield  {author} {\bibinfo {author} {\bibfnamefont {D.}~\bibnamefont
  {Langreth}}\ and\ \bibinfo {author} {\bibfnamefont {J.}~\bibnamefont
  {Perdew}},\ }\bibfield  {title} {\enquote {\bibinfo {title} {The
  exchange-correlation energy of a metallic surface},}\ }\href {\doibase
  10.1016/0038-1098(75)90618-3} {\bibfield  {journal} {\bibinfo  {journal}
  {Solid State Communications}\ }\textbf {\bibinfo {volume} {17}},\ \bibinfo
  {pages} {1425--1429} (\bibinfo {year} {1975})}\BibitemShut {NoStop}%
\bibitem [{\citenamefont {Olsen}\ and\ \citenamefont {Thygesen}(2013)}]{RPA_2}%
  \BibitemOpen
  \bibfield  {author} {\bibinfo {author} {\bibfnamefont {T.}~\bibnamefont
  {Olsen}}\ and\ \bibinfo {author} {\bibfnamefont {K.~S.}\ \bibnamefont
  {Thygesen}},\ }\bibfield  {title} {\enquote {\bibinfo {title} {{Random phase
  approximation applied to solids, molecules, and graphene-metal interfaces:
  From van der Waals to covalent bonding}},}\ }\href {\doibase
  10.1103/PhysRevB.87.075111} {\bibfield  {journal} {\bibinfo  {journal} {Phys.
  Rev. B}\ }\textbf {\bibinfo {volume} {87}},\ \bibinfo {pages} {075111}
  (\bibinfo {year} {2013})}\BibitemShut {NoStop}%
\bibitem [{C2D()}]{C2DB}%
  \BibitemOpen
  \href@noop {} {\enquote {\bibinfo {title} {{Computational 2D Materials
  Database (C2DB)}},}\ }\bibinfo {howpublished}
  {\url{https://cmr.fysik.dtu.dk/c2db/c2db.html}}\BibitemShut {NoStop}%
\end{thebibliography}%

\end{document}